\documentclass[fleqn,usenatbib]{mnras}

\usepackage{newtxtext,newtxmath}

\usepackage[T1]{fontenc}

\DeclareRobustCommand{\VAN}[3]{#2}
\let\VANthebibliography\thebibliography
\def\thebibliography{\DeclareRobustCommand{\VAN}[3]{##3}\VANthebibliography}

\usepackage{graphicx}	
\usepackage{amsmath}	

\newcommand{\be}{\begin{equation}}
\newcommand{\ee}{\end{equation}}
\newcommand{\ba}{\begin{eqnarray}}
\newcommand{\ea}{\end{eqnarray}}
\newcommand{\ci}{\mbox{[\,C\,{\sc i}\,}]\,}
\newcommand{\cii}{\mbox{[\,C\,{\sc ii}\,}]\,}
\newcommand{\msolm}{{\rm M}_\odot}

\title[The Multi-Phase ISM in the First Two Billion Years]{A High-Resolution Investigation of the Multi-Phase ISM in a Galaxy during the First Two Billion Years}

\author[S. Dye et al.]{
S. Dye,$^{1}$\thanks{E-mail: simon.dye@nottingham.ac.uk}
S. A. Eales,$^{2}$
H. L. Gomez,$^{2}$
G. C. Jones,$^{3,4}$
M.W.L. Smith,$^{2}$
E. Borsato,$^{5}$
A. Moss,$^{1}$
\newauthor
L. Dunne,$^{2}$
J. Maresca,$^{1}$
A. Amvrosiadis,$^{6}$
M. Negrello,$^{2}$
L. Marchetti,$^{7,8}$
E. M. Corsini,$^{5,9}$
R. J. Ivison,$^{10}$
\newauthor
G. J. Bendo,$^{11}$
T. Bakx,$^{12,13}$
A. Cooray,$^{14}$
P. Cox,$^{15}$
H. Dannerbauer,$^{16,17}$
S. Serjeant,$^{18}$
D. Riechers,$^{19}$
\newauthor
P. Temi,$^{20}$
C. Vlahakis$^{21}$
\\
$^{1}$School of Physics and Astronomy, University of Nottingham, University Park, Nottingham, NG7 2RD, UK\\
$^{2}$School of Physics and Astronomy, Cardiff University, The Parade, Cardiff, CF24 3AA, UK\\
$^{3}$Cavendish Laboratory, University of Cambridge, 19 J. J. Thomson Ave., Cambridge, CB3 0HE, UK\\
$^{4}$Kavli Institute for Cosmology, University of Cambridge, Madingley Road, Cambridge, CB3 0HA, UK\\
$^{5}$Dipartimento di Fisica e Astronomia "G. Galilei", Universit\`a di Padova, vicolo dell'Osservatorio 3, I-35122 Padova, Italy\\
$^{6}$ICC, Department of Physics, Durham University, South Road, Durham DH1 3LE, UK\\
$^{7}$Department of Astronomy, University of Cape Town, Private Bag X3, Rondebosch 7701, Cape Town, South Africa\\
$^{8}$ INAF - Instituto di Radio astronomia, via Gobetti 101, I-40129 Bologna, Italy\\
$^{9}$INAF - Osservatorio Astronomico di Padova, vicolo dell'Osservatorio 2, I-35122 Padova, Italy\\
$^{10}$European Southern Observatory, Karl-Schwarzschild-Strasse 2, D-85748 Garching, Germany\\
$^{11}$UK ALMA Regional Centre Node, Jodrell Bank Centre for Astrophysics, Department of Physics and Astronomy, The University of Manchester, \\ \hspace{2.5mm}Oxford Road, Manchester M13 9PL, UK\\
$^{12}$Division of Particle and Astrophysical Science, Graduate School of Science, Nagoya University, Aichi 464-8602, Japan.\\
$^{13}$National Astronomical Observatory of Japan, 2-21-1, Osawa, Mitaka, Tokyo 181-8588, Japan\\
$^{14}$Department of Physics and Astronomy, University of California, Irvine, CA92697, USA\\
$^{15}$IRAM, 300 rue de la piscine, F-38406 Saint-Martin d’H\`{e}res, France\\
$^{16}$Instituto de Astrof\'{i}sica de Canarias (IAC), E-38205 La Laguna, Tenerife, Spain\\
$^{17}$Universidad de La Laguna, Dpto. Astrof\'{i}sica, E-38206 La Laguna, Tenerife, Spain\\
$^{18}$School of Physical Sciences, The Open University, Milton Keynes, MK7 6AA, UK\\
$^{19}$Cornell University, Space Sciences Building, Ithaca, NY 14853, USA\\
$^{20}$Astrophysics Branch, NASA Ames Research Center, Moffett Field, CA 94035, USA\\
$^{21}$National Radio Astronomy Observatory, 520 Edgemont Road, Charlottesville, VA 22903-2475, USA\\
}

\date{Accepted XXX. Received YYY; in original form ZZZ}

\pubyear{2022}

\begin{document}
\label{firstpage}
\pagerange{\pageref{firstpage}--\pageref{lastpage}}
\maketitle

\begin{abstract}
We have carried out the first spatially-resolved investigation of the multi-phase interstellar medium (ISM) at high redshift, using the $z=4.24$ strongly-lensed sub-millimetre galaxy H-ATLASJ142413.9$+$022303 (ID141). We present high-resolution (down to $\sim$350 pc) ALMA observations in dust continuum emission and in the CO(7-6), $\rm H_2O (2_{1,1} - 2_{0,2})$, \ci(1-0) and \ci(2-1) lines, the latter two allowing us to spatially resolve the cool phase of the ISM for the first time. Our modelling of the kinematics reveals that the system appears to be dominated by a rotationally-supported gas disk with evidence of a nearby perturber. We find that the \ci(1-0) line has a very different distribution to the other lines, showing the existence of a reservoir of cool gas that might have been missed in studies of other galaxies. We have estimated the mass of the ISM using four different tracers, always obtaining an estimate in the range $\rm 3.2-3.8 \times 10^{11}\ M_{\odot}$, significantly higher than our dynamical mass estimate of $\rm 0.8-1.3 \times 10^{11}\ M_{\odot}$. We suggest that this conflict and other similar conflicts reported in the literature is because the gas-to-tracer ratios are $\simeq$4 times lower than the Galactic values used to calibrate the ISM in high-redshift galaxies. We demonstrate that this could result from a top-heavy initial mass function and strong chemical evolution. Using a variety of quantitative indicators, we show that, extreme though it is at $z=4.24$, ID141 will likely join the population of quiescent galaxies that appears in the Universe at $z\sim 3$.

\end{abstract}



\section{Introduction}

Sub-millimetre galaxies (SMGs) are among the brightest objects observed at far-infrared (far-IR) and sub-millimetre (submm) wavelengths in the early universe. SMGs are typically highly optically-obscured by dust that originates from low/intermediate mass stars and supernovae \citep[e.g.,][]{rowlands2014} in regions undergoing extreme rates of star formation. These stellar nurseries are most abundant at redshifts around $z=2-4$ when the Universe was most rapidly forming the galaxies and clusters we see around us today. SMGs have therefore become a valuable probe of the physics of the growth of these structures, particularly at the more massive end of the galaxy mass function \citep[e.g.,][]{casey2014,hodge2020}.

The most extreme SMGs are those discovered in the wide-area surveys carried out with {\it Herschel} \citep{negrello2017,bakx2018}, the South Pole Telescope \citep{vieira2013,reuter2020} and {\it Planck} \citep{canameras2015}, including ones that have star-formation rates many thousands of solar masses per year, which have been found out to a redshift of $\sim6$ \citep[e.g.,][]{fudamoto2017}. An obvious question is, what do these extreme objects evolve into at more recent epochs? In particular, do these galaxies evolve into the population of quiescent galaxies that appear in the Universe at $z \sim 3$? \citep[e.g.,][]{glazebrook2017,girelli2019,valentino2020,forrest2020a,forrest2020b,merlin2019}. 

These huge star-formation rates must be fueled by large amounts of gas and many studies have found that SMGs house substantial molecular gas reservoirs of masses generally in the range $10^{10}-10^{11}\,\msolm$ \citep[e.g.,][]{ivison2011,bothwell2013,huynh2017,dye2018,kaasinen2019}. The bulk constituent of this gas, molecular hydrogen, is not rotationally excited at the typical temperatures in the molecular phase in galaxies. As such, investigations have been forced to use tracers, either dust or a variety of molecular and atomic species, thus requiring knowledge of the gas-to-tracer ratio. There is a complex hierarchy of calibrations but most of the values assumed for these ratios are ultimately based on observations in the Milky Way, the only place where it is possible to measure both the luminosity in the tracer and the mass of the underlying gas \citep{dunne2021a}.

As for galaxies in the nearby Universe, the most frequently used tracer of the molecular phase of the ISM, because it has so many bright spectral lines in atmospheric windows, is the CO molecule \citep{tacconi2018,birkin2021}. As a tracer of the underlying gas, CO has some well known disadvantages \citep{bolatto2013}, in particular the fact that the J=1-0 line, the best for tracing the coldest gas, is optically thick. In addition, observations with {\it Fermi} \citep{abdo2010}, {\it Planck} \citep{planck2011} and {\it Herschel} \citep{pineda2013} all suggest that approximately one third of the molecular gas in the Galaxy does not emit CO, probably because of photo-dissociation of the CO molecule \citep{planck2011}. For high-resolution studies of the ISM in high-redshift galaxies there is the additional problem that the Atacama Large Millimetre Array (ALMA) is only able to observe the higher-excitation CO lines, making it possible that these studies have missed much of the gas (although lower-frequency ALMA bands will become available in the near future).

The other bright molecular lines in high-redshift galaxies are those of the water molecule, which have been used to observe both SMGs \citep{omont2013,spilker2014,yang2016,jarugula2019} and AGN \citep[e.g.][]{vanderwerf2011}. The water lines are excited by far-IR photons (compared to collisional excitations of CO by H$_2$),  explaining the correlation between line luminosity and far-infrared luminosity over three orders of magnitude \citep{yang2013}. The water lines are therefore probably more useful as a tracer of star formation than the ISM \citep{jarugula2019}.

Arguably the best tracers of the molecular phase of the ISM in high-redshift galaxies are the fine-structure lines of atomic carbon, despite being generally fainter. The \ci($^3P_1-^3P_0$) and \ci($^3P_2-^3P_1$) lines have similar excitation temperatures and critical densities to the low-J CO lines, but they are optically thin and they are emitted in all molecular regions including CO-dark regions \citep{papadop2004,papadop2004b,dunne2021a}.

The high-resolution configurations of ALMA, plus the magnification (often greater than a factor of 10) provided by gravitational lensing, have huge potential for investigating the physical conditions in the ISM in galaxies in the early Universe. This potential has been demonstrated quite dramatically using observations of the gravitationally-lensed SMG SDP81 that were carried out in the Science Verification Phase of the ALMA long baselines \citep{vlahakis2014}. After being corrected for gravitational lens distortion, the reconstructed spectral-line and continuum images had a resolution of $\sim$50 pc \citep{dye2015,rybak2015a,rybak2015b}. The images revealed a smoothly-rotating but clumpy disk, with a Toomre Q-parameter  suggesting that the disk is in a state of collapse. The resolution, approximately the size of a giant molecular cloud, made it possible to compare the Larson relations for the clumps with those in the Milky Way \citep{swinbank2015}.

The drawback of the observations of SDP81, however, is that they were in high-J CO lines, making it uncertain whether a cool molecular phase had been missed; it is possible that these high-J CO clumps simply trace regions in the disk where the gas is particularly warm rather than genuine agglomerations of gas. In addition, at the redshift of SDP81, several spectral lines do not fall within atmospherically-transparent windows. To circumvent these shortfalls, we have therefore embarked on a second high-resolution study of the ISM using a galaxy at a redshift such that all of the key spectral lines are visible to ALMA.

One of the brightest strongly-lensed SMGs, H-ATLAS J$142413.9+022303$, hereafter referred to as 'ID141' following its label assigned by \cite{cox2011} was discovered in the {\it Herschel} Astrophysical Terrahertz Large-Area Survey \citep{eales2010,cox2011,negrello2017}. ID141 is not only bright because it is strongly-lensed but also because it is intrinsically very luminous. After correction for the lens magnification factor of $\sim 6$, the star-formation rate in ID141, estimated from the bolometric luminosity of the dust \citep[and assuming the initial mass function of][]{kroupa2003}, is $\rm \simeq 2400\ M_{\odot}\ yr^{-1}$, five times higher than SDP81. With a redshift of $z=4.24$ \citep{cox2011}, ID141 is also one of the most distant strongly-lensed SMGs. At this redshift, both \ci lines, the bright \cii\ line at 158 $\mu$m, multiple CO lines (including the CO(1-0) line in the near future) and water lines, and many other lines are visible to ALMA.

As a result of its interest, ID141 has been observed many times in the submillimetre/millimetre wavebands \citep{cox2011,bussmann2012,bussmann2013,omont2013,dye2018,cheng2020}. Lens modelling by \citet[][D18 hereafter]{dye2018} and \citet{enia2018} revealed evidence of a disturbed morphology in the reconstructed continuum emission of the source (see also \citealt{bussmann2012} who showed that significant image residuals remain after subtracting the best fit lensed image of a single smooth source). A relatively high level of ionisation in the source was inferred from a high far-IR luminosity to gas mass ratio measured by \cite{cox2011}. This, along with very large line widths of $\sim 800$\,km\,s$^{-1}$ observed in CO emission led the authors to suggest the SMG is a merger, like most other SMGs \citep[e.g.,][]{engel2010}.

In this paper, we present high-resolution ($\sim 0.1$\,arcsec) ALMA observations of ID141 in bands 3 and 4, complementing the ALMA band 7 observations presented in D18. The new observations cover the emission lines CO(7-6), $\rm H_2O (2_{1,1} - 2_{0,2})$, \ci($^3P_2-^3P_1$) and \ci($^3P_1-^3P_0$), which have for the first time made it possible to carry out a high-resolution investigation of the ISM in a high-redshift galaxy that includes both the warm and the cool gas. The spectral-line observations have also allowed us to make a detailed kinematic analysis of this system, in particular addressing the question of whether ID141 is a merger. 

One of the key findings of this paper is that the dynamical mass of ID141 is approximately four times lower than the gas mass inferred from Galactic gas-to-tracer ratios. SMGs with inferred gas masses that are equal to or larger than their dynamical mass have been reported in the literature for over a decade \citep[e.g.][]{Tacconi2008,hodge2012,bothwell2013}. More recently, there has been an acceleration in the prevalence of such conflicting systems \citep[e.g.][]{dye2015,yang2017, rivera2018,neeleman2020,mizukoshi2021}. A common approach to reconcile these anomalies is to reduce the conversion factor (based on local Universe observations) that scales CO line intensity to total gas mass. In this work, we find that four different tracers yield the same excessive gas mass. It is unlikely that all four tracer conversion factors would need to be scaled by the same amount and so we have sought a more natural explanation in the form of a chemical evolution model which solves the discrepancy between the gas mass and dynamical mass in ID141.

The layout of this paper is as follows. In Section \ref{sec_data} we describe the acquisition and the reduction of the ALMA data. Section \ref{sec_lens} presents our lens modelling and source reconstruction. Section \ref{sec_src} describes our investigation of the basic properties of the source, including the kinematic modelling and estimates of the mass of the ISM in the galaxy made five different ways. We discuss our findings in Section \ref{sec_discussion} and introduce a chemical evolution model of ID141. Section \ref{sec_summary} is a summary of both our basic observational results and inferences from these results. Throughout this work, we have adopted the flat cosmological model with parameters $h=0.678$, $\Omega_{\rm m}=0.308$ and $\Omega_\Lambda=0.692$ \citep{planck2016}.

\section{Data Acquisition and Processing}
\label{sec_data}

The ALMA data analysed in this paper were acquired under two separate ALMA programmes. The first, 2016.1.00450.S, obtained band 4 spectral imaging on 20th August 2017 in four spectral windows (SPWs) centred on the frequencies 141.2, 143.1, 153.3 and 154.9\,GHz, each with bandwidths of approximately 2\,GHz. The total on-source integration time was 2360s using 44 12m antennas giving a spatial resolution of 0.15\,arcsec, a maximum recoverable scale of 2.1\,arcsec and a sensitivity of 0.34\,mJy\,beam$^{-1}$ assuming 10\,km\,s$^{-1}$ channels. 

The second programme, 2017.1.00029.S, obtained band 3 spectral imaging on the 14th and 16th of November 2017 when 46 and 44 12m antennas were used respectively. Combining the measurement sets from both dates, the resulting images have a spatial resolution of 0.11\,arcsec, a maximum recoverable scale of 2.0\,arcsec and a sensitivity of 0.56\,mJy\,beam$^{-1}$ assuming 10\,km\,s$^{-1}$ channels. These observations were again carried out in four SPWs with central frequencies of 93.9, 95.8, 105.8 and 107.7\,GHz with bandwidths of $\sim$2\,GHz and with a total on-source integration time of 3560s. This configuration of SPWs was chosen specifically to target the emission lines of CO, H$_2$O and \ci (see below). Table~\ref{tab_alma_obs} gives more specific details on the spectral set up.

\begin{table}
\centering
\caption{Configuration of ALMA used for spectral imaging of ID141. The table lists the central frequency, $\nu_{\rm cent}$, the band width, BW, and the number of channels, $N_{\rm chan}$, for each spectral window.}
\label{tab_alma_obs}
\begin{tabular}{cccc}
\hline
Band & $\nu_{\rm cent}$\,(GHz) & BW\,(MHz) & $N_{\rm chan}$ \\
\hline
3 & 93.9 & 1875 & 3840 \\
3 & 95.8 & 2000 & 128 \\
3 & 105.8 & 1875 & 3840 \\
3 & 107.7 & 2000 & 128 \\
4 & 141.2 & 2000 & 128 \\
4 & 143.1 & 2000 & 128 \\
4 & 153.3 & 1875 & 3840 \\
4 & 154.9 & 1875 & 3840 \\
\hline
\end{tabular}
\end{table}

Calibration of both the band 3 and band 4 data was carried out using the scripts provided by the ALMA science archive using version 5.1.1 for band 3 and version 4.7.2 for band 4 of the Common Astronomy Software Applications ({\tt CASA}) software package \citep{mcmullin2007}. Bandpass and phase calibration was provided via the calibrators J1337-1257 and J1410+0203 respectively.
For band 3, the two measurement sets (MSs) were independently calibrated and then combined into one MS using {\tt CASA}. We manually verified that no erroneous visibilities remained after calibration using {\tt CASA}'s {\tt plotms} tool. All observed images of ID141 displayed in this paper were produced by cleaning these MSs using {\tt CASA} with natural weighting and a pixel scale of 0.024\,arcsec. 
To isolate emission lines (see Section \ref{sec_obslines}), we used the {\tt uvcontsub} tool within {\tt CASA} for each SPW to fit a linear spectrum to channels not containing line emission and then subtract this fit from the full visibility data. A separate set of visibilities was created for analysis of the continuum emission by excluding channels containing line emission.

\begin{figure*}
	\includegraphics[width=17.8cm]{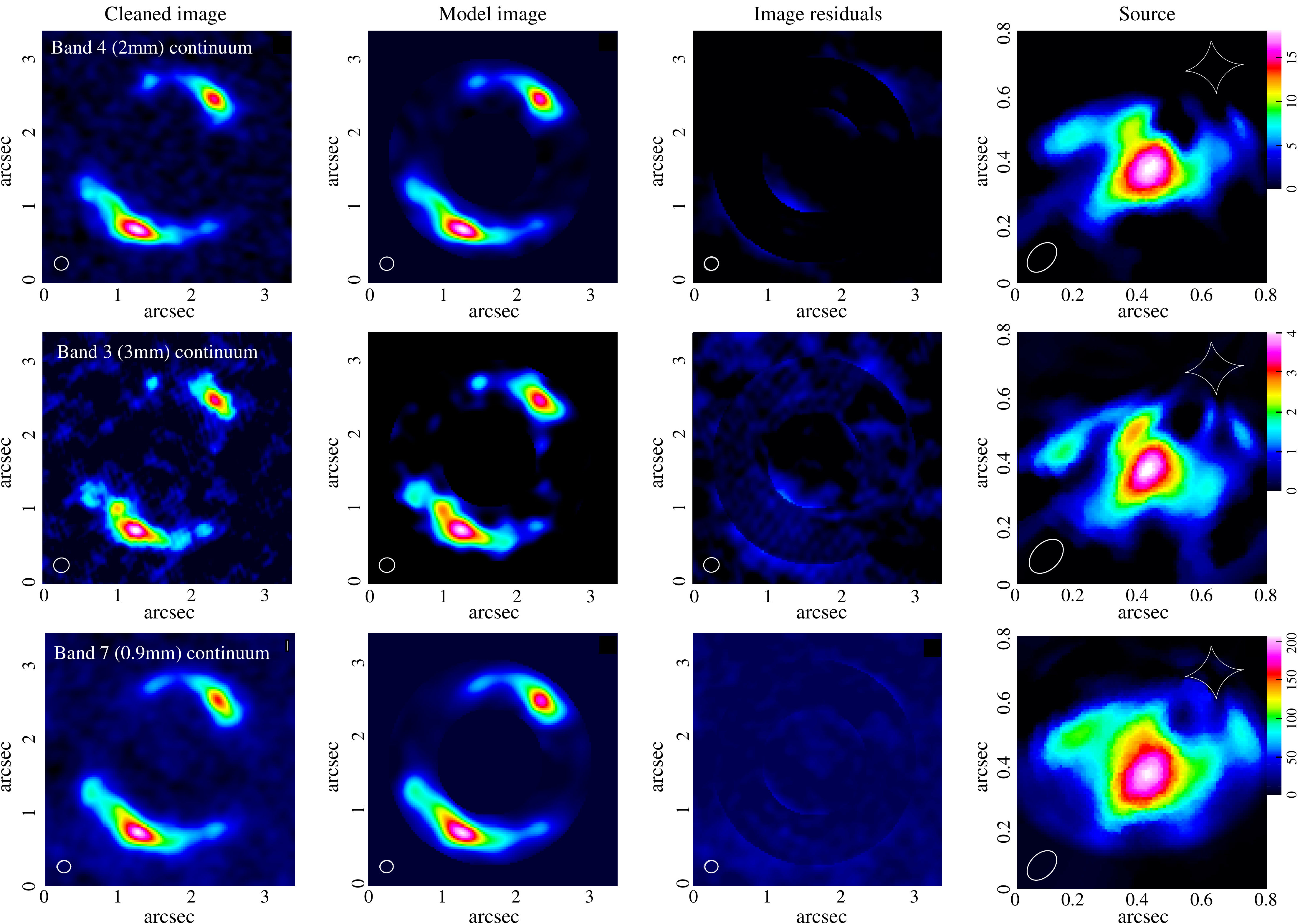}
    \caption{From left to right, columns show observed cleaned images, model lensed images of the reconstructed source emission, residuals (observed minus model) and reconstructed source (from the visibilities -- the caustic is shown in white) of the continuum emission in band 3, 4 and 7. Observed images have been cleaned with a natural weighting with a pixel scale of 0.024\,arcsec. The white ellipse shown in the lower left corners of the plots indicates the FWHM of the beam. (For the source plane, the effective central beam is plotted; see Section 3.2.) The colour wedge at the far right indicates the surface brightness in units of mJy\,arcsec$^{-2}$.}
    \label{cont_images}
\end{figure*}

In addition to the band 3 and 4 ALMA data detailed above, we have also used the ALMA band 7 data acquired under the ALMA programme 2013.1.00358.S and analysed by D18 (see Fig. \ref{cont_images}). The observations in this programme amount to a total on-source integration time of 151s within four spectral windows spanning 336 to 352\,GHz. The spectral windows contain no detectable emission lines and so we use the resulting 0.12\,arcsec resolution continuum image solely for the purposes of morphological comparison of the reconstructed lensed SMG. We refer the reader to D18 for further details regarding these observations. Upon re-cleaning this dataset, we identified an error made in D18 such that the flux was over-estimated by approximately 30 per cent. The updated flux measured from the lensed image is $91\pm 5$\,mJy.

\subsection{Emission lines}
\label{sec_obslines}

\begin{figure*}
	\includegraphics[width=17.8cm]{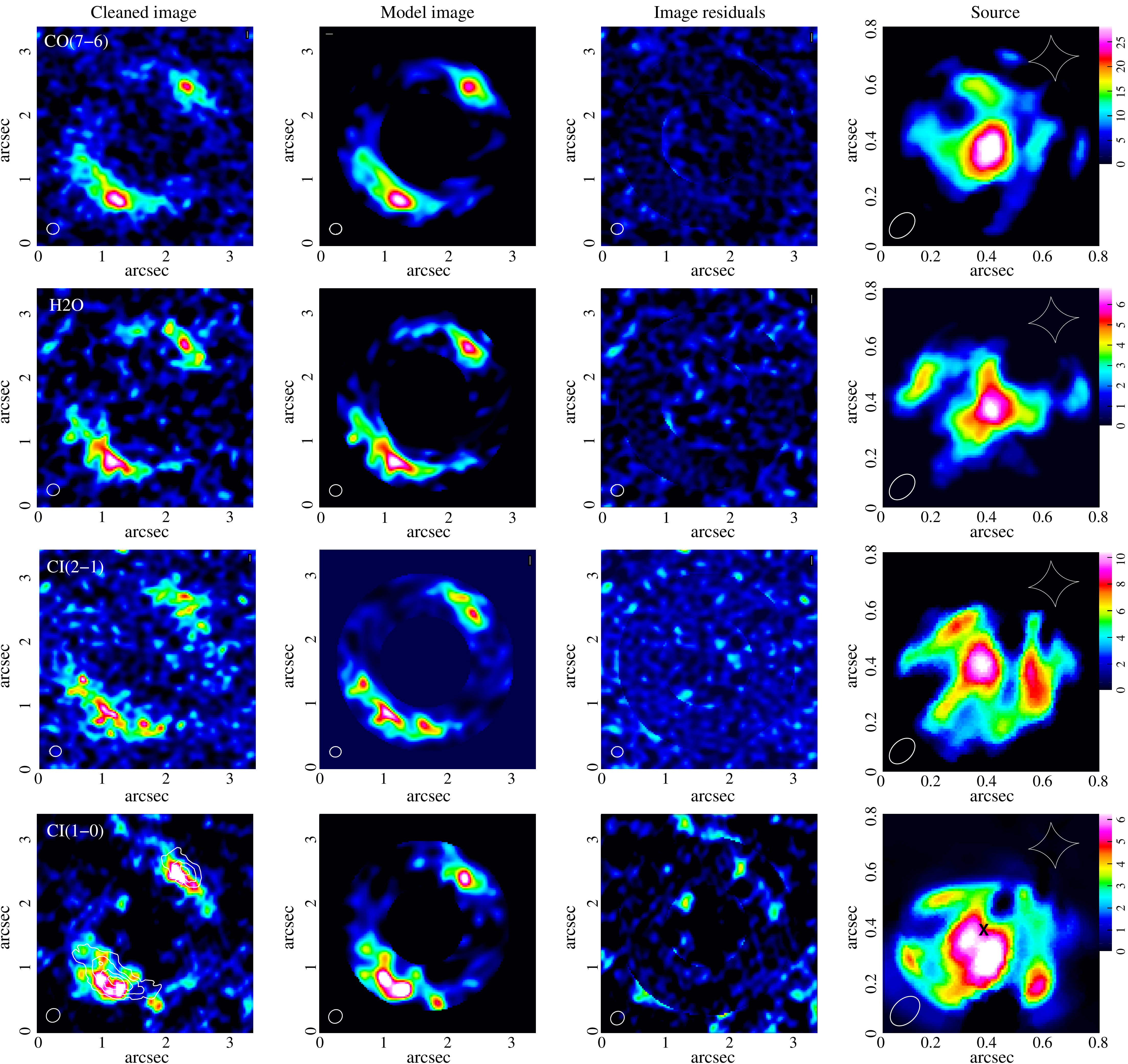}
    \caption{From left to right, columns show observed cleaned images, model lensed images of the reconstructed source emission, residuals (observed minus model) and reconstructed source (from the visibilities -- the caustic is shown in white) of the four emission lines, CO(7-6), H$_2$O, \ci(2-1) and \ci(1-0) identified in our observations. Observed images have been cleaned with a natural weighting with a pixel scale of 0.024\,arcsec. The white ellipse shown in the lower left corners of the plots indicates the FWHM of the beam. (For the source plane, the effective central beam is plotted; see Section 3.2.) The black cross in the panel showing the reconstructed \ci(1-0) line emission (bottom-right) shows the location of the peak reconstructed \ci(2-1) emission. The white contours in the observed \ci(1-0) line map (bottom-left) which follow the band 3 the continuum flux illustrate that the \ci(1-0) is spatially offset. The colour wedge at the far right indicates the line surface brightness in units of Jy\,km\,s$^{-1}$\,arcsec$^{-2}$.}
    \label{line_images}
\end{figure*}

\begin{figure*}
	\includegraphics[width=17.8cm]{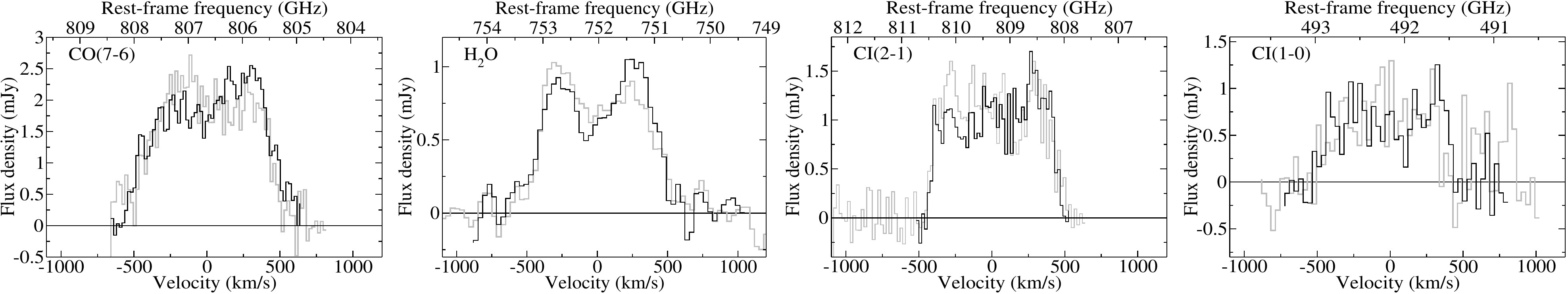}
    \caption{Emission lines, H$_2$O($2_{1,1}-2_{0,2}$), CO(7-6), \ci(2-1) and \ci(1-0), measured within a masked area determined by selecting significant pixels in their cleaned observed images (grey) and the corresponding line profile in their reconstructed source plane cubes (black). Image plane line profiles have been corrected for total lens magnification. Differential magnification causes the lensed image profiles to contain more flux at negative velocities even though intrinsically the source emits more flux at positive velocities.}
    \label{lines}
\end{figure*}

We identified four emission lines: H$_2$O($2_{1,1}-2_{0,2}$) ($\nu_{\rm rest}=752.033$\,GHz), CO(7-6) ($\nu_{\rm rest}=806.652$\,GHz), \ci($^3P_2-^3P_1$) ($\nu_{\rm rest}=809.342$\,GHz, referred to as \ci(2-1) hereafter) and \ci($^3P_1-^3P_0$) ($\nu_{\rm rest}=492.161$\,GHz, referred to as \ci(1-0) hereafter). Observed images of these lines are shown in Fig.~\ref{line_images} and the corresponding observed lensed line profiles (measured within a mask determined by selecting significant pixels) are shown in Fig.~\ref{lines}. 

Table \ref{tab_zsource} lists the four line fluxes measured in the observed (i.e. lensed) images. The CO(7-6), \ci(1-0) and \ci(2-1) lines were also detected by \cite{cox2011}. The line flux measured for the \ci(1-0) line measured by \cite{cox2011} is consistent with our measurement, but the Cox et al. fluxes of the other two lines are nearly a factor of two lower than those measured by ALMA. Although the discrepancies are larger than expected, the errors quoted in \cite{cox2011} do not include any uncertainty due to continuum subtraction which are large enough to eliminate any significant disagreement (R. Neri, private communication).

Fig.~\ref{lines} also shows the emission lines determined from the reconstructed source plane emission (see Section \ref{sec_srcpix}). In this case, we extracted the line emission from the source plane region that, according to the lens model, maps to the image plane mask used to determine the image plane emission. It is clear from the three lines with the highest signal-to-noise ratio (SNR), i.e. H$_2$O, CO(7-6) and \ci(2-1), that emission lines extracted from the image have slightly stronger flux at negative velocities whereas the reverse is true in the lines extracted from the source plane. Our lensing model confirms that this is due to differential magnification; the region of the source responsible for the emission of the negative side of the lines lies closer to the lens caustic where the magnification is higher.

Taking the median observed frequency of each of the H$_2$O, CO(7-6) and \ci(2-1) lines measured from the reconstructed source and calculating the mean of these weighted by their SNR results in an inferred source redshift of $z=4.24417\pm0.00003$ (see Table \ref{tab_zsource} - note that we did not include the \ci(1-0) line in this measurement due to its significantly lower SNR). This measurement is consistent with the redshift of $z=4.243\pm0.001$ determined by \cite{cox2011}.

\begin{table}
\centering
\caption{Emission lines detected in the band 3 and band 4 ALMA data. The rest-frame frequencies, $\nu_{\rm rest}$, are measured from the reconstructed source cubes and therefore eliminate differential magnification of the emission lines. The SNR-weighted mean of the redshifts implied by the H$_2$O, CO(7-6) and \ci(2-1) lines is $z=4.24417\pm0.00003$. The final column gives the line flux in the observed (i.e. lensed) image.}
\label{tab_zsource}
\begin{tabular}{ccccc}
\hline
Line & $\nu_{\rm rest}$ (GHz) & $\nu_{\rm obs}$(GHz) & $z_{\rm inferred}$ & flux (Jy\,km\,s$^{-1}$) \\
\hline
H$_2$O & 752.033 & 143.403 & 4.24419 & $4.1\pm 0.4$ \\
CO(7-6) & 806.652 & 153.820 & 4.24413 & $11.4\pm 1.1$ \\
\ci(2-1) & 809.342 & 154.332 & 4.24416 & $6.2\pm 0.8$ \\
\ci(1-0) & 492.161 & 93.8479 & (4.24424) &  $3.1 \pm 0.4 $\\
\hline
\end{tabular}
\end{table}

\section{Lens modelling}
\label{sec_lens}

To recover an undistorted image of the lensed SMG, we applied the modelling process of D18. Using a given model for the distribution of mass in the foreground lensing galaxy, this process reconstructs a pixelised map of the source surface brightness \citep[see][]{warren2003}. In this paper, we modelled solely the interferometric visibilities. The process works by computing a set of model visibilities from the lensed image of each source pixel. We computed these visibilities using {\tt uvmodel} contained within the {\tt MIRIAD} software suite \citep{sault2011}. The linear combination of the sets of model visibilities that best fits the observed visibilities then recovers the source surface brightness distribution for a given lens model. The lens model is optimised iteratively, repeating the linear inversion with every iteration of the lens model. In this work, lens model optimisation was performed with {\tt NeuralNest} \citep{moss2020} which uses a neural network to generate highly efficient proposals for Markov Chain Monte Carlo sampling.

\subsection{Lens Model}
\label{sec_lensmodel}

Following the lens modelling in D18, we used an elliptical power-law density profile for the foreground lens, using the form given by \cite{kassiola1993}
which has a surface mass density
\be
\kappa=\kappa_0\,({\tilde r}/{\rm 1kpc})^{1-\alpha} \, .
\ee
where $\kappa_0$ is the normalisation surface mass density and $\alpha$ is the power-law index of the volume mass density profile. The elliptical radial co-ordinate, $\tilde r$, is defined in terms of the Euclidean co-ordinates, $(x.y)$ relative to the ellipse centre, and the elongation, $\epsilon$ (i.e., the ratio of semi-major to semi-minor axis length), by $\tilde{r}^2 = x^2 + y^2 / \epsilon^2$. The orientation of the semi-major axis, $\theta$ is measured east of north and the co-ordinates of the centre of the lens in the image plane are $(x_c,y_c)$. Initial lens modeling tests with a model including external shear were attempted but it was found that negligible shear was required in the best fit model. This agrees with the findings of D18 and hence a parameterisation of external shear was not included in the final lens model which comprised six parameters in total. We also tried a lens model comprising two separate mass components as used by \cite{bussmann2012} and as in D18, we found no significant improvement in the fit compared to a single ellipsoidal power-law. Although the two models give rise to a difference in the total source plane magnification of approximately 5 per cent, this is negligible in comparison to the errors on the intrinsic source properties we derive (see Section \ref{sec_ism_vs_gas_mass} for further discussion). Additionally, the two models yield almost identical source morphology and kinematical properties.

To determine our optimal lens model, we simultaneously fitted the visibilities from all four SPWs in the band 4 continuum data from this paper and the band 7 visibilities from D18. Combining bands and including the continuum in this way maximises the signal-to-noise ratio (SNR) to give the best possible lens model. We used the band 7 visibilities as presented in D18 but we time-averaged the more numerous band 4 visibilities into bins of 20\,s to improve computational efficiency. We found that the size of this temporal bin can be changed by a factor of several without making any noticeable difference to the reconstructed lens model or source. We obtained a model that improves upon the model presented in D18 when applied solely to the band 7 visibilities. We attribute this to a combination of the new optimisation strategy (i.e., {\tt NeuralNest}) used in this work and the complicated parameter space that results from randomising the source plane pixelisation (see Section \ref{sec_srcpix}). 

The parameters of our best-fit lens model are: $\kappa_0=(0.60\pm0.01)\times 10^{10}\msolm\,{\rm kpc}^{-2}$, $\epsilon=1.14\pm0.02$, $\theta=(96\pm2)^\circ$, $\alpha=2.01\pm0.03$. Comparing against the model in D18, there are two significant changes; the elongation has increased from 1.07 to 1.14 and the lens orientation has increased from 85$^\circ$ to 96$^\circ$ east of north. The source structure qualitatively looks very similar (see Fig.~\ref{cont_images}) but the total magnification of the source continuum has fallen from 6.4 to 6.1 in band 7. The Einstein radius of $0.97\pm0.04$ arcsec remains the same.

\subsection{Source Reconstruction}
\label{sec_srcpix}

We reconstructed the lensed source on a $0.8 \times 0.8$ arcsec source plane consisting of 400 Voronoi pixels formed using the method described in D18. The Voronoi tessellation is formed within the source plane region traced by mapping the region enclosed within an annular mask applied to the ring images as can be seen in Figs.~\ref{cont_images} and \ref{line_images}. When optimising the lens model, we computed seven randomisations of the source plane pixelisation for each trial set of lens model parameters and returned the median likelihood of these seven randomisations to {\tt NeuralNest}. This alleviates the effect reported in \cite{nightingale2018} whereby lens model optimisation can occasionally stagnate when exceptionally good pixelisations occur by chance.

When reconstructing source cubes, we extracted the set of visibilities for each channel and applied the best-fit lens model determined in Section \ref{sec_lensmodel}. We reconstructed the visibilities corresponding to each slice of each cube 50 times, randomising over source plane pixelisations every time. All 50 random Voronoi pixelisations were then individually binned onto a fine grid of $100 \times 100$ square pixels across the source plane and their mean computed to arrive at the final source reconstruction for a given cube slice.

\begin{figure}
	\includegraphics[width=\columnwidth]{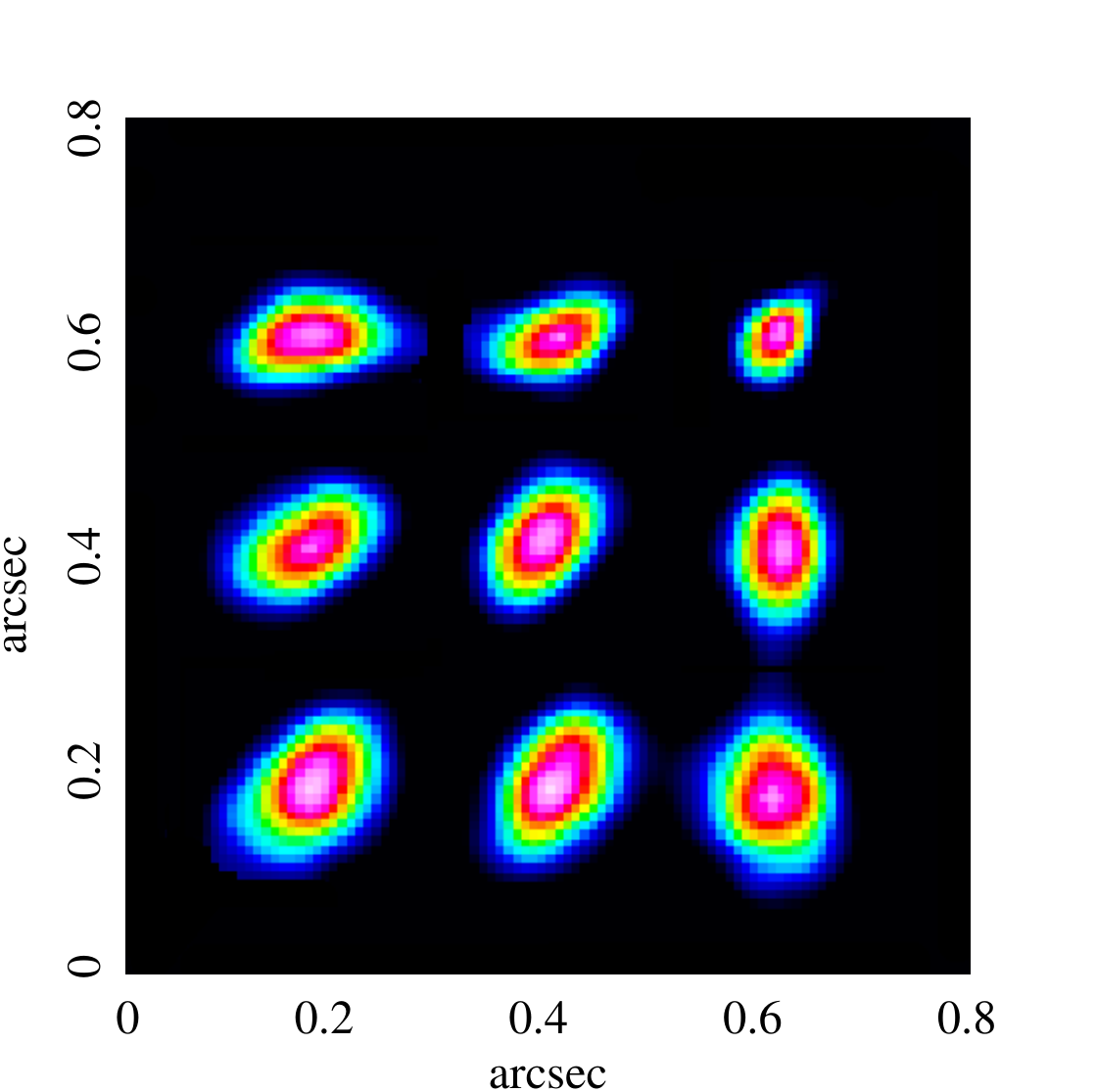}
    \caption{Variation of PSF across reconstructed source plane for band 4 data. The figure is a superposition of the reconstruction of nine individual point sources placed at one of nine positions in a $3\times 3$ grid in the source plane. For each point source, we made a model set of visibilities and reconstructed the set with the same source pixelisation scheme used for reconstructing the lensed SMG (see main text for more details). The beam at the centre of the grid has a FWHM of 0.12 and 0.08\,arcsec along its major and minor axis respectively. For the band 3 data, the PSF shows the same qualitative variation across the source plane but has a linear scale that is approximately 15 per larger than the band 4 PSF.}
    \label{src_psf}
\end{figure}

The reconstruction scheme applied to cube slices results in an effective point spread function (PSF) that varies with lens magnification and therefore with position in the source plane. To quantify the resolution of our reconstructions, we computed the spatial variation of this PSF in the following way. First, we generated a set of model visibilities of a lensed image of a point source created using the best fit lens model. These visibilities were obtained using {\tt MIRIAD}'s {\tt uvmodel} task which evaluates them at the same uv co-ordinates as the observed visibilities. We then applied our reconstruction method, including averaging the 50 random Voronoi source planes as described above to arrive at a reconstructed point source. This procedure was repeated for a grid of $3 \times 3$ point source positions in the source plane (arranged to cover the actual observed source light) and the resulting superposition of these recovered point sources is shown in Fig.~\ref{src_psf}. The dimensions of the PSFs obtained in this way for band 3 and 4 are very similar and these are $0.17\times0.10$\,arcsec, $0.08\times0.05$\,arcsec and $0.12\times0.08$\,arcsec, in the south-eastern, north-western and middle of the source respectively where the dimensions quoted are full widths at half maximum (FWHM) along the major and minor axes. At the source redshift (where 1\,arcsec corresponds to a physical scale of 6.90\,kpc), the smallest physical scale we are able to probe is therefore approximately 350\,pc. The median resolution across the source, computed as the geometric mean of the PSF FWHMs, is approximately 550\,pc.

To determine the unlensed continuum flux of ID141, for each band, we applied the source reconstruction method to the emission line-subtracted visibilities obtained as outlined in section \ref{sec_data}. The reconstructed source continuum in bands 3, 4 and 7 is shown in Fig.~\ref{cont_images}. Note that the same morphological features seen in the band 7 continuum source shown in this figure are also seen in D18, despite the different lens model and the different source pixelisation scheme. The quantitative difference is a 5 per cent reduction in magnification compared to D18 which results in a visually imperceptible increase in the linear scale of the source shown in Fig.~\ref{cont_images} of 2--3 per cent. To determine the unlensed fluxes of the four identified emission lines, we reconstructed the source using the visibilities containing line emission, for each line. These are shown in Fig.~\ref{line_images}. 

With the source reconstructed in this way, the flux was then measured in the source plane aperture that maps to the image plane annulus for both the continua and emission lines. Line fluxes are given in Table \ref{tab_line_measurements} and the continuum fluxes measured in this way are $0.26\pm0.04$\,mJy, $1.16\pm0.14$\,mJy and $14.9\pm0.8$\,mJy in band 3, 4 and 7 respectively. The continuum and line magnifications were computed in more restricted apertures. We formed an image-plane mask by selecting all pixels with a significance of at least $3\sigma$ and then mapped this to the source plane. Magnifications were then determined as the ratio of the image flux to source flux in these apertures. Table \ref{tab_magns} lists the magnifications obtained in this way for the emission lines and band 3, 4 and 7 continua.

\begin{table}
\centering
\caption{Total lens magnifications, $\mu_{\rm tot}$ of the continuum in bands 3, 4 and 7 and the emission lines identified in the ALMA data.}
\label{tab_magns}
\begin{tabular}{cccc}
\hline
Band: & 3 & 4 & 7 \\
$\mu_{\rm tot}$ : & 5.5 & 5.9 & 6.1 \\
\end{tabular}
\begin{tabular}{ccccc}
\hline
Line: & CO(7-6) & H$_2$O & \ci(2-1) & \ci(1-0) \\
$\mu_{\rm tot}$ : & 6.2 & 5.8 & 6.3 & 5.6 \\
\hline
\end{tabular}
\end{table}

\section{The properties of the lensed source}
\label{sec_src}

In this section we give consideration to the reconstructed morphology of ID141, its spectral energy distribution (SED) including the inferred quantities of dust mass, dust temperature and star formation rate and the gas mass of the ISM. Section \ref{sec_srckin} presents our kinematical analysis. We discuss the implications of these results in Section \ref{sec_discussion}. 

\subsection{Source Morphology}

The reconstructed emission line images of ID141 in Fig.~\ref{line_images} show clear morphological differences to the reconstructed continuum emission in Fig.~\ref{cont_images}. Qualitatively, the line emission is more irregular than the continuum emission. This is partly an effect of the lower SNR of the reconstructed line flux but not entirely; the lensed emission line images in the leftmost column of Fig.~\ref{line_images} clearly have higher noise than the lensed continuum images in Fig.~\ref{cont_images} but many of the clumpy features seen in the reconstructed line images in the rightmost column of Fig.~\ref{line_images} are multiply-imaged in the observed ring and can be seen at different wavelengths.

One particularly striking feature is an offset in the peak emission seen in the \ci(1-0) line relative to the other lines and relative to the continuum emission. This is apparent in both the reconstructed source and the lensed image. The bottom-left panel of Fig.~\ref{line_images} illustrates this offset by over-plotting contours of the band 3 continuum flux on the line emission map. Since \ci(1-0) has a lower excitation temperature than the other lines, this suggests the presence of cooler gas in the southern parts of the galaxy. We return to this point in Section \ref{sec_tempgrad}.

\subsection{Far-IR Spectral Energy Distribution}
\label{sec_sed}

\begin{figure}
	\includegraphics[width=\columnwidth]{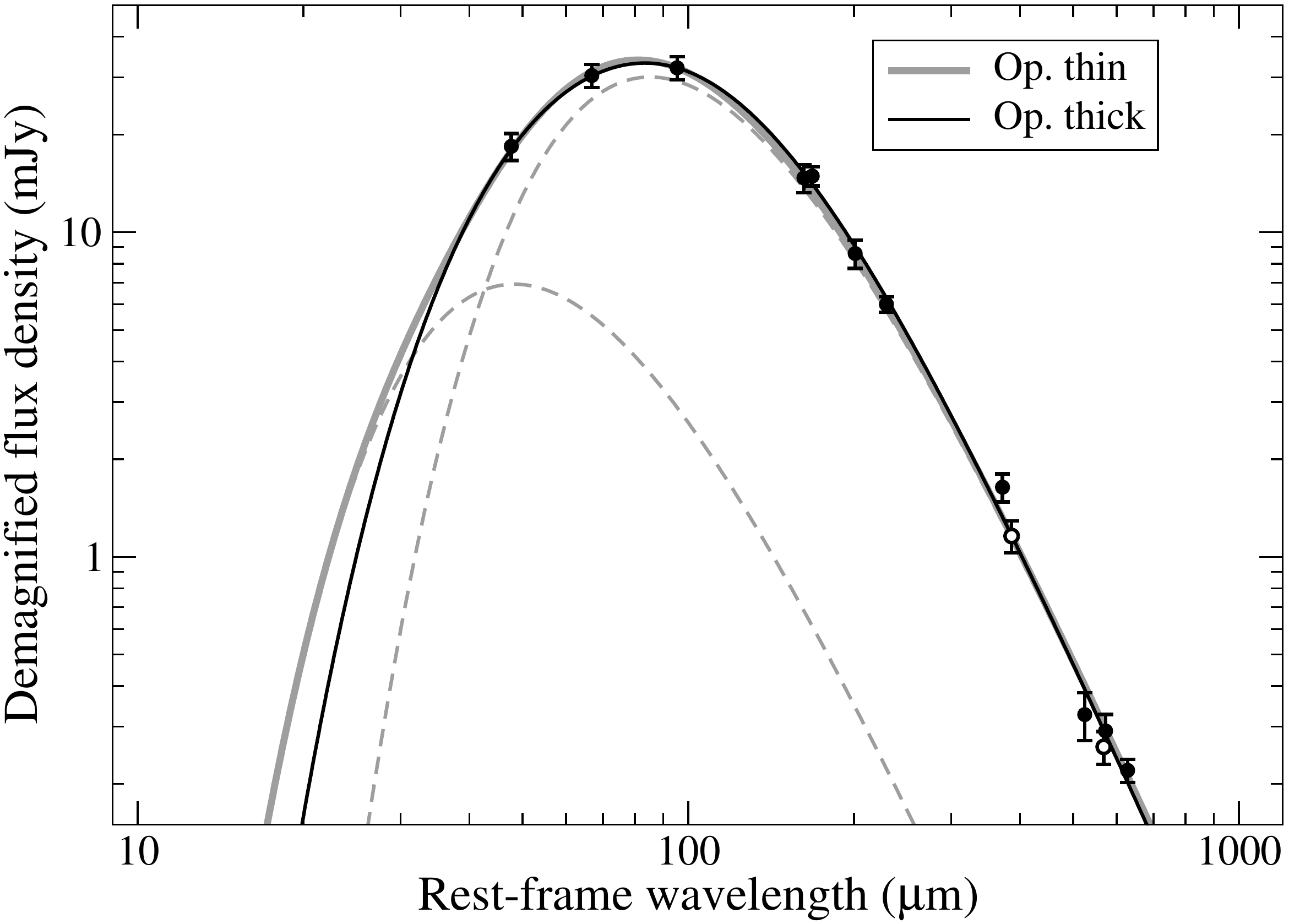}
    \caption{Spectral energy distribution of ID141. Data points with filled circles are from D18 (but adjusted for the new lens model) and \citet{cheng2020}. The open circles are the band 3 and band 4 continuum fluxes from the present work. The solid grey and black lines show an optically thin SED with two temperature components and an optically thick single-temperature SED fit respectively (see text for more details). The dashed lines are the two separate components that comprise the optically thin SED.}
    \label{sed}
\end{figure}

Fig.~\ref{sed} plots measurements of the continuum flux of ID141 at different rest-frame wavelengths. The filled points are taken from D18 but reduced by 5 per cent to account for the updated lens magnification and from \citet{cheng2020}. We also updated the 850\,$\mu$m flux originally taken from \citet{bakx2018} with the flux of 90\,mJy following \citet{bakx2020}. The new band 3 and 4 continuum fluxes are shown in this figure with empty circles. 

Following D18, we fitted the rest-frame photometry with a modified black-body SED (referred to hereafter as the 'optically thick SED' ) comprising a single dust temperature component as well as an optically thin SED with two dust temperature components. This is an over-simplistic treatment of the expected continuous range of dust temperatures which we incorporate in later SED fitting (see Section \ref{sec_stellar_mass}). Nevertheless, these two simple SED models provide a good estimate of the range of possible dust masses. We measured the dust mass predicted by each SED using the method described in \cite{dunne2011}. For this, we took the observed ALMA 880\,$\mu$m flux density of $91\pm5$\,mJy from the band 7 data and a dust mass absorption coefficient at the corresponding rest-frame wavelength, 168\,$\mu$m, of 1.7\,m$^2$kg$^{-1}$ from \cite{james2002}. Both SEDs were fitted with a fixed emissivity index of 2.0 \citep{smith2013}. 

We allowed the temperature and normalisation of both components to vary in the fit for the optically thin SED giving four degrees of freedom. For the optically thick SED, temperature, normalisation and the 100\,$\mu$m-opacity, $\tau_{100}$, were varied in the fit, giving three degrees of freedom. The best-fit parameters are $T_{\rm warm}=60\pm 6$\,K and $T_{\rm cold}=34\pm 3$\,K with a ratio of cold-to-warm dust mass of $74\pm13$ for the optically thin SED. For the optically thick SED, we found $T=57\pm 1$\,K and $\tau_{100}=4.8$. The predicted dust masses are $(1.4\pm 0.3)\times 10^9$\,M$_\odot$ and $(7.4\pm 0.8)\times 10^8$\,M$_\odot$ for the optically thin and optically thick SEDs respectively.

Integrating these SEDs from 8-1000\,$\mu$m gives a far-IR luminosity, $L_{\rm FIR}$, of $1.6\times10^{13}$\,L$_\odot$ and $1.5\times10^{13}$\,L$_\odot$ for the optically thin and thick SEDs respectively. These values are consistent with \citet{cox2011} once corrected for lens magnification and places ID141 in the category of hyper-luminous infra-red galaxies \citep{rowanrobinson2000}. Taking the mean of these luminosities and converting it to a star formation rate (SFR) according to the prescription given by \cite{kennicutt2012}, which uses the initial mass function given by \citet{kroupa2003}, gives a value of $2400\pm500$\,M$_\odot$yr$^{-1}$.

This is consistent with the SFR calculated in D18 but is 5 per cent higher due to the 5 per cent reduction in magnification of the new lens model. As discussed in D18, given its dust mass, this places ID141 significantly above the range of SFRs for $z<0.5$ H-ATLAS galaxies and in the upper envelop of high redshift SMGs \citep{rowlands2014}.

\subsection{The Temperature Distribution in the ISM}
\label{sec_tempgrad}

There is a significant offset of the peak emission seen in the \ci(1-0) line compared to the peak of emission seen in the other three lines. Initially concerned that this may simply be the result of an offset in the world co-ordinate system between the band 3 data (where the \ci(1-0) line is located) and band 4 data, we confirmed that the continuum images between both bands align perfectly. Indeed, the \ci(1-0) emission is offset significantly from the band 3 continuum (see Fig.\,\ref{line_images}) and this is clearly apparent in both the reconstructed source and the lensed image. The offset between the \ci(1-0) emission and the \ci(2-1) immediately suggests the temperature of the ISM is varying across the galaxy.

The advantages of the \ci lines for tracing the cool ISM over the traditional CO method has become increasingly recognised in the last few years \citep[][Dunne et al. {\it in preparation} ]{papadop2004,dunne2021a}. The \ci lines have similar excitation temperatures and critical densities to the low-J CO lines, but their advantages over the CO lines are that the \ci lines are mostly optically thin and atomic carbon is present everywhere in the molecular phase \citep{papadop2004}. In contrast, a significant and variable fraction of the molecular phase in galaxies appears to contain no CO, presumably because of photo-dissociation \citep{abdo2010,pineda2013,planck2011,bigiel2020,dunne2021a}.

\begin{figure}
	\includegraphics[width=\columnwidth]{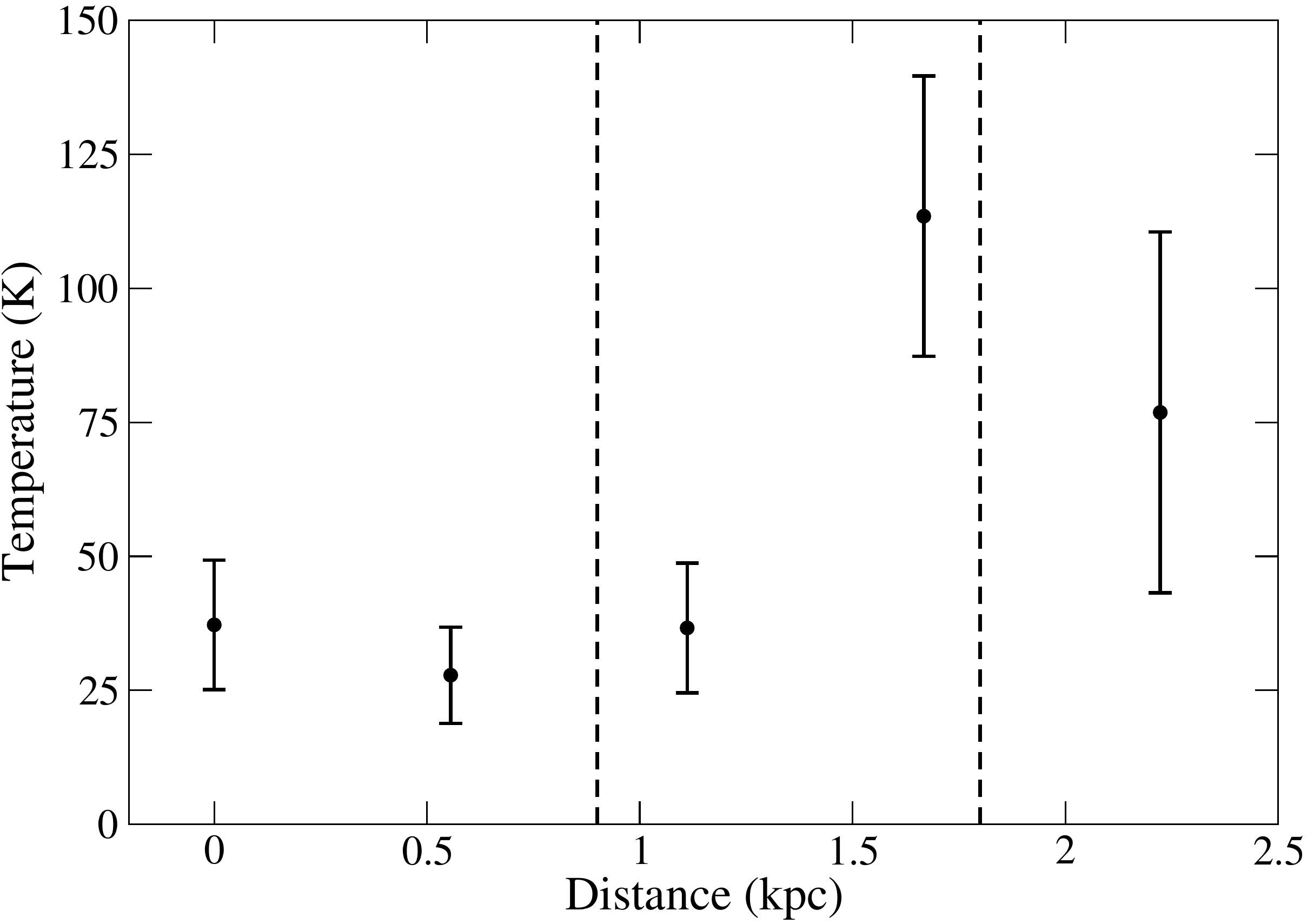}
    \caption{Gas temperature as a function of distance moving northwards across ID141. The temperature has been estimated from the ratio of the luminosity of the \ci(2-1) and \ci(1-0) lines in east-west slices of length 0.2 arcsec and width 0.08 arcsec, on the assumption of local thermodynamic equilibrium. The vertical dashed line on the left shows where the \ci(1-0) emission peaks (see also Fig. \ref{line_images}) and the vertical dashed line on the right shows where the \ci(2-1) emission peaks.}
    \label{temperature}
\end{figure}

\begin{figure}
	\includegraphics[width=\columnwidth]{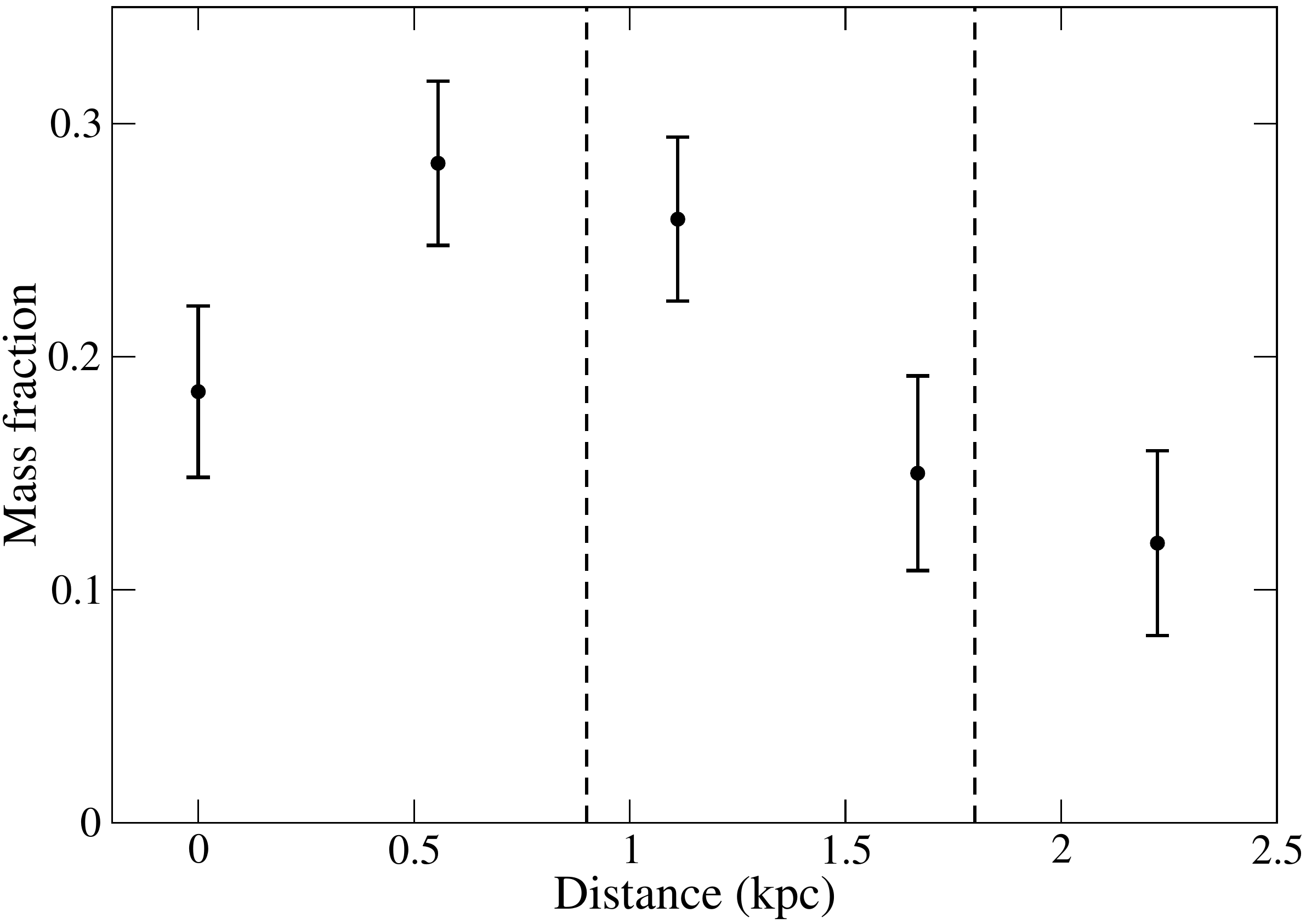}
    \caption{Fraction of the total ISM mass in the molecular phase as a function of distance moving northwards across ID141. Each data point has been determined from the \ci(1-0) flux in a series of east-west slices of length 0.2 arcsec and width 0.08 arcsec arranged across ID141 and the excitation temperature estimated from the ratio of the luminosity in the \ci(1-0) and \ci(2-1) lines (Fig. \ref{temperature}). The vertical dashed line on the left shows where the \ci(1-0) emission peaks (see also Fig. \ref{line_images}) and the vertical dashed line on the right shows where the \ci(2-1) emission peaks.}
    \label{gas_mass_slices}
\end{figure}

Since the \ci lines are optically thin, it is reasonably straightforward to use the ratio of the \ci(2-1) and \ci(1-0) lines to investigate the variation of the temperature in the ISM. Therefore, we divided the \ci(2-1) and \ci(1-0) source images in Fig. \ref{line_images} into a stack of five slices running east-west, each 0.2 arcsec long and 0.08 arcsec wide, with the stack positioned to cover the main \ci emission. We calculated the ratio of the luminosity \citep[measured in K km s$^{-1}$ pc$^2$,][]{solomon2005} of the two lines, $R=L'(2-1)/L'(1-0)$, in each of these slices. Assuming local thermodynamic equilibrium, we then converted the line ratio into an excitation temperature using T$_{\rm ex}=38.8{\rm K}/\ln(2.11/R_{\rm CI})$ \citep{stutzki1997}. 

Fig. \ref{temperature} shows the variation of temperature in the stacked slices moving northwards across ID141. A $\simeq 2.5 \sigma$ increase in temperature is seen rising from $\simeq$40\,K in the three southern bins to $\gtrsim 100$\,K in the two northern bins. This trend is in accord with the other images in Fig. \ref{line_images} because the water line and the CO(7-6) line also peak in the north, and both lines are likely to be from high-temperature gas \citep{papadop2012,liu2017}. The excitation temperature of the galaxy as a whole, estimated from the global fluxes for the \ci lines (Table \ref{tab_line_measurements}) and the equation above, is 33\,K. To put this into context, in the sample of 49 galaxies of Dunne et al. ({\it in preparation}), ID141 would rank as the fifth highest \ci excitation temperature.

The mass of the molecular phase of the ISM is proportional to $L'(\ci(1-0))/Q_{10}$, where $Q_{10}$ is the partition function for the \ci(1-0) line \citep{solomon2005}, which depends on the excitation temperature. Fig. \ref{gas_mass_slices} shows the fraction of the molecular mass in each of our five slices calculated from the \ci(1-0) flux and our estimate of the excitation temperature in each slice, based on the assumption of local thermodynamic equilibrium. This demonstrates that although the gas is much hotter in the north, most of the mass of the molecular phase is in the colder gas to the south.

\subsection{First Estimates of the Mass of the ISM}
\label{sec_gasmass}

Any estimate of the mass of the molecular ISM in a galaxy other than our own requires the use of an ISM tracer such as CO molecules or dust. This in turn requires an estimate of the constant of proportionality (the calibration factor) between the luminosity of the tracer and the mass of the ISM. In this section, we estimate the mass of the ISM in ID141 using four different tracers and five different methods. The uncertainties quoted on the ISM mass estimates include the flux uncertainty and the uncertainty of the conversion factors used.

\subsubsection{The gas mass from CO} 

We have two measurements of CO lines, a CO(7-6) line luminosity which has been corrected by the lensing model derived in this paper, and a CO(1-0) line luminosity which has not been corrected for lensing because the observations did not have high enough resolution (Dunne et al.  {\it in preparation}). Since most of the work on calibrating the CO method has used the CO(1-0) line and since the ratio of high-J to low-J line flux has a large dispersion \citep{canameras2018}, we have estimated the gas mass from the CO(1-0) measurement. 

The total CO(1-0) line flux is $0.61\pm 0.09$\,Jy\,km\,s$^{-1}$ (Dunne et al.  {\it in preparation}). We have corrected this using the magnification factor of 5.5 derived for band 3. \cite{dunne2021a} have derived the CO(1-0) calibration factor from a sample of H-ATLAS galaxies with $z \sim 0.35$ with ALMA measurements of CO(1-0), \ci(1-0) and the continuum dust emission. By combining the measurements, they derive calibration factors for three methods of estimating the mass of molecular hydrogen. All three of these methods scale linearly with the mean gas-to-dust ratio (GDR) in the Universe, which they assume is 135, similar to the estimates of the value in the Milky Way. They derive a calibration factor $\alpha_{\rm CO}= 3.0\pm0.5\,$M$_{\odot}$ (K\,km\,s$^{-1}$pc$^2$)$^{-1}$. Using this value and including a factor of 1.36 to allow for the elements other than hydrogen gives an estimate of the mass of the ISM of $(3.3\pm0.7) \times 10^{11}$M$_{\odot}$ (Table \ref{tab_gas_mass}). This mass is also consistent with \citet{cox2011} after correcting for lens magnification and using the value of $\alpha_{\rm CO}$ adopted above.

\subsubsection{The gas mass from [CI]} 

A major problem with using the CO molecule as a tracer of the ISM is the mounting evidence that a significant fraction of the molecular phase of the ISM does not contain any CO molecules, probably as the result of photo-dissociation. For example, there is evidence from {\it Fermi} \citep{abdo2010}, {\it Planck} \citep{planck2011} and {\it Herschel} \citep{pineda2013} that roughly one third of the molecular gas in the Milky Way does not contain any CO. There is also increasing evidence that the fraction of CO-dark molecular gas varies from galaxy to galaxy \citep{bigiel2020,dunne2021a}. The \ci(1-0) line is an attractive alternative to CO because it does not suffer from the problem of photo-dissociation, is optically thin, and has a similar critical density and excitation temperature to CO \citep{papadop2004,geach2012,tomassetti2014,bourne2019}. We have estimated the gas mass from the line flux of the \ci($\rm ^3P_1 - ^3P_0$) transition, corrected with our lensing model (Table \ref{tab_line_measurements}). Using their calibration sample (see above), \cite{dunne2021a} have derived a value for the \ci calibration factor of $\rm \alpha_{[CI]} = 18.8\ M_{\odot} (K\ km\ s^{-1} pc^2)^{-1}$. The excitation function for the line ($Q_{10}$) varies only weakly with temperature \citep{dunne2021a} and so we have used this calibration factor rather than trying to make any correction to it using our estimate of the temperature from the ratio of the two [CI] lines (Section 4.3). We estimate that the mass of the ISM in ID141, including all elements, is $(3.8 \pm 0.8)\times10^{11}$\,M$_\odot$ (Table \ref{tab_gas_mass}).

\subsubsection{The gas mass from [CII]}

There is a strong correlation between the luminosity of the fine-structure line of \cii at 158\,$\mu$m and the mass of molecular gas in the ISM \citep{zanella2018}. Zanella et al. derive a \cii to $H_2$ conversion ratio of $\alpha_{\rm CII} \simeq 30$\,M$_{\odot}/$L$_{\odot}$. We do not have a high-resolution map of the \cii line and therefore no specifically-derived lensing model for it. Instead, we simply used the global line flux given by \cite{cox2011}, corrected it for the magnification factor of 5.9 we measured for the dust continuum emission in band 4, the closest in wavelength to the \cii line, and then applied the calibration factor of Zanella et al. We obtained a total gas mass of $(3.2\pm0.6)\times10^{11}$\,M$_\odot$ (Table \ref{tab_gas_mass}).

\subsubsection{The gas mass from dust emission}
\label{sec:gas_dust_scov}

Many studies have made the case that dust grains are a better way of tracing the ISM than the CO molecule \citep[e.g.][]{eales2012,scoville2014,scoville2016,scoville2017}. Some of the advantages are that the dust grains do not suffer photo-dissociation like CO molecules, the dust emission is optically thin, and the relationship between the calibration factor and metallicity appears to be simpler \citep{eales2012,eales2018}. We used the calibration factor between dust luminosity at 850\,$\mu$m and ISM mass derived by \cite[][their Eq. A11]{scoville2016}:
\begin{equation}
{\rm \alpha_{850 \mu m} = <L_{\nu}/M_{ISM} > 6.7 \pm 1.7 \times 10^{19} ergs\,s^{-1}\,Hz^{-1} \msolm^{-1}} \, .
\end{equation}

We used the global flux density measured by \cite{cox2011} at 3290\,$\mu$m of $1.2\pm0.1$ mJy, which makes it the closest flux measurement in rest-frame wavelength (628\,$\mu$m) to the wavelength used in the calibration factor above, minimising the uncertainty in extrapolating over wavelength. We estimated the luminosity at rest-frame 850\,$\mu$m, using a dust temperature of 25K and an emissivity index of ($\beta=1.8$), the same values used by \cite{scoville2016}. We then calculated the ISM mass using the magnification factor of 5.5 derived from our band 3 lensing model and the calibration factor above, obtaining a total ISM mass of $(3.6\pm0.7) \times 10^{11}$\,M$_{\odot}$ (Table \ref{tab_gas_mass}).

\subsubsection{Combining Three Methods}

\citet{dunne2021a} have used several samples of calibration galaxies to derive an optimal technique for estimating the calibration factors ($\alpha_{\rm CO}$, $\alpha_{[CI]}$, $\alpha_{\rm dust}$) for an individual galaxy and for producing a best estimate of the ISM mass in the galaxy that uses all three. After correcting for lens magnification, using this technique on ID141 gives an estimate of the ISM mass of $(3.6\pm0.7) \times 10^{11}$\,M$_\odot$ (Table~\ref{tab_gas_mass}).

\subsubsection{Results}

Table 5 lists the ISM mass estimates from the five different methods.  We note a few important points. One caveat is that we could have used the lower value of $\alpha_{\rm CO}$ that has been claimed for ultra-luminous infrared galaxies. This would lower the ISM mass estimated from CO by a factor of a few, although there are now some compelling arguments that the claim that ultra-luminous infrared galaxies have a different $\alpha_{\rm CO}$ than other galaxies is incorrect \citep[][Dunne et al. {\it in preparation}]{scoville2016}. A second caveat is that the only estimate in which the observations of the tracer have been corrected for gravitational lensing from the observations themselves is the estimate from the \ci(1-0) line. For every other estimate, we have corrected the tracer using a total magnification value derived at some other wavelength. There is therefore the possibility of differential magnification \citep[see, for example][]{serjeant2012,omont2013}. However, the similarity between the line profiles measured in the image plane and those in the source plane, together with the similarity of the seven magnification values presented in Table~\ref{tab_magns} (from four line images and three continuum images), indicate that this is only a minor effect. We have quantified a typical size of the effect by computing the magnification factor of a source that has twice the diameter of the band 3 source shown in Fig.~\ref{cont_images}, finding this to be only 3 per cent lower than the original magnification. Another important point is that all ISM estimates are only for the molecular phase of the ISM because of the way the methods are calibrated. Finally, \citet{daCunha2013} and \citet{Zhang2016} have shown that the effect of the cosmic microwave background radiation will make it more difficult to detect the full extent of the ISM in high-redshift galaxies, with this effect depending on the frequency of the observations.

The mass estimates lie between $3.2 \times 10^{11}$\ M$_{\odot}$ and $3.8 \times 10^{11}$\ M$_{\odot}$, which are very large values and suggest we are unlikely to be missing much of the ISM. The agreement between the values is also quite remarkable considering that four of the methods use different tracers of the ISM and we are observing the galaxy only 1.5 billion years after the Big Bang. However, none of the methods are truly independent of each other because they are all calibrated from observations of the Milky Way. We discuss this further in Section~\ref{sec_ism_vs_gas_mass}.

\begin{table*}
\caption{Emission line fluxes used in the estimation of the gas mass of ID141. The CO(7-6), H$_2$O and two \ci lines are from the present work and have been determined in the source plane. Errors quoted for these lines include uncertainty due to the lens model and continuum subtraction. The CO(1-0) and \cii lines were taken from Dunne et al. {\it in preparation} and \citet{cox2011} respectively and have been corrected for lens magnification using the band 3 and band 4 magnification factors respectively.}
\label{tab_line_measurements}
\begin{tabular}{llllll}
\hline
Line & $\nu_{\rm obs}$ & Line Flux & L & L$'$  & Reference  \\
&  (GHz) & (Jy\,km\,s$^{-1}$) & $(10^9 {\rm L}_\odot)$ & ($10^{10}$\,K\,km\,s$^{-1}$\,pc$^2$)   & \\
\hline
CO(1-0) & 21.991 & $0.111\pm0.016$ & $0.0040\pm0.0007$ & $8.0\pm1.6$   & Dunne et al. {\it in preparation} \\
CO(7-6) & 153.876 & $1.84\pm0.31$ & $0.45\pm0.07$ & $2.7\pm0.5$    & this paper \\
$\rm \ci (^3P_2 - ^3P_1)$ & 154.339 & $0.98\pm0.19$ & $0.224\pm0.045$ & $1.3\pm0.2$   & this paper \\
$\rm \ci (^3P_1 - ^3P_0)$ & 93.870 & $0.55\pm0.10$ & $0.076\pm0.014$ & $2.0\pm0.4$    & this paper \\
$\rm \cii (^3P_{3/2} - ^2P_{1/2})$ & 362.45 & $18.1\pm 2.9$ & $10.4\pm1.7$ & $4.8\pm0.8$ & \citet{cox2011} \\
H$_2$O & 143.436 & $0.70\pm0.14$ & $0.16\pm0.03$ & $1.14\pm0.23$  & this paper \\
\hline
\end{tabular}
\end{table*}

\begin{table}
\caption{Gas mass estimates determined from five different methods (see main text for details).}
\label{tab_gas_mass}
\begin{tabular}{lcl}
\hline
Method & ISM mass ($10^{11}$M$_{\odot}$) & Calibration reference \\
\hline
CO(1-0) & $\rm 3.3\pm0.7$  & \citet{dunne2021a}\\
\ci(1-0) & $\rm 3.8\pm0.8$ & \citet{dunne2021a} \\
\cii 158 $\mu$m & $\rm 3.2\pm0.7$  & \citet{zanella2018}\\
dust & $\rm 3.6\pm0.7$ & \citet{scoville2016} \\
CO/\ci/dust & $\rm 3.6\pm0.7$ & \citet{dunne2021a} \\
\hline
\end{tabular}
\end{table}

\subsection{The stellar mass}
\label{sec_stellar_mass}

An additional means of establishing the evolutionary stage of a galaxy is by measuring its mass in stars. We return to the theme of stellar mass in Section \ref{sec_chem_evol} where we consider a chemical evolution model for ID141, but in this section, we outline our estimation of stellar mass of ID141 from photometry following the procedure outlined in \cite{hopwood2011}. The key ingredient is the mid-IR source flux density which we have determined using Spitzer Space Telescope observations (proposal ID. 80156) acquired with the Infra Red Array Camera (IRAC) as described below.

To begin, we measured the surface brightness profiles of the lensing mass from the $K_{\rm s}$-band data presented in \cite{bussmann2012} using {\tt GALFIT} \citep{peng2002}. As \cite{bussmann2012} found, subtracting the {\tt GALFIT} model light profiles from the observed $K_{\rm s}$-band image showed negligible residual lens light and no evidence of the lensed source image. We then used these profiles to construct model IRAC 3.6\,$\mu$m and 4.5\,$\mu$m surface brightness profiles and subtracted these from the IRAC images to leave residuals resembling the ALMA image (see Fig.~\ref{irac_resids}). We repeated this removal procedure for three other galaxies selected to have similar morphology and flux in the IRAC data and found only negligible residuals. This demonstrates that an accurate IRAC PSF was obtained and that the near-IR traces the mid-IR well in these three systems, although we are unable to rule out the possibility of there being a mis-match in the specific case of the lensing mass in ID141. A final measure of the lensed source IRAC flux was then extracted from an annulus surrounding the observed residual flux (see Fig.~\ref{irac_resids}) resulting in fluxes of $8.7\pm2.2 \, \mu$Jy and $8.1\pm1.8 \, \mu$Jy at 3.6\,$\mu$m and 4.5\,$\mu$m respectively. The errors here do not include any uncertainty arising from lens light removal.

In the final part of the process, following \cite{hopwood2011}, we used version 2 of the SED fitting code {\tt MAGPHYS} \citep{battisti2020,dacunha2008} and supplied this with the two IRAC fluxes as described above, a limiting $K_{\rm s}$-band flux, the far-IR fluxes as described in Section \ref{sec_sed} and Wide-field Infrared Survey Explorer fluxes at 12\,$\mu$m and 22\,$\mu$m \citep[WISE;][]{lang2016}. We refer the reader to Section \ref{sec_app1} for further details. 

To de-magnify the quantities output by {\tt MAGPHYS}, we assumed a magnification of 5.8, equal to the mean of the continuum magnifications given in Table~\ref{tab_magns}. In addition, since {\tt MAGPHYS} assumes a Chabrier initial mass function \citep{chabrier2003}, we scaled the stellar mass, dust mass and  SFR output by {\tt MAGPHYS} by a factor of 1.41 for comparison with the values determined in Section \ref{sec_sed}. This factor of 1.41 was determined from the scalings derived in \citet{madau2014}. The resulting stellar mass is M$_\star = 4.8^{+2.5}_{-2.3} \times 10^{11}$\,M$_\odot$. In addition to stellar mass, {\tt MAGPHYS} also gives a dust mass of $1.8\pm0.3 \times 10^9$\,M$_\odot$ and a SFR of $2100\pm500$\,M$_\odot$\,yr$^{-1}$. The $1\sigma$ errors quoted are taken from the distributions output by {\tt MAGPHYS} (see Appendix~\ref{sec_app1} for more details). These results are consistent with those obtained in Section \ref{sec_sed} based on our own SED derived from solely the far-IR photometry. 

\subsection{The kinematics of the source}
\label{sec_srckin}

The double-peaked nature of our highest SNR emission line profile, the H$_2$O line, shown in Fig.~\ref{lines} suggests that either the source is a rotating disk-like system with a prominent outer ring \citep[see, for example][]{stewart2014}, or that it is composed of two or more individual merging components.

To better understand the dynamical characteristics of the source, we investigated the spatial extent of the kinematics using the kinematic modelling code {\tt $^{\mathrm{3D}}$Barolo} \citep{teodoro2015}. Instead of fitting a kinematical model to the 2D velocity field, this code fits directly to the full datacube. This avoids the effect of artificial flattening of velocity fields due to inadequate spatial resolution and allows for the fitting of additional galaxy properties (i.e., velocity dispersion, disk scale height, brightness profile).

\subsubsection{The kinematical model}

First, we identified the signal in the data cube by using the `SEARCH' routine of {\tt $^{\mathrm{3D}}$Barolo}, which is based on the {\tt DUCHAMP} source finder \citep{whit12}. In short, this routine automatically identifies the noise level of the data cube, searches each channel for spaxels that are above a user-provided SNR (here 3.5$\sigma$), creates a mask by extrapolating outwards to a second user-provided S/N threshold (here 2.0$\sigma$), and combines overlapping masks in different channels into discrete `sources'. All sources that are too spectrally thin (i.e., possessing only one channel) or spatially compact (i.e., smaller than the synthesized beam of the data cube) are rejected. This final three-dimensional mask contains all significant emission, while excluding noise peaks that would be included using a flat S/N threshold.

With the signal identified, we next performed a preliminary analysis of the masked data cube in order to produce physically motivated estimates for further dynamical fitting. The {\tt CASA} toolkit task {\tt image.moments} was applied to the masked cubes to create maps of integrated intensity (zeroth-moment), line-of-sight velocity (first-moment), and velocity dispersion (second-moment). The zeroth-moment map was fitted with a two-dimensional Gaussian using the {\tt CASA} toolkit task {\tt image.fitcomponents}, resulting in a central position, morphological position angle, and FWHM of the minor and major axes. The initial estimate of the rotational velocity at this central position was then extracted from the first-moment map.

Before performing the fitting, it is crucial to determine the size of each ring, as well as the maximum radius of the model. In order to avoid over-resolving the data, we adopted a minimum ring width of a factor of 0.4 times the FWHM of the minor axis of the restoring beam. The factor of 0.4 here is comparable to factors used in other similar works (e.g., \citealt{tali18,fan19}).
Similarly, we adopted a maximum model radius of a factor of 0.8 times the FWHM of the major axis of the 2D Gaussian fitted to the zeroth-moment map. The factor of 0.8 is the largest value that does not return non-physical results at large radii.

With these initial parameters and estimates in hand, we engaged the {\tt 3DFIT} routine of {\tt $^{\mathrm{3D}}$Barolo} to fit the masked data cube with a tilted ring model comprising a series of concentric rings which each rotate with an independent velocity. The number of rings and ring width were determined as detailed in the previous paragraph, while the model center was fixed to the central position based on the 2D Gaussian fitted to the zeroth-moment map. We assumed a thin disk (i.e., $z_o=0.01$\,arcsec). This routine begins by populating a model physical space (i.e., X, Y, Z) with gas clumps, so that these initial estimates of galaxy properties are replicated. This physical space is then translated into an observed-space data cube (i.e., RA, Dec, velocity), and the variable galaxy properties (i.e., rotational velocity, velocity dispersion, inclination, position angle, and systemic velocity) are perturbed until a minimum residual is met. This process is repeated for each ring, after which most variables are averaged, and the fitting process is repeated with a smaller number of variables (i.e., rotational velocity and velocity dispersion). This process results in best-fit values of each parameter and a model data cube that may be compared to the original data cube on a pixel-by-pixel basis.

The normalisation of the flux density within each ring can be either `local', meaning that the integral along the spectral axis of a given model pixel in the ring must match that of the corresponding datacube pixel or, `azimuthal',  whereby the model is normalised to the azimuthally-averaged flux in each ring. Local normalisation allows for non-axially symmetric distributions of emission across the disk that can bias fitting of azimuthally-averaged fits. In this work, we have used both normalisation types for comparison. 

We  applied {\tt $^{\mathrm{3D}}$Barolo} to the two emission lines with the highest SNR, namely the CO(7-6) and H$_2$O  lines.

\subsubsection{The results of the kinematic model}
\label{bbres}

\begin{figure*}
\centering
\includegraphics[width=\textwidth]{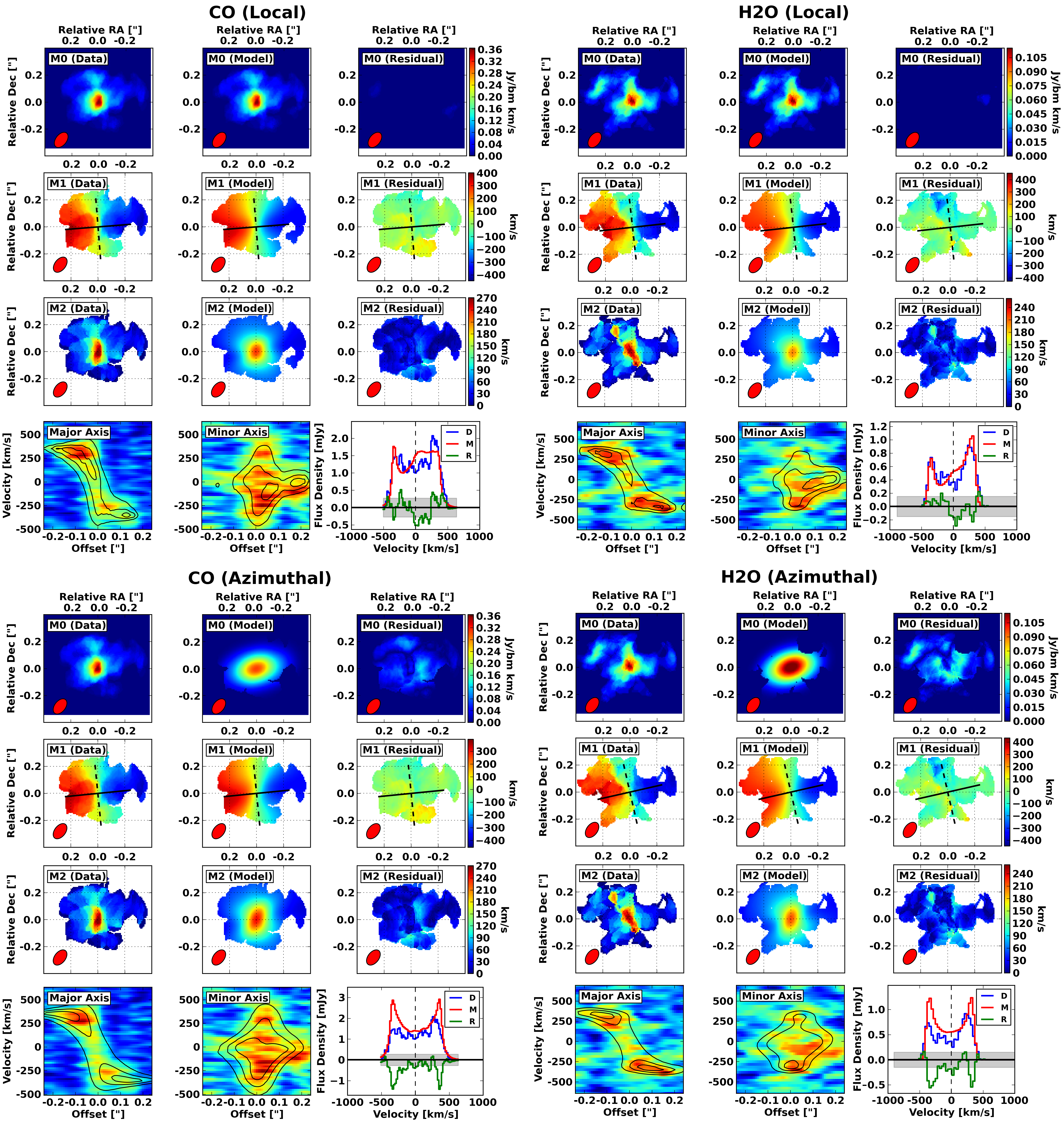}
\caption{Comparison of observed emission lines and best-fit {\tt $^{\mathrm{3D}}$Barolo} models for local normalization (top row) and azimuthal normalization (bottom row), as well as both CO (left column) and H$_2$O (right column). In each set of panels, we show the zeroth-moment (i.e., integrated intensity) maps (top row), first-moment (i.e., line-of-sight velocity) maps (second row; solid and dashed line indicates major and minor kinematical axis respectively), and second-moment (i.e., velocity dispersion) maps (third row) for the data (left column), model (middle column), and the residual between the two (right row). We also show the position-velocity diagrams taken along the major axis (bottom row, left column) and minor axis (bottom row, middle column) of the data cube (colour) and model cube (contours). Finally, we compare the spectra of the data (blue), model (red), and residual (green) in the lower right panel, showing the $1\sigma$ noise level by the shaded grey region.}
\label{12plots}
\end{figure*}

\begin{figure*}
\centering
\includegraphics[width=0.47\textwidth]{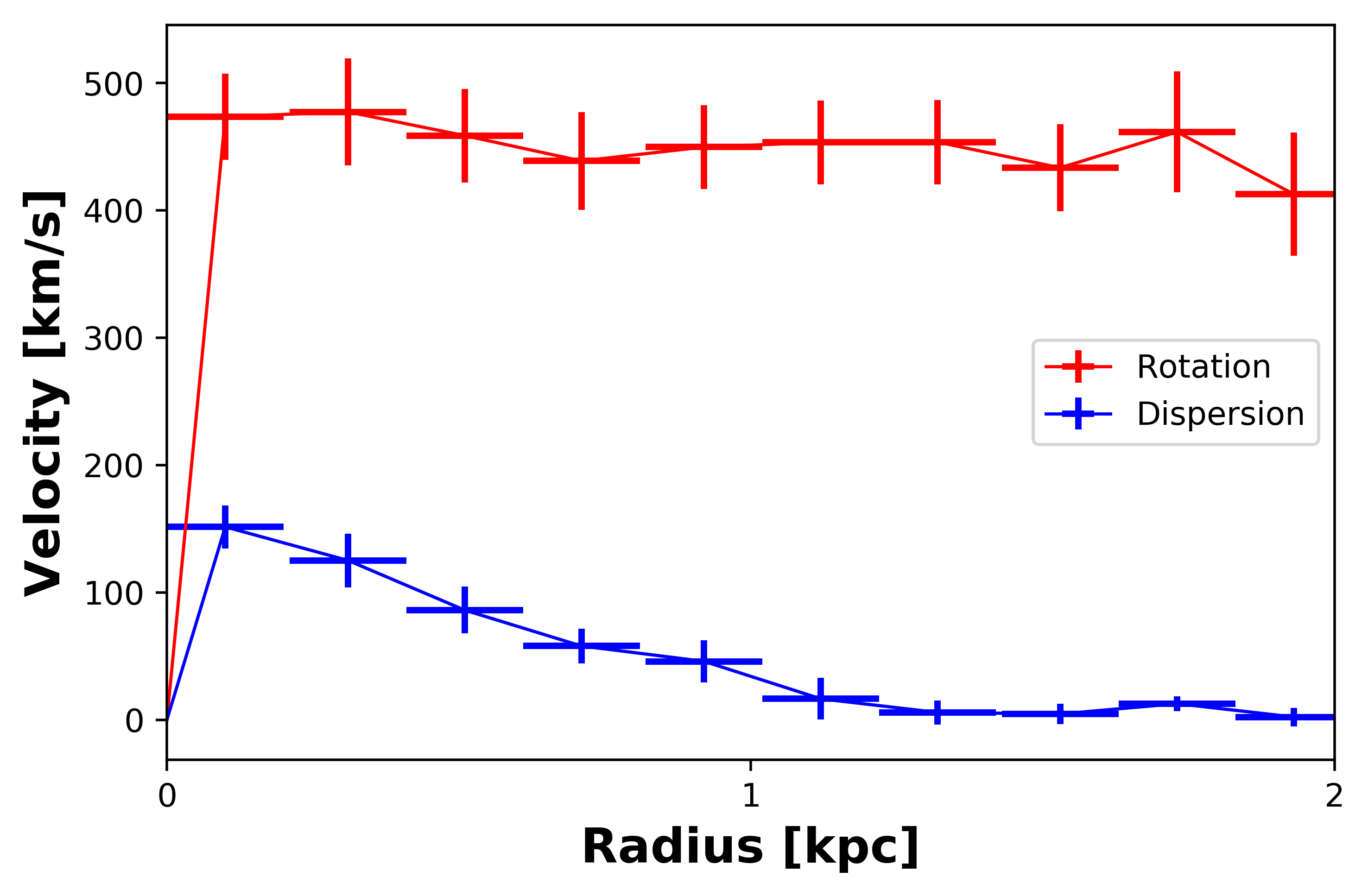}
\includegraphics[width=0.47\textwidth]{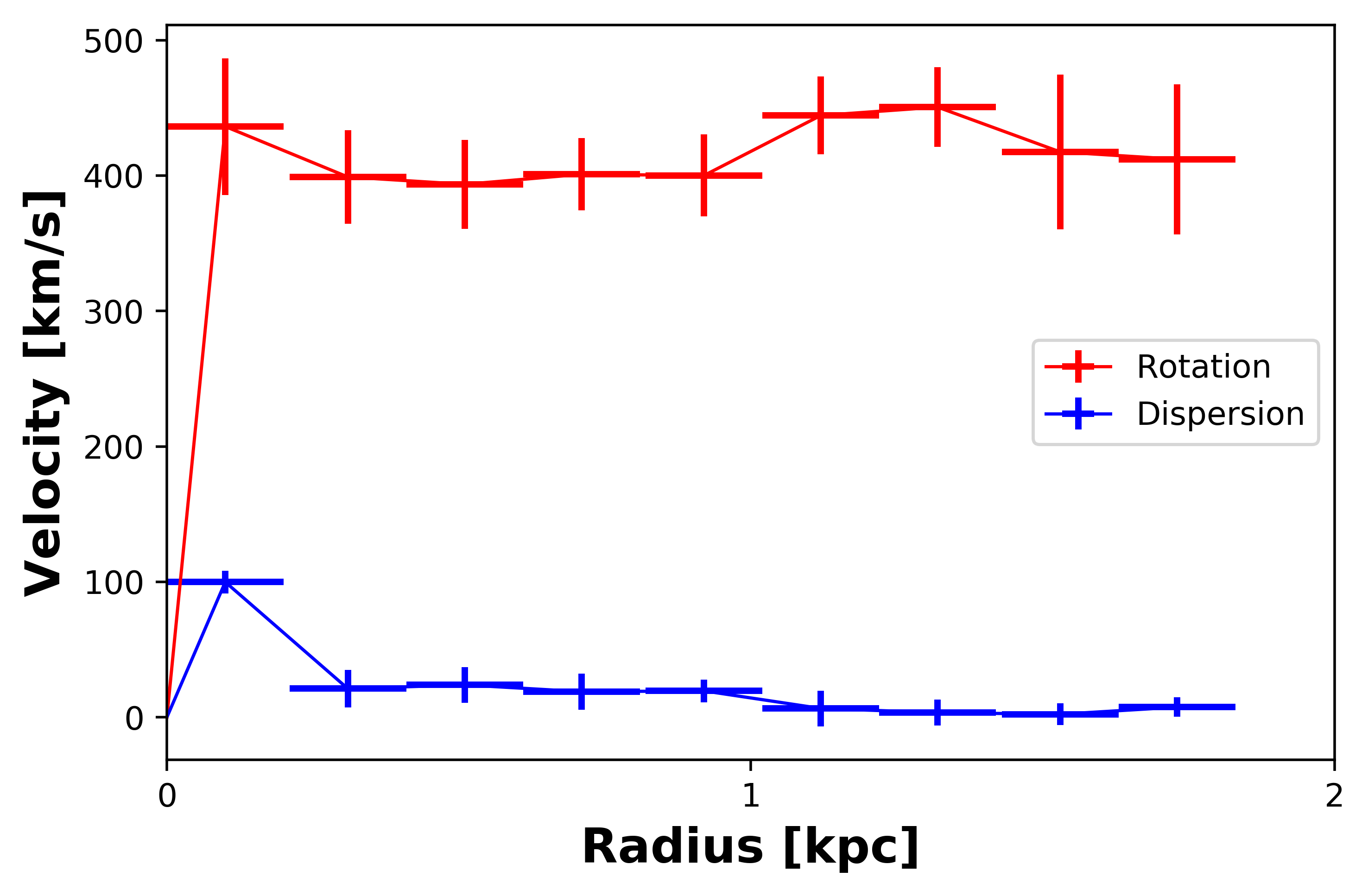}
\includegraphics[width=0.47\textwidth]{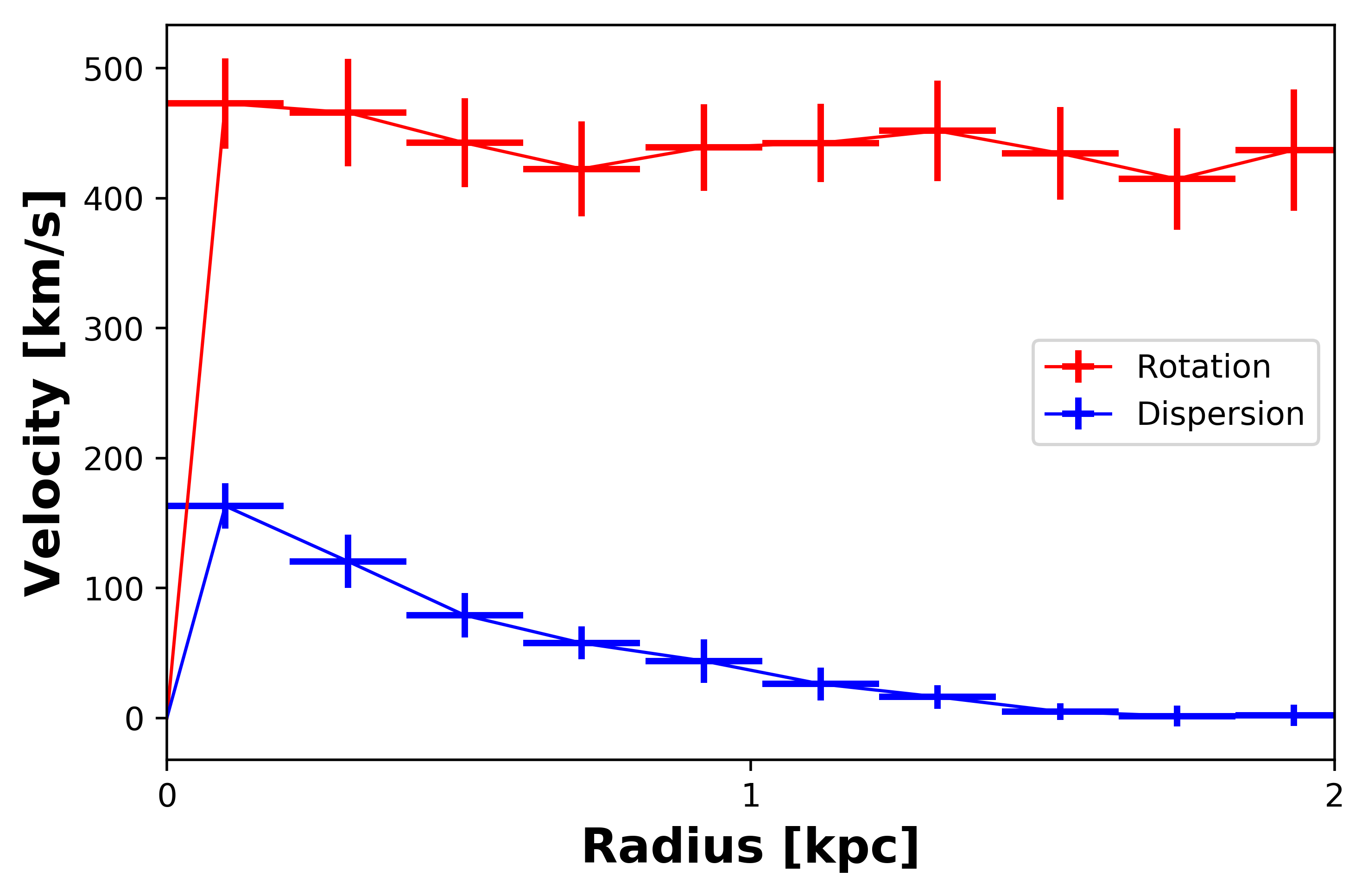}
\includegraphics[width=0.47\textwidth]{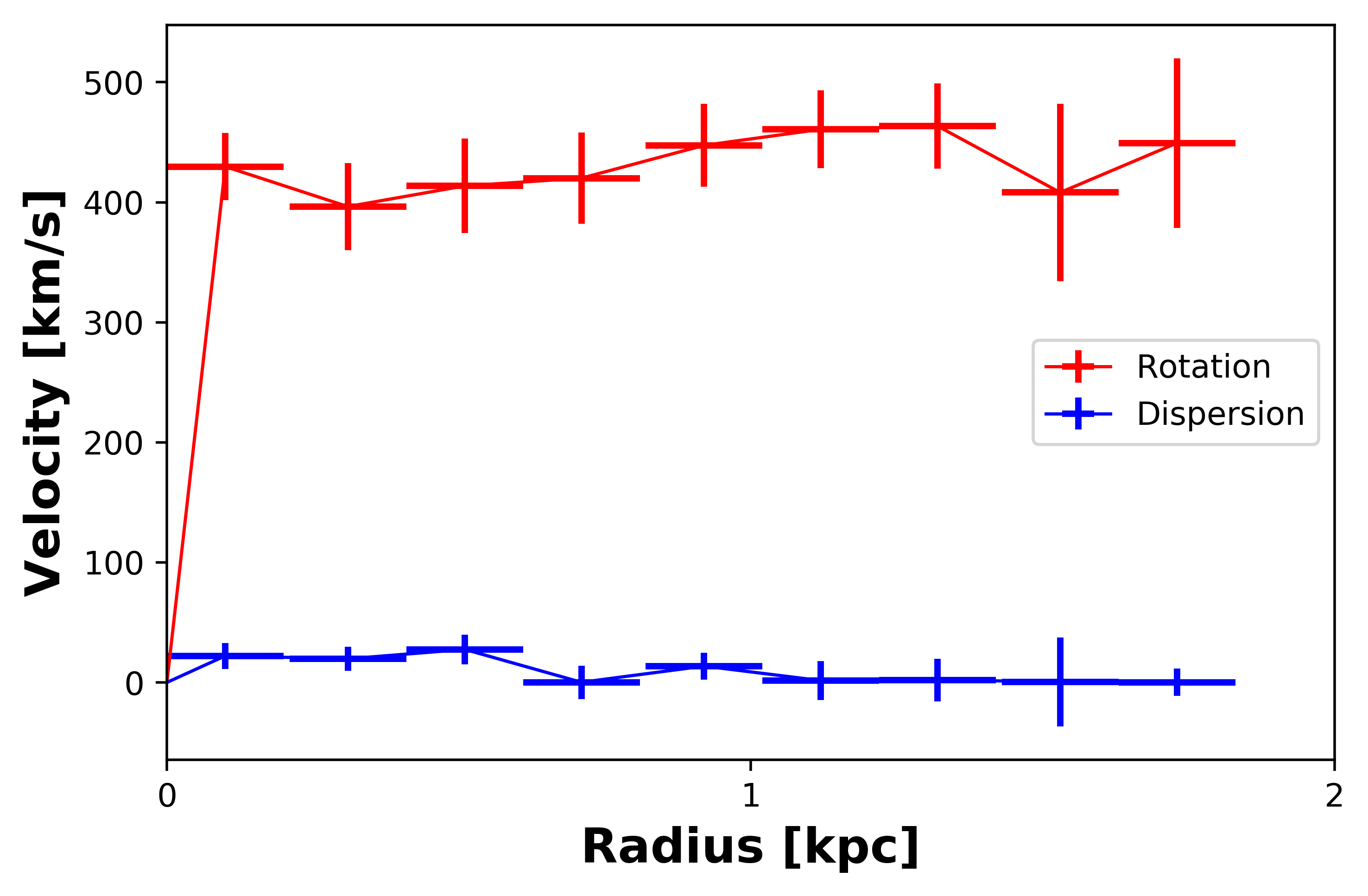}
\caption{Comparison of best-fit rotation curves (red) and velocity dispersion profiles (blue) for local normalization (top row) and azimuthal normalization (bottom row), as well as both CO (left column) and H$_2$O (right column).}
\label{RCVD}
\end{figure*}

By applying this tilted ring fitting code to the CO(7-6) and H$_2$O data cubes, we found good fits (as seen in the comparisons of Fig. \ref{12plots}). This implies that ID141 is a strongly rotation-dominated disk, a fact that is also apparent from the rotation curves shown in Fig. \ref{RCVD}. The resulting dynamical masses and galaxy properties are listed in Table \ref{bbout} and show close agreement with each other, regardless of the line or normalization employed (local or azimuthal). Comparing these dynamical mass estimates with our ISM mass estimates in the preceding section, we find a very significant discrepancy to the extent that the ISM masses are a factor of approximately four times those of the dynamical mass estimates. We discuss possible remedies for this disparity in Section \ref{sec_discussion}.

One of the strengths of three-dimensional dynamical characterization is the ability to compare the data and best-fit model on a pixel-by-pixel basis. Inspecting Fig. \ref{12plots}, it is clear that both CO and H$_2$O are poorly fit by models using azimuthal averaging (note the strong residuals in zeroth-moment maps and spectra). However, the velocity fields, velocity dispersion maps, and general shape of each position-velocity diagram are well-reproduced. This raises an interesting issue: the kinematics of the galaxy are well-fit by a rotating disk, but the distribution of emission (i.e., morphology) is not. This feature will be discussed in Section \ref{1or2} where we consider the possibility of ID141 being a merger of two or more galaxies.

\begin{table*}
\caption{Best-fit morphological (from a 2-D Gaussian fit to each zeroth-moment map) and kinematic (from {\tt $^{\mathrm{3D}}$Barolo} fitting) parameters.}
\centering
\begin{tabular}{cc||c|cc|cc|c}
\hline
Line & Normalization & Size & $\mathrm{i_M}$ & $\mathrm{i_K}$ & $\mathrm{PA_M}$ & $\mathrm{PA_K}$ & $\mathrm{M_{dyn} (10^{10} M_{\odot})}$\\
 & & (arcsec $\times$ arcsec) & ($^{\circ}$) & ($^{\circ}$) & ($^{\circ}$) & ($^{\circ}$) &\\  \hline 
 H$_2$O & Local & $(0.36\pm0.06)\times(0.23\pm0.05)$ & $50\pm12$ & $61\pm11$ & $66\pm17$ & $96\pm10$ & $6.9\pm1.8$\\
 H$_2$O & Azimuthal & $-$ & $-$ & $57\pm12$ & $-$ & $103\pm5$ & $8.1\pm2.4$\\ \hline
 CO & Local & $(0.39\pm0.05)\times(0.33\pm0.04)$ & $33\pm16$ & $54\pm10$ & $80\pm31$ & $95\pm6$ & $7.7\pm1.7$\\
 CO & Azimuthal & $-$ & $-$ & $65\pm9$ & $0$ & $96\pm7$ & $8.6\pm2.0$\\ \hline
\end{tabular}
\label{bbout}
\end{table*}

\section{Discussion}
\label{sec_discussion}

\subsection{One galaxy or several?}
\label{1or2}

While the strong emission lines in this paper (i.e., CO and H$_2$O) are well-fit by {\tt $^{\mathrm{3D}}$Barolo} (see Section \ref{bbres}) and return similar dynamical masses, the source morphologies are complex, resulting in imperfect fits. One key piece of information in interpreting this issue is the lack of symmetry between the velocity field structure to the north and south. While the velocity field to the north is a continuation of the central gradient, it extends to the east, and lacks a strong southern counterpart. When combined with the strong line emission and the H$_2$O dispersion peak in the northern region, this raises the possibility that the northern emission is not an extension of ID141, but a separate galaxy. 

\begin{figure*}
\centering
\includegraphics[width=\textwidth]{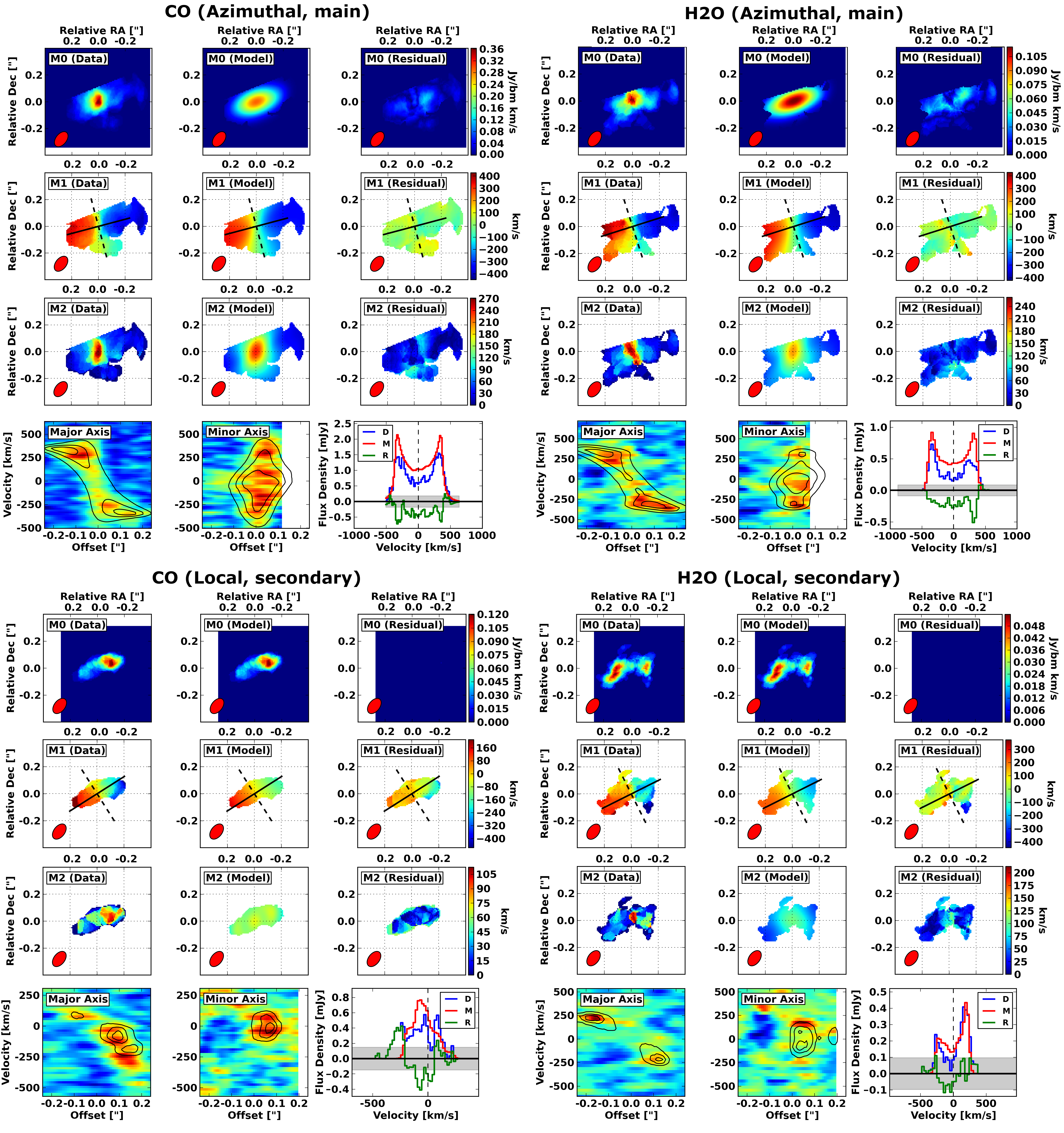}
\caption{Comparison of observed emission lines and best-fit {\tt $^{\mathrm{3D}}$Barolo} models for masked CO (left column) and H$_2$O (right column) data cubes. In each set of panels, we present the zeroth-moment (i.e., integrated intensity) maps (top row), first-moment (i.e., line-of-sight velocity) maps (second row), and second-moment (i.e., velocity dispersion) maps (third row) for the data (left column), model (middle column), and the residual between the two (right row). We also present the position-velocity diagrams taken along the major axis (bottom row, left column) and minor axis (bottom row, middle column) of the data cube (colour) and model cube (contours). Finally, we compare the spectra of the data (blue), model (red), and residual (green) in the lower right panel, showing the $1\sigma$ noise level by the shaded grey region.}
\label{main12}
\end{figure*}

To investigate this possibility, we masked the northern section of the CO and H$_2$O data cubes, and fitted this ``main'' galaxy using {\tt $^{\mathrm{3D}}$Barolo}. In each case, we assumed azimuthal normalization and the same model central position as in the unmasked data cubes. The fit to these masked data significantly improve upon the azimuthal fit to the unmasked data with a disk-like model, as shown in Fig \ref{main12}. 

Next, we took the best-fit models of these masked CO and H$_2$O data cubes and subtracted them from the unmasked data cube ({\tt CASA} task {\tt immath}) to isolate the signal to the north (hereafter referred to as the `secondary'). As seen in the lower row of Fig. \ref{main12}, the secondary signal shows a continuous velocity gradient. The corresponding dynamical mass of this secondary is $(3.9\pm0.9)\times 10^{10}$\,M$_\odot$ and when added to the dynamical mass of the main galaxy, increases the total dynamical mass by approximately 50 per cent compared to the single-galaxy model. The secondary exhibits an offset velocity dispersion peak and two separate components in each zeroth-moment map. Because of these properties, {\tt $^{\mathrm{3D}}$Barolo} returns a poor fit using azimuthal normalization, and a better but still unsatisfactory fit using local normalization (which we also show in Fig. \ref{main12}). While {\tt $^{\mathrm{3D}}$Barolo} is able to reproduce the brightness distribution and overall velocity gradient of each data cube, the significant residuals in each moment map and position-velocity diagram suggest that this secondary source is not a single, well-ordered rotating galaxy.

Taking these findings into account, there are three obvious possibilities for the nature of ID141:
\begin{enumerate}
    \item The system may simply be a single rotating disk. In this case, the asymmetry to the north would be a minor disturbance.
    \item The source to the north may be a separate, rotating galaxy that is closely associated to ID141. This is supported by the continuous velocity gradients shown in the unmasked data cube (see Fig. \ref{12plots}) and reasonable fits to the H$_2$O emission from the `main' and `secondary' sources (see right column of Fig. \ref{main12}).
    \item Because the `secondary' source shows two kinematically-separated clumps in both CO and H$_2$O emission, it is conceivable that the ID141 system is composed of three significant clumps: the main galaxy, a source to the north that is strong in CO emission, and a third source to the northeast that is strong in H$_2$O emission. 
\end{enumerate}

Due to the diversity in observed continuum (Fig, \ref{cont_images}) and line (Fig. \ref{line_images}) morphologies, we are unable to ascertain which of these scenarios, if any\footnote{The strong water emission may also be indicative of an Eddington-limited starburst in which case the complex morphology may, in part, be an outflow. See \cite{vanderwerf2011} for example}, is the most likely without further observations although most bright SMGs are known to comprise mergers \citep[e.g.][]{engel2010}.

\subsection{Why do the ISM mass estimates agree so well and why are they greater than the total mass?}
\label{sec_ism_vs_gas_mass}

The five gas masses derived in Section \ref{sec_gasmass} (see Table \ref{tab_gas_mass}) have a remarkably small range of $(3.2-3.8) \times 10^{11}$\,M$_\odot$, especially since four of the methods use different tracers. They are also remarkably larger than the total mass estimated from the kinematics of the galaxy; both application of {\tt $^{\mathrm{3D}}$Barolo} and a simple estimate from $v^2r/G$ with velocity $v=430$\,km\,s$^{-1}$ and radius $r=2$\,kpc give a dynamical mass of $\sim 0.8 \times 10^{11}$\,M$_\odot$. We first consider possible observational explanations of the mass discrepancy and then suggest an astrophysical explanation of both points.

One possible explanation for the mass discrepancy is that the lens model and therefore the lens magnification is incorrect. In this case, the derived intrinsic source properties will be systematically biased. We can estimate the extent to which such an error will propagate into the ratio of gas mass to dynamical mass. Regarding the gas mass, since $M_g\propto L_{\rm int} \propto L_{\rm obs}/\mu$, where $L_{\rm int}$ and $L_{\rm obs}$ are the intrinsic and observed (i.e., lensed) line luminosities, the gas mass scales inversely proportionally with magnification. Under the assumption of circular orbits used by {\tt $^{\mathrm{3D}}$Barolo}, the dynamical mass scales as $M_{\rm dyn} \propto v_c^2 r_{\rm int}$, where $v_c$ is the circular velocity and $r_{\rm int}$ is some intrinsic physical length scale. Since the observed $v_c$ is invariant under lensing and the intrinsic physical length scales as $r_{\rm int} \propto r_{\rm obs}/\sqrt{\mu}$, the dynamical mass scales with magnification as $M_{\rm dyn} \propto \mu^{-1/2}$ and so $M_g / M_{\rm dyn} \propto \mu^{-1/2}$. Therefore, the magnification we have obtained from our lens modelling would have to be underestimated by a factor of 16 to obtain a gas mass that is equal to the dynamical mass. 

Although the statistical uncertainties we quote are orders of magnitude beneath this, larger errors can arise if the lens model we have adopted is not a faithful representation of the true lensing mass. As previously stated, changing our lens model by adding an additional mass profile to match a possible second lensing galaxy only modifies the magnification by approximately 5 per cent. An alternative is to use a mass profile that does not follow a power law. \citet{Schneider2013} present an example of a composite lens model comprising two significantly different density profiles with a resulting magnification that differs from the power-law model by 40 per cent but still broadly reproduces the observed lensed images. Even this rather extreme example is over an order of magnitude beneath the size of shift required to reconcile our dynamical and gas mass estimates.

Another possible explanation for the discrepancy is that the total (dynamical) mass has been estimated from the CO (7-6) and water lines while the ISM estimates have been obtained from observations of the continuum and different spectral lines. The \ci(1-0) line, which is probably the best tracer of the cool ISM but which is too weak for us to perform a full kinematic analysis, does have a different spatial distribution from the lines used to investigate the kinematics (Fig.~\ref{line_images}). However, the velocity extent of all three lines is very similar (Fig.~\ref{lines}) and since the kinematic analysis extends out to a similar radius to the \ci(1-0) emission (Fig.~\ref{lines}), it is unlikely that our estimate of the total mass is too low. A related issue that has been argued in the literature \citep[e.g.,][]{daCunha2013,Zhang2016} is that the mass estimates of high-redshift galaxies may be too low because the increasing energy density in the cosmic microwave background with redshift makes it increasingly hard to detect extended gas. This is unlikely to remedy such a large discrepancy because, to first order, this effect simultaneously reduces our measurements of the total mass (by causing an under-estimation of dynamical radius) and the ISM mass.

An additional concern to consider is in the kinematic analysis from Section~\ref{1or2}. We were ultimately unable to confirm that ID141 is a single rotating disk rather than several galaxies. The strongly-lensed SMG SMM\,J02399-0136, initially identified as a large disk galaxy \citep{genzel2003} and subsequently resolved into a complex system of several components \citep{ivison2010} is an example of why caution must be taken. Nevertheless, in the case of ID141, this is unlikely to explain the large discrepancy. If we consider option (ii) in Section~\ref{1or2}, that ID141 consists of two rotating disks, this gives the largest of our dynamical mass estimates of $1.3 \times 10^{11}$\,M$_\odot$ which is still significantly short of explaining the discrepancy. 

Finally, we note that the discrepancy becomes even worse if we include stellar mass. Our estimate is that the stellar mass in this system is M$_\star = 4.8^{+2.5}_{-2.3} \times 10^{11}$\,M$_\odot$, which in itself is greater than the dynamical mass. However, the uncertainty on this measurement is high and furthermore, does not include any uncertainty arising from removal of the lens light.

We now consider a possible astrophysical solution, which might also partly explain the surprising agreement between the mass estimates made with four different tracers, a solution which arises because none of the methods is really independent. The dust method of \cite{scoville2016} is calibrated using CO(1-0) measurements of galaxies and is based on the CO(1-0) calibration factor derived for giant molecular clouds in the Milky Way; the calibration factor for \cii \citep{zanella2018} is based on dust measurements, which are again ultimately based on the CO(1-0) calibration factor from galactic giant molecular clouds; and all of the calibration factors discussed in \citet{dunne2021a} are based on the assumption that the GDR is the same as in the Milky Way. All the methods are ultimately benchmarked to the same astrophysical source, for the obvious reason that it is only in the Milky Way (and to a much lesser extent in galaxies in the Local Group) where it is possible to make the direct measurements of the ISM masses that are needed to calibrate the tracer methods.

A possible solution is if the calibration factors for ID141 are very different to the Galactic ones. To explain the agreement observed between the different mass estimates, all of the gas-to-tracer ratios would have to be $\simeq$4 times lower than in the Milky Way. Since all of the tracers are composed of heavy elements, the most direct way to bring some insight to the problem is with a chemical evolution model.

\subsection{Chemical Evolution in the Epoch of Galaxy Formation}
\label{sec_chem_evol}

In this section, we describe the results of a chemical evolution model we have constructed to investigate the likely evolution of the interstellar dust, gas and stellar mass of ID141. The details of this model can be found in Appendix \ref{app_chem_evol}. We have used this model to investigate the evolution of the GDR. Although we have not modelled the evolution of the other tracers (gas-to-CO etc.), it is likely that the results will be very similar since all of the tracers must depend on the abundances of the metals in the galaxy. The similarity of the ISM estimates for four different tracers provides strong circumstantial evidence in support of this.

In the model, which is an extension of the model in \citet{DeVis2017b}, the galaxy starts as a cloud of gas with no heavy elements that is  gradually converted into stars. The model incorporates gas flowing in to the galaxy via accretion from the cosmic web and out from it via stellar feedback and active galactic nuclei. Dust is produced by stars, both asymptotic giant branch stars and supernovae, and grain growth and it is destroyed by astration, supernova shocks and outflows. 

We considered two star formation histories, one with continuous star formation starting at some formation redshift, $z_{\rm f}$, and a second where we are currently observing 40\,per\,cent of the way through a burst lasting $200\rm \,Myr$\footnote{The duration of high-$z$ starbursts are thought to range from 50-250\,Myr, e.g. \citet{Danielson2017}; typical duty cycles of high-redshift SMGs are thought to range from 40-100\,Myr \citep{Coppin2008,Toft2014,Narayanan2015}.}.  Hereafter, we refer to these as the `conSF' and the `SB' model. 

To account for possible variations in the IMF (Appendix~\ref{app_chem_evol}), we used three functions that span most of the range of IMFs considered in the literature, namely a Salpeter IMF \citep{Salpeter1955}, the Chabrier IMF \citep{chabrier2003} and the more top-heavy IMF proposed by \citet{Zhang2018}, required to explain isotope ratios at high redshifts\footnote{Note that the so-called top heavy IMF in this work is different to the top heavy IMF evoked by \citet{baugh2005} to explain number counts of bursty submillimetre galaxies (Appendix~\ref{app_chem_evol}).}. In Section \ref{sec_sed} we estimated a SFR for ID141 from its far-infrared luminosity of 2400 $\pm$ 500\ M$_\odot {\rm yr^{-1}}$, based on the assumption of an IMF from \citet{kroupa2003}. If we instead use the IMFs considered here, we obtain an SFR of 2800, 1800 and 1300 M$_\odot {\rm yr^{-1}}$ for the Salpeter, Chabrier and top-heavy IMF. In our model we set $z_{\rm f} = 4.8$, $5.5$ and $7$ respectively (the higher observed current-day SFRs for the Salpeter and Chabrier IMFs leads to a more rapid conversion of gas into stars, and so their star formation needs to begin later). The parameters of each model have been chosen so that at the redshift of ID141 ($z = 4.24$) the SFR predicted by the model is similar to the SFR estimated for ID141 from its far-infrared luminosity.

For the conSF star formation model, we assume a SFR that follows the Kennicutt-Schmidt law \citep{Kennicutt1998}, i.e., $\psi(t) = \epsilon M_g^k(t)$ where $\epsilon$ is the star formation efficiency, $M_g$ is the total gas mass and $k$ is set to unity.  $\epsilon$ is estimated using $\psi = \epsilon M_g/t_{\rm dyn}$ where $t_{\rm dyn}$ is the dynamical (free-fall) timescale $t_{\rm dyn} = \sqrt{r^3/2GM}$ \citep{Walter2009,Swinbank2010} and $\psi$ is the observed SFR (see also Appendix~\ref{app_chem_evol}). Given an effective galaxy radius of 1.4\,kpc (measured from the reconstructed source continuum) and an observed ISM mass of $3.5\times 10^{11}$\,M$_{\odot}$, the maximum SFE is 1\,per\,cent for the Salpeter IMF. If we instead assume that the true ISM mass must be less than this (because of the limit from the dynamical mass) and set the ISM mass equal to the dynamical mass of the one-disk model ($M_{\rm dyn}\simeq0.8\times10^{11}$\,M$_\odot$), the range of maximum SFE obtained is 5 and 2\,per\,cent for the Salpeter and top-heavy IMFs respectively. Although ID141 is one of the most extreme objects in the universe in terms of its SFR, both estimates of SFE suggest that it is not forming stars at a significantly more efficient rate than normal systems. The SFE is similar to other high redshift SMGs \citep{Swinbank2010,Riechers2013} but an order of magnitude lower than the $z=6.4$ hyper-starburst from \citet{Walter2009}. Using the current observed level of SFR and assuming that $M_g \approx M_{\rm dyn}$ when the galaxy started forming stars, the gas reservoir would be exhausted within 40\,Myr for a Salpeter IMF if there is no additional replenishment of gas via mergers or inflows. 

In Table~\ref{tab:model}, we compare the model outputs predicted at the redshift of ID141 of $z=4.24$. The Salpeter IMF models produce lower dust masses (since there are fewer high mass stars producing metals and dust and there is more gas being locked up in stellar remnants) and higher stellar masses than the Chabrier and top heavy IMF models. The resulting gas fractions are also lower, indicative of a much later evolutionary stage. \citet{Dave2011} predict gas fractions of $\sim 0.4$ at $z=4$ for a current day halo mass of $M_h(z=0) = 10^{14}\,M_{\odot}$. This is larger than the Salpeter IMF results but more in line with the results from the starburst, top heavy IMF model.  The total baryonic mass ($M_g +M_*$) predicted by the conSF models is $\sim 5$ times larger than the observed dynamical mass of ID141, though the starburst models predict a more comparable mass ($\sim 1-3\times$).

\begin{table*}
    \centering
    \begin{tabular}{l|l|l|l|l|l|l|l|l} \\ \hline
    Model & $f_g$ & SFR  & $ M_g$ & $M_\star$ & $M_{\rm tot}$ & $Z$ & $M_d$ & GDR  \\  
     & & $({\rm M_\odot yr}^{-1})$  & $(10^{11}{\rm M}_{\odot})$ & $(10^{11}{\rm M}_{\odot})$ & $( 10^{11}{\rm M}_{\odot})$ & & $(10^{9}{\rm M}_{\odot})$ &  \\ \hline 
   {\bf Continuous (conSF)} & & & && & & &  \\
    Top Heavy &0.24 & 1336 & 1.16 & 3.66 & 4.83 & 0.06 & 4.10 & 28 \\
     Chabrier & 0.20 & 1754 & 1.14 & 4.61 & 5.91 & 0.03 & 1.65 & 69 \\
    Salpeter & 0.23 & 2790 & 1.30 & 4.39 & 5.69 & 0.02 & 1.30 & 100 \\ \hline
     {\bf Starburst (SB)}  & & & && & & &  \\
     Top Heavy& 0.50 & 1300 & 0.79 & 0.80 & 1.60 & 0.04 & 1.84 & 43 \\
    Chabrier & 0.25 & 1800 & 0.48 & 1.41 & 2.71 & 0.03 & 0.71 & 68 \\
    Salpeter  & 0.15 & 2800 & 0.41 & 2.39 & 2.80 & 0.02 & 0.51 & 80 \\
    \hline
    \end{tabular}
    \caption{ The values predicted by our chemical evolution models for the
    properties of our ID141-like galaxy at the redshift at which we observe
    ID141: $z=4.24$. We give results for two different star-formation histories  - starburst (`SB') and continuous (`conSF') - and for the Salpeter, Chabrier and top heavy initial mass functions. $f_g$ is the gas fraction. GDR denotes the gas-to-dust ratio and M$_{\rm tot}$ is the total (gas mass + stellar mass) returned by the model. The metallicity, $Z$, is defined as the ratio of the metal mass and gas mass. By design, the SFR predicted by each model is similar to the SFR estimated for ID141 from its far-infrared luminosity.}
    \label{tab:model}
\end{table*}

\subsubsection{Evolution of the Gas to Dust Ratio}

\begin{figure}
    \centering
    \includegraphics[width=\columnwidth]{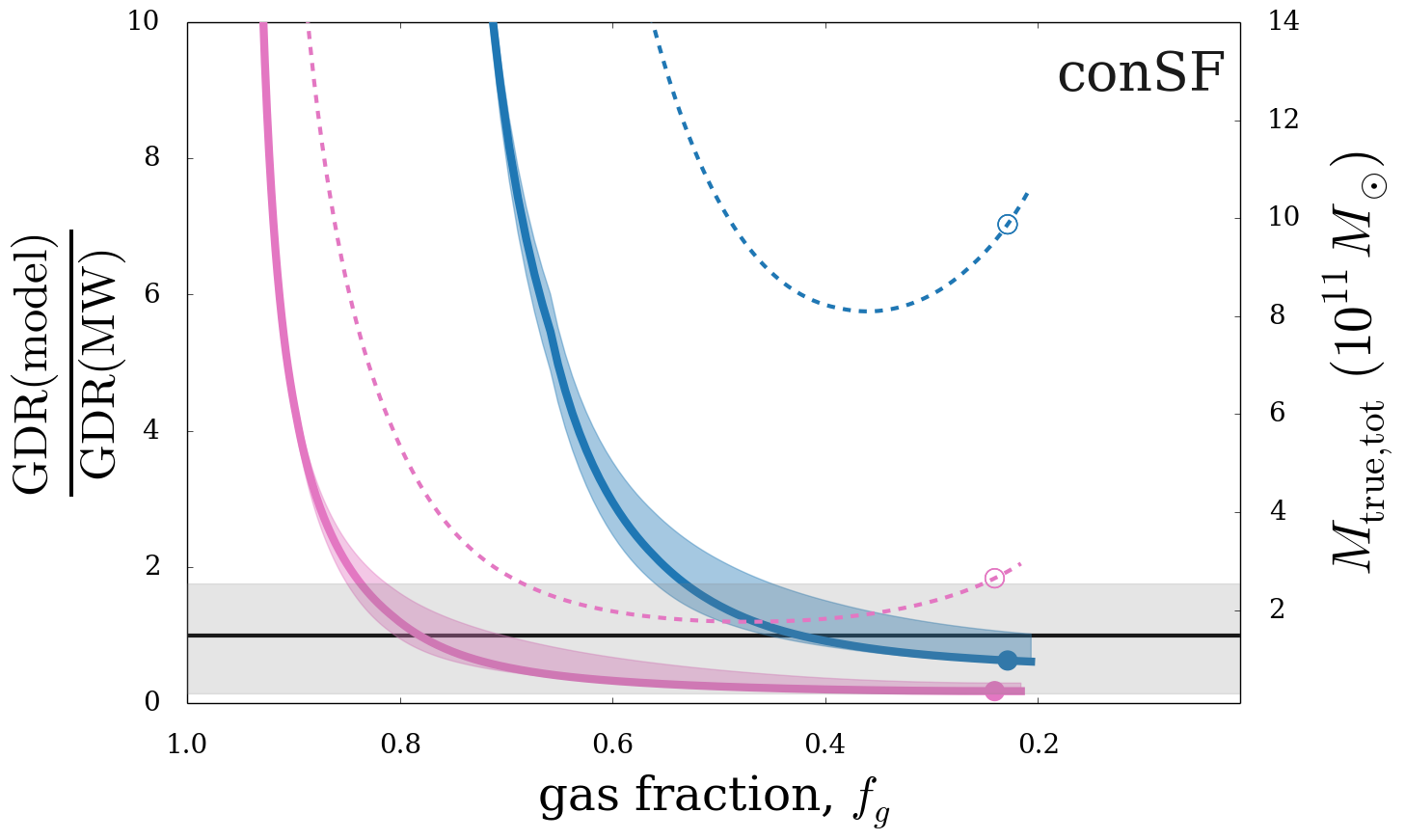}
    \includegraphics[width=\columnwidth]{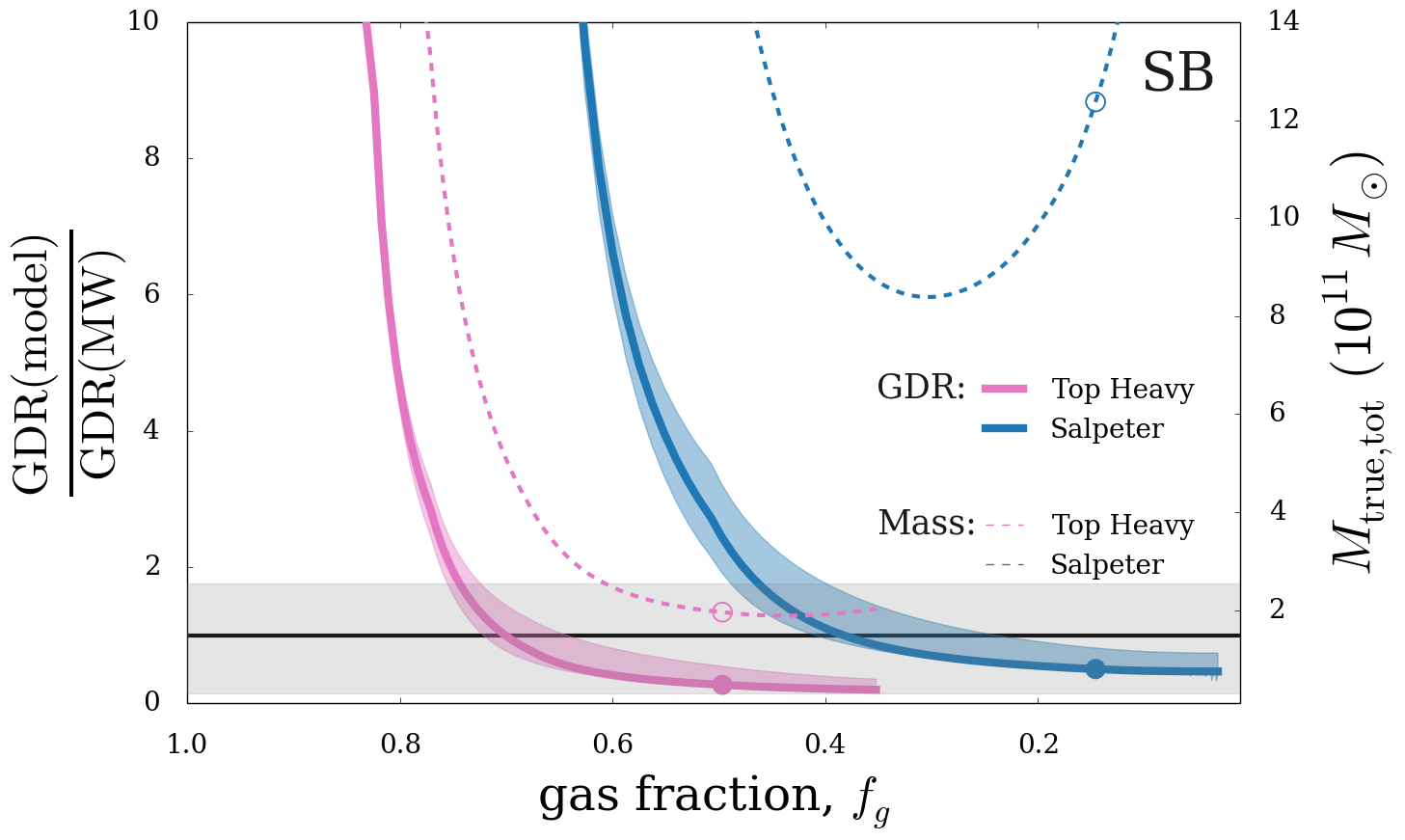}
    \caption{The gas-to-dust ratio (solid lines), as a fraction of the Milky Way value, predicted by standard dust-evolution models plotted against $f_g$, the fraction of baryon mass in the form of gas for {\it top:} the continuous star forming (conSF) model and {\it bottom}: the starburst (SB) model. The pink and blue colours denote the top-heavy and Salpeter IMFs respectively and the circles (filled and solid) show the values at $z=4.24$. The dashed lines show the `true' total mass ($M_{\rm true, tot}$) derived from the GDRs from the models and the observed gas mass assuming a MW-like GDR (Eqn.~\ref{eq:mtrue_tot}) as the galaxy evolves. The black horizontal line indicates a GDR equal to the Milky Way.  The shaded grey region indicates $M_g + M_\star \lesssim M_{\rm dyn}$, where we use the maximal value ($M_{\rm dyn} = 1.3\times 10^{11}\,\rm M_{\odot}$) with a $3\,\sigma$ uncertainty.}
    \label{fig:GDR}
\end{figure}

In Fig.~\ref{fig:GDR} we show the evolution of the model GDR as a ratio of the Milky Way GDR, $(M_g/M_d)_{\rm model}/(M_g/M_d)_{\rm MW}$, for each model plotted against $f_g=M_g/(M_g+M_\star)$, the fraction of baryonic mass that is gas. For clarity, we only show and discuss the results for the Salpeter and top heavy IMFs as the results assuming a Chabrier IMF sits mid-way between the two. The shaded area around each line shows the effect of changing the parameter governing the rate of dust grain growth in the dense clouds (Appendix~\ref{app_chem_evol}) from 700 to 15,000, where the former value is appropriate for the Milky Way today and the latter indicates extremely rapid grain growth \citep{Mattsson2012,Asano2013,DeVis2017b}. The GDR is high at high gas fractions because the abundance of metals (and therefore dust) is still low. The GDR decreases as the gas is converted into stars and as dust is formed in stellar winds, supernovae and, after reaching some critical metallicity, via grain growth.

We have assumed a value for the GDR ratio of the Milky Way that is appropriate for the \citet{scoville2016} calibration used in Section~\ref{sec:gas_dust_scov} to convert dust mass and submm continuum luminosity to gas mass. We have estimated this by assuming $\beta=1.8$, $T_{\rm d}=25\,\rm K$, $\kappa_{850} = 0.077\,\rm m^2\,kg^{-1}$ \citep{james2002}, and $\kappa_{\rm ISM} = 4.8\times 10^{-4}\rm \,m^2\,kg^{-1}$ \citep{scoville2014}. Since the models account for the total gas, whereas the calibration of dust-to-gas ($\alpha_{850}$) provided by Scoville et al. is for the molecular component only, we follow \citet{scoville2014} and assume that there are roughly equal amounts of H{\sc i} and $\rm H_2$. We also account for Helium.  This gives a GDR for the Milky Way of 160 given the assumed $\kappa_{850}$ (a factor of $\sim$1.2 higher than that assumed in \citealt{dunne2021a}). Table 7 shows that, for all models, the GDR reached by redshift 4.24 is $<160$ (i.e., the ratio of the model GDR to the Milky Way GDR is less than unity), suggesting \textit{it is not appropriate to assume a Milky Way calibration when estimating gas masses from the dust emission in SMGs like ID141.} Such low GDRs have been measured in other high redshift SMGs \citep[see, for example,][and references therein]{berta2021, yang2017}.

Fig.~\ref{fig:GDR} also shows that the GDR can vary by over an order of magnitude during the SMG phase of a galaxy. Therefore it is also dangerous to assume a constant value for the GDR when estimating the ISM mass from the submm continuum emission. However, we can use the model to our advantage: the results in Fig.~\ref{fig:GDR} provide a way to use the model itself to estimate the true GDR, and thus the mass of the ISM. We can also use it to probe how far through ID141 is in its conversion of gas into stars i.e., its evolutionary phase. Defining $M_{g}({\rm obs})$ to be the ISM mass we have estimated on the assumption of the Galactic GDR, the true ISM mass, including the contribution of the atomic gas, is given by:
\begin{equation}
    M_{g}({\rm true}) = M_{g}({\rm obs}) \dfrac{(M_g/M_d)_{\rm model}}{(M_g/M_d)_{\rm MW}}\, .
\end{equation}

\noindent The total mass of the model galaxy is then
given by:
\begin{equation}
    M_{\rm true,tot} =\dfrac{M_{g}({\rm true})}{f_g}.
    \label{eq:mtrue_tot}
\end{equation}

\noindent We have made the assumption that there is no contribution from
dark matter \citep[see][]{genzel2017}, but if there is a contribution from
dark matter it will strengthen the conclusions below. 

We show the evolution of $M_{\rm true,tot}$ in Fig. \ref{fig:GDR} as dashed lines. The shaded horizontal region indicates where the total baryonic mass (gas plus stars) is consistent with the dynamical mass ($M_g + M_\star \leq M_{\rm dyn}$) within the uncertainty, defined as 3\,$\sigma$. 

We can immediately draw the following conclusions:
\begin{enumerate}
    \item it is only possible to make the masses consistent if the IMF is top-heavy, adding to the evidence from isotopic abundances \citep{Zhang2018} that the IMF is different in high-redshift star-forming galaxies. 
    
    \item  It is possible to make the masses consistent when the
    galaxy is roughly halfway through its evolution ($f_g \sim 0.5$). Although it is often assumed that galaxies like ID141 must have very high gas fractions, the discrepancy between the mass estimates gets rapidly worse above $f_g \sim 0.6$.
\end{enumerate}

We also notice that the true total mass calculated using Eq.~\ref{eq:mtrue_tot} overestimates $M_{\rm dyn}$ in comparison to the model $M_{\rm tot}$ when assuming a Salpeter IMF. This is due to a combination of the faster consumption of gas due to the higher SFR and the lower dust masses formed per stellar population with a Salpeter IMF. For example, to reach a total mass $\lesssim 1\times 10^{11}\,\rm M_{\odot}$ using  Eq.~\ref{eq:mtrue_tot}, the conSF Salpeter model would require a ${\rm GDR} / f_{\rm g} \lesssim 50$, or alternatively, for a galaxy with $f_{\rm g} < 0.5$, a GDR of $\lesssim 25$ is needed.  Such low GDRs are not reached in the Salpeter IMF models.  

Although the conSF top heavy IMF appears to result in Eqn.~\ref{eq:mtrue_tot} producing a total mass within the shaded region, this is misleading. The variation of the GDR ratio with $f_g$ from this model is low enough that scaling the observed gas mass produces a true total mass from Eqn.~\ref{eq:mtrue_tot} that lies below the dynamical mass limit (in the shaded region). However, with the continuous SF model, such low GDRs can only be produced in a model that has already consumed a lot of its gas and built up a large reservoir of stars, hence violating the mass criteria we are trying to satisfy ($M_{\rm tot}$ (conSF, top heavy) $> 4\times 10^{11}\,\rm M_{\odot}$ in Table~\ref{tab:model}). 

Finally, we note that if we increase the amount of gas accreted in the model, although this would increase the gas fraction, it would also increase both the GDR and the stellar mass, hence would not solve this issue.

\subsubsection{Metallicity Evolution}

\begin{figure*}
    \centering
    \includegraphics[width=\columnwidth]{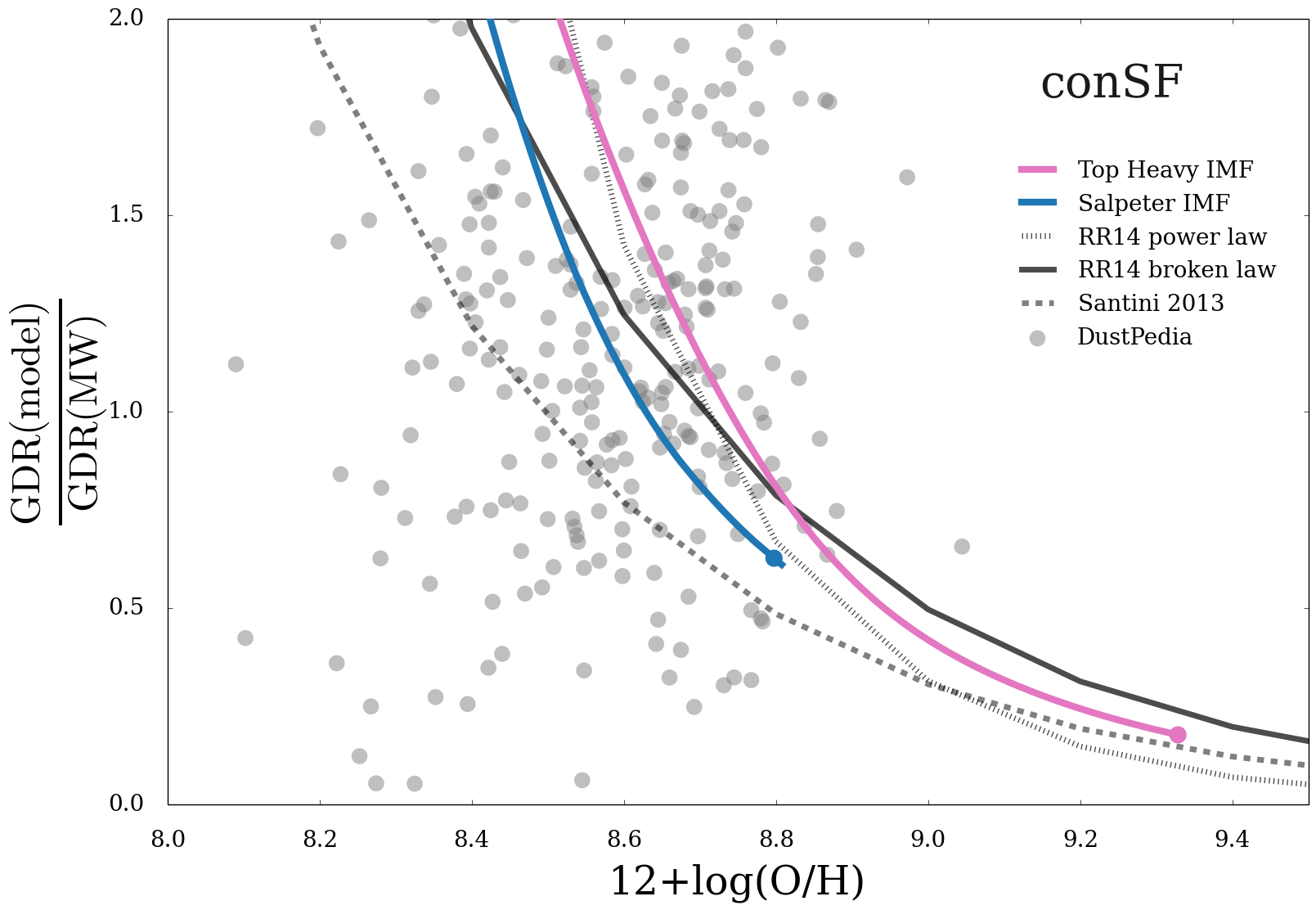}
    \includegraphics[width=\columnwidth]{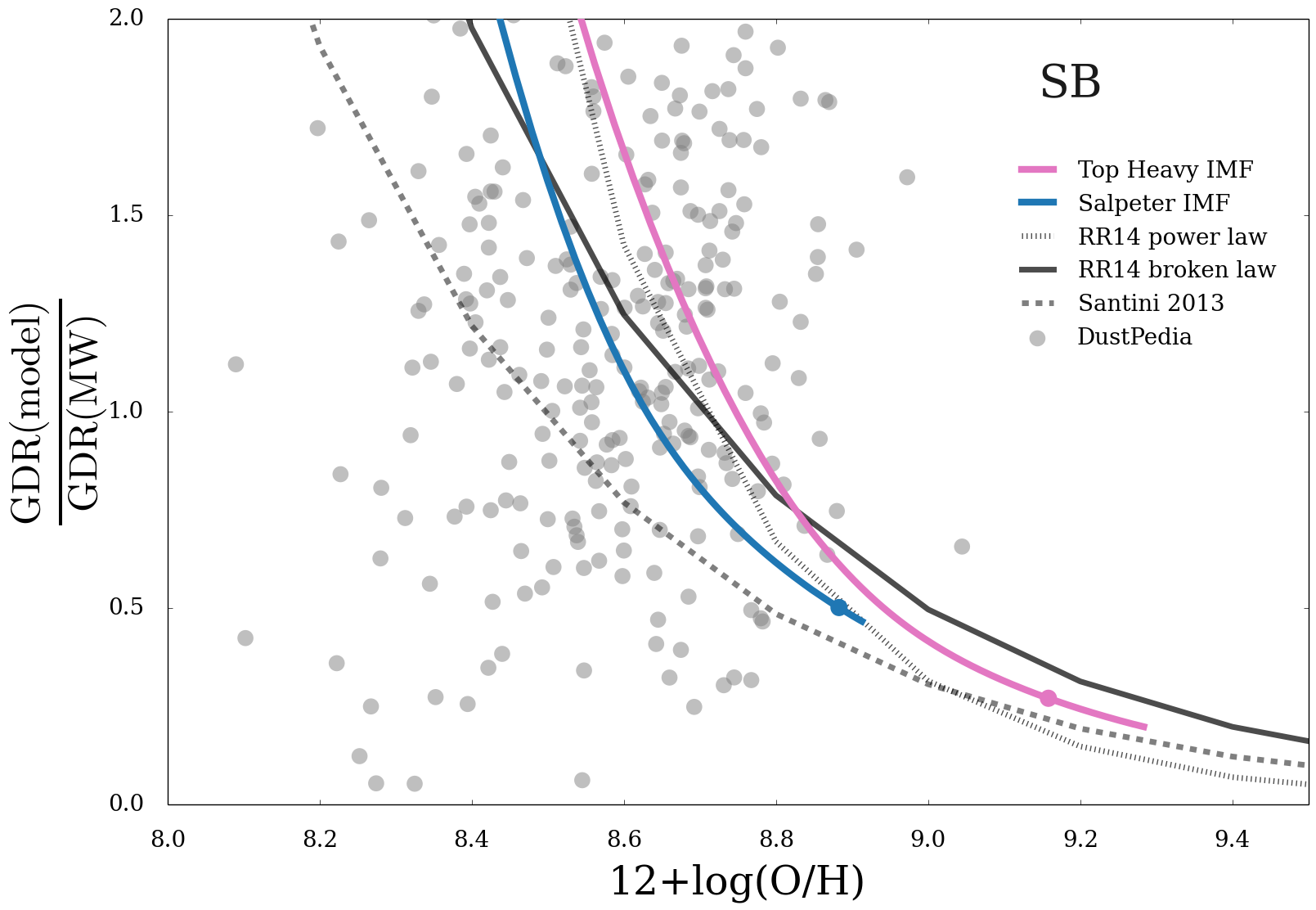}
    \caption{The gas-to-dust ratio (solid lines), as a fraction of the Milky Way value, predicted by the dust-evolution models as a function of metallicity. {\it Left:} the continuous star forming (conSF) model and {\it right}: the starburst (SB) model. The pink and blue colours denote the top-heavy and Salpeter IMFs respectively and the coloured circles show the values at $z=4.24$. The black lines show the predictions from (i) \citet{remyruyer2014} (their empirical fits to GDR and metallicity for local galaxies, note that their broken power laws calibrated to MW $X_{\rm CO,MW}$ or assuming $X_{{\rm CO},Z}$ are the same in this regime), (ii) \citet{santini2014}. The grey scatter points show the observed values for 454 local galaxies with metallicity measurements from DustPedia \citep{devis2019}(assuming metallicities calibrated using the O3N2 method, and gas masses derived using their Eqs. 7-9). 
    }
    \label{fig:metals}
\end{figure*}

The low GDRs we predict for ID141 imply these galaxies will reach supersolar metallicities by $z=4.24$ (where $Z=M_{\rm metals}/M_g = 1.5 - 4.4\,Z_{\odot}$\footnote{Solar metallicity taken from \citet{asplund2009}.}, Table~\ref{tab:model}) depending on the assumed IMF and SFH. To check if such high metallicities are expected in these galaxies, we next compare the evolution of the gas-to-dust ratios (Fig.~\ref{fig:GDR}) with metallicity in Fig.~\ref{fig:metals}. We convert the model oxygen abundance using $12 + {\rm log(O/H)} = 12 + {\rm log(X_{\rm O}/X_g)}$ where $X_{\rm O} = M_{\rm O}/16$ and $X_g = M_g/1.32$.  In these units, the metallicity reached by the model assuming a top-heavy IMF is $12 + {\rm log(O/H)}=9.33$ and $9.16$ for the constant SF and starburst respectively.

Local GDR-$Z$ scaling relations from {\it Herschel} surveys, e.g. \citet{remyruyer2014,devis2019}, are shown in Fig.~\ref{fig:metals}. These relationships imply values of $12+{\rm log(O/H)}$ that are consistent with the predicted model GDRs of 25-50 for ID141. At the same metallicity as the SB model with a top heavy IMF, the \citet{remyruyer2014} relationships predict GDRs ranging from 28-55. The properties of local galaxies from the DustPedia sample of \citet{devis2019} also reach low gas-to-dust ratios, but at lower metallicities than those predicted for ID141 (Fig.~\ref{fig:metals}). This sample is made up mostly of late type galaxies, with stellar masses $>10^{8}\,M_{\odot}$ making up only 4\,per\,cent.

We also show the relationship from \citet{santini2014} derived by combining observed dust masses of galaxies out to $z=2.5$ with metallicities derived from the stellar mass-SFR-metallicity relationship of \citet{mannucci2010}. Although a number of mass-metallicity relations exist in the literature (eg \citealt{bothwell2013,peng2015}, \citealt{maiolino2019} and references therein) we caution here that there are very few studies carried out at high enough stellar mass, at redshifts $>3$, and with SFRs $>100\,\rm M_{\odot}\,yr^{-1}$, making it difficult to apply these relationships to ID141.

What we can say, is that if we assume that ID141 will shortly turn into a massive quiescent early-type galaxy with very little future evolution, then our models predict high mass galaxies with super solar metallicities $\sim 1.3-3 \,Z_{\odot}$ at present day.  This is in agreement with observations of massive ellipticals in the local universe (\citealp{gallazzi2005,peng2015}, see also \citealp{maiolino2019}, their Fig.~12).

\subsection{Broader Implications of Low Gas-to-Dust Ratios}

The conflict between dynamical mass and ISM mass when estimating the latter using CO has been previously reported in the literature \citep[e.g.,][]{Tacconi2008,hodge2012,bothwell2013,rivera2018}. A common conclusion is to reduce the CO conversion factor, $\alpha_{\rm CO} \simeq 1$ \citep[but, in contrast, see][]{boogaard2021}. However, even with this low value, the tension between the dynamical mass and the ISM mass is sometimes still seen, requiring an additional explanation \citep[e.g. in changing the structure of the galaxy][]{bothwell2013}. Low values of $\alpha_{\rm CO}$ seemed plausible given the evidence for a low value in low-redshift Ultraluminous Infrared Galalaxies  (ULIRGs) \citep{downes1998}. More recently, this evidence has weakened \citep[][also see Dunne et al. {\it in preparation}]{papadop2012a,papadop2012b}, which makes the low $\alpha_{\rm CO}$ explanation of the mass tension given in these previous studies less likely. The advantage of this work is that we have used four different tracers of the ISM: \ci, \cii, dust and CO. Therefore, although we might resolve the tension in our ISM estimate from the CO tracer by reducing $\alpha_{\rm CO}$ by a factor of $\sim 4$, we would also have to reduce the three other calibration factors by the same amount. 

Given the quite startling agreement between the ISM mass estimates for ID141, which has been noticed for other galaxies \citep[e.g.,][]{dye2015}, it seems more natural to assume that the true explanation of the mass discrepancy is due to chemical evolution rather than the assumption that the physical conditions in the ISM are producing a low value of $\alpha_{\rm CO}$. The former naturally explains why the calibration factors for all four gas tracers are lower in ID141.

This may seem counter-intuitive at first since galaxies in the first few billion years might be expected to have higher GDRs than in low-redshift galaxies because there has been less time for heavy elements to form \citep[e.g., see][]{peroux2020}. There is one strong argument why, in this situation, common sense is probably wrong. All the high-redshift galaxies observed have been found because they contain a lot of dust or because they already contain a large mass of stars \citep{scoville2014,scoville2017,tacconi2018,millard2020}, and so there is a strong selection effect that will mean that samples of high-redshift galaxies are likely to contain galaxies in which the stellar mass or dust mass is already high. The simplest closed-box models for the formation of dust (\citealt{eales1996}, \citealt{Dunne2003}), for example, predict that the dust mass in a galaxy is at a peak when $f_g \sim 0.5$, and thus the galaxies in a sample compiled from a submm continuum survey are likely to have high dust masses and to have already turned much of their gas into stars.

If our conclusion that the gas-to-tracer ratio is $\simeq$4 times lower than in the Milky Way is true for high redshift galaxies in general, it would have a dramatic effect on our assessment of their evolutionary states, reducing all ISM mass estimates by a factor of 4. However, the variation in the GDR seen in our chemical-evolution model (Fig.~\ref{fig:GDR}) shows that calibrating all the tracers using ID141 would be just as risky as using the Galactic calibrations. The indication from Fig.~\ref{fig:GDR} is that there is no perfect calibration, and it is better to think of a gas-to-tracer ratio as being a new useful tool for assessing the evolutionary state of a galaxy.

\subsection{How quickly will ID141 become a quiescent galaxy?}

In this section, we consider the evolution of ID141 itself. In particular, we examine whether the evolution might be fast enough for it to join the population of quiescent galaxies that is seen at $z \sim 3$ \citep{glazebrook2017,girelli2019,valentino2020,forrest2020a,forrest2020b,merlin2019}. For reference, the time interval between the two redshifts is $\rm 7.2 \times 10^8\ yr$.

To begin addressing this question, we turn to our kinematic analysis (Section \ref{sec_srckin}) which shows evidence for a large rotating disk, with possibly one or two additional merging systems. Since high-redshift galaxies are often separated into two classes - disks or mergers - the kinematics of ID141 suggest that both are often going on at the same time. The escape velocity of the system is given by:
\begin{equation}
v_{\rm esc} = \sqrt{\frac{2 G M}{R}}.
\end{equation}
\noindent Given a dynamical mass of $\rm 1.3 \times 10^{11}\ M_{\odot}$ and a radius of 1.4 kpc, the escape velocity is $\rm \sim 910\ km\ s^{-1}$. This is similar to the velocity range of the system (Fig.~\ref{lines}), which suggests that most of the material we see is destined to end up in a single galaxy.

There are two other observational results that are relevant to the dynamical evolution of the system. First, the \ci(1-0) image (Fig.~\ref{cont_images}), which likely gives us the most unbiased view of the gas, shows that the gas is divided into a number of clumps. The existence of clumps in cleaned interferometric data needs to be treated with care \citep[e.g., see][]{ivison2020} since there are observational effects that can lead to them being falsely identified. However, when clumps are multiply-imaged in a strong lensing system such as ID141, these observational effects are mitigated. A particularly striking example of this is in the system SDP81 which unequivocally shows multiply-imaged clumps in high resolution, high signal-to-noise ALMA data \citep{dye2015}. 

Clumps are expected to migrate into the centre of the galaxy through dynamical friction on a timescale given by \citep{genzel2011}:
\begin{equation}
t_{\rm inspiral} \simeq \left( \frac{v_c}{\sigma_0} \right)^2 t_{\rm dyn}\, ,
\end{equation}
\noindent where $\sigma_0$ is the velocity dispersion and $v_c$ is the circular velocity.  The ratio of $v_c$ and $\sigma_0$ is $\simeq$ 10, giving a migration timescale of $t_{\rm inspiral} \simeq 100 t_{\rm dyn} \simeq 2.3\times10^8\ {\rm yr}$.

Alternatively, we can consider the dynamical evolution of the system by calculating the Toomre Q-parameter \citep{toomre1964}, which quantifies the stability of rotating, thin symmetric disks against gravitational collapse \citep{toomre1964}. For a flat  rotation curve, the Q-parameter is given by \citep{genzel2011}:
\begin{equation}
Q_{\rm gas} =  \sqrt{2} \, \frac{\sigma_0}{v_c} \, 
\frac{M_{\rm tot}}{M_{\rm gas}}\, .
\end{equation}
Using the results of our chemical evolution model, we assume $M_{\rm gas}/M_{\rm tot} \simeq 0.5$. With this value of the mass ratio, the results from the kinematical modelling of the CO line (Fig. \ref{RCVD} - left panels) imply that $Q\simeq$0.6 in the central few hundred parsecs of the galaxy, falling to a lower value beyond a radius of 1 kpc. The results from the kinematical modelling of the H$_2$O line (Fig. \ref{RCVD} - right panels) imply much lower values of $Q_{\rm gas}$. Our results therefore show that the gas disk is unstable. We reached the same conclusion for the disk of the other gravitationally-lensed {\it Herschel} source that has been observed at high resolution with ALMA, SDP81.

All of these results show that the dynamic evolution of ID141 is very fast and suggest that it could easily turn into a single fairly undisturbed galaxy by $z \sim 3$.

Turning now to the evolution of ID141's gas, the only chemical evolution models that produce galaxies consistent with the tight mass limits are those with a top-heavy IMF and a starburst (Table~\ref{tab:model}). The conSF model has a gas mass and star-formation rate at $z = 4.24$ of $1.16 \times 10^{11}$\ M$_{\odot}$ and $1336$\ M$_{\odot}\ \rm yr^{-1}$ respectively, but the model predicts a stellar mass greater than the dynamical mass limit. However, if we assume that the galaxy continues to form stars at the same rate, and that the galaxy accretes no fresh gas, the gas will be consumed in $\simeq 9\times 10^7\ {\rm yr}$. If we perform the same calculation with the SB model, we obtain a timescale of $\simeq 6 \times 10^7\ {\rm yr}$.

In conclusion, both the dynamics and the other properties of ID141 are consistent with the idea that, although it is an extreme system at $z=4.24$, by a redshift of $z \sim 3$ it will have evolved into a single, fairly undisturbed galaxy, containing little gas.


\section{Conclusions}
\label{sec_summary}

ID141 is a gravitationally-lensed hyper-luminous SMG at a redshift of 4.24 with a star-formation rate of 2400 $\rm M_{\odot}\ yr^{-1}$, making it an extreme example of an already extreme class. The system was chosen as a target for a comprehensive high-resolution investigation of the phases of the ISM in high-redshift galaxies because at its redshift, the key ISM spectral lines fall in the ALMA bands. In this paper we have presented high-resolution observations in the \ci(1-0) line, the \ci(2-1) line, the CO(7-6) line and the $\rm H_2O (2_{1,1} - 2_{0,2})$ lines. After the lensing reconstruction, the images in the source plane reach down to a resolution of $\rm \simeq350\ pc$. In this section, we outline the  observational results and our inferences from these results. 

Our observational results are:
\begin{enumerate}

    \item The gas is distributed over a region of $\simeq 3$\,kpc. The \ci(1-0)  emission is offset by $\simeq 1$\,kpc from the other lines, including the \ci(2-1) line. This shows that the galaxy contains a large reservoir of cool gas that is missed by the other spectral lines that we have observed with ALMA, although the estimates of the ISM mass from the \ci(1-0) and CO(1-0) lines
    are remarkably similar (see below).
    
    \item We have used the ratio of the \ci(2-1) and \ci(1-0) lines to measure the variation in excitation temperature over the galaxy. We find that the temperature increases from $T_{\rm ex} \sim 40$K where the \ci(1-0) emission peaks to $T_{\rm ex} \gtrsim 100$K where the CO and water lines are at a peak.
    
    \item The gas disk appears to be broken into clumps.
    
    \item We have used the CO and water lines to carry out a kinematic analysis of the galaxy. We find strong evidence of a rotating system but cannot distinguish between there being a single rotating galaxy or there also being one or two additional galaxies. Our estimate of the dynamical mass depends on which of these scenarios we choose, but our estimates lie in the range $0.8-1.3 \times 10^{11}$\ M$_{\odot}$.
    
    \item We have estimated the ISM mass in the galaxy five ways: (1) from the \ci(1-0); (2) from the dust continuum emission: (3) from a measurement of the global CO(1-0) flux; (4) from a measurement of the global flux in the \cii 158-$\mu$m line; (5) from a method that combines (1), (2) and (3). Our mass estimates lie in the range $\rm 3.2-3.8 \times 10^{11}$\ M$_{\odot}$.
    
    \item After subtracting the lens from a {\it Spitzer} image, we estimate that the stellar mass of ID141 is M$_\star = 4.8^{+2.5}_{-2.3} \times 10^{11}$\,M$_\odot$.
    
\end{enumerate}

From these basic observational results we make the following inferences:
\begin{enumerate}

\item All five of our ISM estimates are very similar and are substantially larger than the dynamical mass of the system (without even considering the stellar mass and any dark matter). Inconsistencies like this have been observed before and have always been attributed to a low value of $\alpha_{\rm CO}$, which is then attributed to physical conditions in the ISM of an SMG probably being different from the Milky Way. This can not be the explanation in ID141 because we are using four different tracers, each of which will be affected by different physical conditions in the ISM in different ways. The similarity of the ISM mass estimates using the four different tracers suggests that chemical evolution could explain the mass discrepancy because the abundances of all the tracers will depend on the abundances of heavy elements in the galaxy.

\item We have used a chemical evolution model to show that the GDR does change substantially during the evolution of an SMG. It can reach values much lower than the Milky Way value as long as the IMF is top-heavy in SMGs for which there is now good evidence from spectroscopy of molecules containing rare isotopes \citep{Zhang2018}.

\item Although there is a complex web of calibrations, all estimates of the ISM masses in high-redshift galaxies are ultimately calibrated in the Milky Way. Our conclusion that the gas-to-tracer ratio for four different tracers is four times lower than in the Milky Way puts all these estimates into doubt. On a positive note, we suggest that the gas-to-tracer ratio may be a useful new tool for assessing the evolutionary state of a galaxy. 

\item We have used ID141 to test whether extreme SMGs like this might evolve into the population of quiescent galaxies that is seen at $z \sim 3$. Several indicators (the Toomre Q-parameter, the inspiral time for the clumps, and the escape velocity system of the system) suggest that ID141 should be a single, fairly undisturbed galaxy by $z \sim 3$. The depletion time for the gas also implies that, on the assumption of no gas infall, the galaxy will be devoid of gas by this redshift. Therefore, extreme though it is at $z=4.2$, it seems plausible that ID141 will evolve into an unexceptional galaxy by $z \sim 3$.

\end{enumerate}

\section*{Acknowledgements}

SD is supported by a UK Science and Technology Facilities Council (STFC) Rutherford Fellowship. SAE and MWLS acknowledge support from STFC. GCJ acknowledges ERC Advanced Grant 695671 ``QUENCH'' and support by the STFC. HLG, LD and MWLS acknowledge support from the European Research Council in the form of Consolidator Grant 647939 CosmicDust. EB and EMC acknowledge support by Padua University grants DOR1885254/18, DOR1935272/19, and DOR2013080/20 and by MIUR grant PRIN 2017 20173ML3WW-001. HD acknowledges financial support from the Spanish Ministry of Science, Innovation and Universities (MICIU) under the 2014 Ram\'{o}n y Cajal program RYC-2014-15686 and from the Agencia Estatal de Investigaci\'{o}n del Ministerio de Ciencia e Innovaci\'{o}n (AEI-MCINN) under grant (La evoluci\'{o}n de los c\'{i}umulos de galaxias desde el amanecer hasta el mediod\'{i}a c\'{o}smico) with reference (PID2019-105776GB-I00/DOI:10.13039/501100011033).This paper makes use of the following ALMA data: ADS/JAO.ALMA \#2013.1.00358.S, ADS/JAO.ALMA \#2016.1.00450.S and ADS/JAO.ALMA \#2017.1.00029.S. ALMA is a partnership of ESO (representing its member states), NSF (USA) and NINS (Japan), together with NRC (Canada), MOST and ASIAA (Taiwan), and KASI (Republic of Korea), in cooperation with the Republic of Chile. The Joint ALMA Observatory is operated by ESO, AUI/NRAO and NAOJ. In addition, publications from NA authors must include the standard NRAO acknowledgement: The National Radio Astronomy Observatory is a facility of the National Science Foundation operated under cooperative agreement by Associated Universities, Inc.

\section*{Data Availability}

All data presented in this paper can be obtained from the ALMA science archive under the project IDs 2013.1.00358.S, 2016.1.00450.S and 2017.1.00029.S.


\bibliographystyle{mnras}
\bibliography{ID141_paper} 

\begin{thebibliography}{}
\makeatletter
\relax
\def\mn@urlcharsother{\let\do\@makeother \do\$\do\&\do\#\do\^\do\_\do\%\do\~}
\def\mn@doi{\begingroup\mn@urlcharsother \@ifnextchar [ {\mn@doi@}
  {\mn@doi@[]}}
\def\mn@doi@[#1]#2{\def\@tempa{#1}\ifx\@tempa\@empty \href
  {http://dx.doi.org/#2} {doi:#2}\else \href {http://dx.doi.org/#2} {#1}\fi
  \endgroup}
\def\mn@eprint#1#2{\mn@eprint@#1:#2::\@nil}
\def\mn@eprint@arXiv#1{\href {http://arxiv.org/abs/#1} {{\tt arXiv:#1}}}
\def\mn@eprint@dblp#1{\href {http://dblp.uni-trier.de/rec/bibtex/#1.xml}
  {dblp:#1}}
\def\mn@eprint@#1:#2:#3:#4\@nil{\def\@tempa {#1}\def\@tempb {#2}\def\@tempc
  {#3}\ifx \@tempc \@empty \let \@tempc \@tempb \let \@tempb \@tempa \fi \ifx
  \@tempb \@empty \def\@tempb {arXiv}\fi \@ifundefined
  {mn@eprint@\@tempb}{\@tempb:\@tempc}{\expandafter \expandafter \csname
  mn@eprint@\@tempb\endcsname \expandafter{\@tempc}}}

\bibitem[\protect\citeauthoryear{{ALMA Partnership} et~al.,}{{ALMA Partnership}
  et~al.}{2015}]{vlahakis2014}
{ALMA Partnership} et~al., 2015, \mn@doi [\apjl] {10.1088/2041-8205/808/1/L4},
  \href {https://ui.adsabs.harvard.edu/abs/2015ApJ...808L...4A} {808, L4}

\bibitem[\protect\citeauthoryear{{Abdo} et~al.,}{{Abdo}
  et~al.}{2010}]{abdo2010}
{Abdo} A.~A.,  et~al., 2010, \mn@doi [\apj] {10.1088/0004-637X/710/1/133},
  \href {https://ui.adsabs.harvard.edu/abs/2010ApJ...710..133A} {710, 133}

\bibitem[\protect\citeauthoryear{{Asano}, {Takeuchi}, {Hirashita}  \&
  {Inoue}}{{Asano} et~al.}{2013}]{Asano2013}
{Asano} R.~S.,  {Takeuchi} T.~T.,  {Hirashita} H.,   {Inoue} A.~K.,  2013,
  \mn@doi [Earth, Planets, and Space] {10.5047/eps.2012.04.014}, \href
  {http://adsabs.harvard.edu/abs/2013EP%26S...65..213A} {65, 213}

\bibitem[\protect\citeauthoryear{{Asplund}, {Grevesse}, {Sauval}  \&
  {Scott}}{{Asplund} et~al.}{2009}]{asplund2009}
{Asplund} M.,  {Grevesse} N.,  {Sauval} A.~J.,   {Scott} P.,  2009, \mn@doi
  [\araa] {10.1146/annurev.astro.46.060407.145222}, \href
  {https://ui.adsabs.harvard.edu/abs/2009ARA&A..47..481A} {47, 481}

\bibitem[\protect\citeauthoryear{{Bakx} et~al.,}{{Bakx}
  et~al.}{2018}]{bakx2018}
{Bakx} T. J.~L.~C.,  et~al., 2018, \mn@doi [\mnras] {10.1093/mnras/stx2267},
  \href {https://ui.adsabs.harvard.edu/abs/2018MNRAS.473.1751B} {473, 1751}

\bibitem[\protect\citeauthoryear{{Bakx} et~al.,}{{Bakx}
  et~al.}{2020}]{bakx2020}
{Bakx} T. J.~L.~C.,  et~al., 2020, \mn@doi [\mnras] {10.1093/mnras/staa658},
  \href {https://ui.adsabs.harvard.edu/abs/2020MNRAS.494...10B} {494, 10}

\bibitem[\protect\citeauthoryear{{Battisti}, {da Cunha}, {Grasha}  \& {Cosmos
  Collaboration}}{{Battisti} et~al.}{2020}]{battisti2020}
{Battisti} A.~J.,  {da Cunha} E.,  {Grasha} K.,   {Cosmos Collaboration} 2020,
  in American Astronomical Society Meeting Abstracts \#236. p. 110.05

\bibitem[\protect\citeauthoryear{{Baugh}, {Lacey}, {Frenk}, {Granato}, {Silva},
  {Bressan}, {Benson}  \& {Cole}}{{Baugh} et~al.}{2005}]{baugh2005}
{Baugh} C.~M.,  {Lacey} C.~G.,  {Frenk} C.~S.,  {Granato} G.~L.,  {Silva} L.,
  {Bressan} A.,  {Benson} A.~J.,   {Cole} S.,  2005, \mn@doi [\mnras]
  {10.1111/j.1365-2966.2004.08553.x}, \href
  {https://ui.adsabs.harvard.edu/abs/2005MNRAS.356.1191B} {356, 1191}

\bibitem[\protect\citeauthoryear{{Behroozi}, {Wechsler}  \&
  {Conroy}}{{Behroozi} et~al.}{2013}]{Behroozi2013}
{Behroozi} P.~S.,  {Wechsler} R.~H.,   {Conroy} C.,  2013, \mn@doi [\apj]
  {10.1088/0004-637X/770/1/57}, \href
  {https://ui.adsabs.harvard.edu/abs/2013ApJ...770...57B} {770, 57}

\bibitem[\protect\citeauthoryear{{Belfiore}, {Vincenzo}, {Maiolino}  \&
  {Matteucci}}{{Belfiore} et~al.}{2019}]{Belfiore2019}
{Belfiore} F.,  {Vincenzo} F.,  {Maiolino} R.,   {Matteucci} F.,  2019, \mn@doi
  [\mnras] {10.1093/mnras/stz1165}, \href
  {https://ui.adsabs.harvard.edu/abs/2019MNRAS.487..456B} {487, 456}

\bibitem[\protect\citeauthoryear{{Berta} et~al.,}{{Berta}
  et~al.}{2021}]{berta2021}
{Berta} S.,  et~al., 2021, \mn@doi [\aap] {10.1051/0004-6361/202039743}, \href
  {https://ui.adsabs.harvard.edu/abs/2021A&A...646A.122B} {646, A122}

\bibitem[\protect\citeauthoryear{{Bigiel} et~al.,}{{Bigiel}
  et~al.}{2020}]{bigiel2020}
{Bigiel} F.,  et~al., 2020, \mn@doi [\apj] {10.3847/1538-4357/abb677}, \href
  {https://ui.adsabs.harvard.edu/abs/2020ApJ...903...30B} {903, 30}

\bibitem[\protect\citeauthoryear{{Birkin} et~al.,}{{Birkin}
  et~al.}{2021}]{birkin2021}
{Birkin} J.~E.,  et~al., 2021, \mn@doi [\mnras] {10.1093/mnras/staa3862}, \href
  {https://ui.adsabs.harvard.edu/abs/2021MNRAS.501.3926B} {501, 3926}

\bibitem[\protect\citeauthoryear{{Bolatto}, {Wolfire}  \& {Leroy}}{{Bolatto}
  et~al.}{2013}]{bolatto2013}
{Bolatto} A.~D.,  {Wolfire} M.,   {Leroy} A.~K.,  2013, \mn@doi [\araa]
  {10.1146/annurev-astro-082812-140944}, \href
  {https://ui.adsabs.harvard.edu/abs/2013ARA&A..51..207B} {51, 207}

\bibitem[\protect\citeauthoryear{{Boogaard} et~al.,}{{Boogaard}
  et~al.}{2021}]{boogaard2021}
{Boogaard} L.~A.,  et~al., 2021, arXiv e-prints, \href
  {https://ui.adsabs.harvard.edu/abs/2021arXiv210512489B} {p. arXiv:2105.12489}

\bibitem[\protect\citeauthoryear{{Bothwell}, {Maiolino}, {Kennicutt}, {Cresci},
  {Mannucci}, {Marconi}  \& {Cicone}}{{Bothwell} et~al.}{2013}]{bothwell2013}
{Bothwell} M.~S.,  {Maiolino} R.,  {Kennicutt} R.,  {Cresci} G.,  {Mannucci}
  F.,  {Marconi} A.,   {Cicone} C.,  2013, \mn@doi [\mnras]
  {10.1093/mnras/stt817}, \href
  {https://ui.adsabs.harvard.edu/abs/2013MNRAS.433.1425B} {433, 1425}

\bibitem[\protect\citeauthoryear{{Bourne}, {Dunlop}, {Simpson}, {Rowland s},
  {Geach}  \& {McLeod}}{{Bourne} et~al.}{2019}]{bourne2019}
{Bourne} N.,  {Dunlop} J.~S.,  {Simpson} J.~M.,  {Rowland s} K.~E.,  {Geach}
  J.~E.,   {McLeod} D.~J.,  2019, \mn@doi [\mnras] {10.1093/mnras/sty2773},
  \href {https://ui.adsabs.harvard.edu/abs/2019MNRAS.482.3135B} {482, 3135}

\bibitem[\protect\citeauthoryear{{Bussmann} et~al.,}{{Bussmann}
  et~al.}{2012}]{bussmann2012}
{Bussmann} R.~S.,  et~al., 2012, \mn@doi [\apj] {10.1088/0004-637X/756/2/134},
  \href {https://ui.adsabs.harvard.edu/abs/2012ApJ...756..134B} {756, 134}

\bibitem[\protect\citeauthoryear{{Bussmann} et~al.,}{{Bussmann}
  et~al.}{2013}]{bussmann2013}
{Bussmann} R.~S.,  et~al., 2013, \mn@doi [\apj] {10.1088/0004-637X/779/1/25},
  \href {https://ui.adsabs.harvard.edu/abs/2013ApJ...779...25B} {779, 25}

\bibitem[\protect\citeauthoryear{{Ca{\~n}ameras} et~al.,}{{Ca{\~n}ameras}
  et~al.}{2015}]{canameras2015}
{Ca{\~n}ameras} R.,  et~al., 2015, \mn@doi [\aap]
  {10.1051/0004-6361/201425128}, \href
  {https://ui.adsabs.harvard.edu/abs/2015A&A...581A.105C} {581, A105}

\bibitem[\protect\citeauthoryear{{Ca{\~n}ameras} et~al.,}{{Ca{\~n}ameras}
  et~al.}{2018}]{canameras2018}
{Ca{\~n}ameras} R.,  et~al., 2018, \mn@doi [\aap]
  {10.1051/0004-6361/201833625}, \href
  {https://ui.adsabs.harvard.edu/abs/2018A&A...620A..61C} {620, A61}

\bibitem[\protect\citeauthoryear{{Cai}, {De Zotti}  \& {Bonato}}{{Cai}
  et~al.}{2020}]{Cai2020}
{Cai} Z.-Y.,  {De Zotti} G.,   {Bonato} M.,  2020, \mn@doi [\apj]
  {10.3847/1538-4357/ab7231}, \href
  {https://ui.adsabs.harvard.edu/abs/2020ApJ...891...74C} {891, 74}

\bibitem[\protect\citeauthoryear{{Calistro Rivera} et~al.,}{{Calistro Rivera}
  et~al.}{2018}]{rivera2018}
{Calistro Rivera} G.,  et~al., 2018, \mn@doi [\apj] {10.3847/1538-4357/aacffa},
  \href {https://ui.adsabs.harvard.edu/abs/2018ApJ...863...56C} {863, 56}

\bibitem[\protect\citeauthoryear{{Casey}, {Narayanan}  \& {Cooray}}{{Casey}
  et~al.}{2014}]{casey2014}
{Casey} C.~M.,  {Narayanan} D.,   {Cooray} A.,  2014, \mn@doi [\physrep]
  {10.1016/j.physrep.2014.02.009}, \href
  {https://ui.adsabs.harvard.edu/abs/2014PhR...541...45C} {541, 45}

\bibitem[\protect\citeauthoryear{{Chabrier}}{{Chabrier}}{2003}]{chabrier2003}
{Chabrier} G.,  2003, \mn@doi [\pasp] {10.1086/376392}, \href
  {https://ui.adsabs.harvard.edu/abs/2003PASP..115..763C} {115, 763}

\bibitem[\protect\citeauthoryear{{Cheng} et~al.,}{{Cheng}
  et~al.}{2020}]{cheng2020}
{Cheng} C.,  et~al., 2020, \mn@doi [\apj] {10.3847/1538-4357/ab980b}, \href
  {https://ui.adsabs.harvard.edu/abs/2020ApJ...898...33C} {898, 33}

\bibitem[\protect\citeauthoryear{{Coppin} et~al.,}{{Coppin}
  et~al.}{2008}]{Coppin2008}
{Coppin} K.~E.~K.,  et~al., 2008, \mn@doi [\mnras]
  {10.1111/j.1365-2966.2008.13553.x}, \href
  {https://ui.adsabs.harvard.edu/abs/2008MNRAS.389...45C} {389, 45}

\bibitem[\protect\citeauthoryear{{Cox} et~al.,}{{Cox} et~al.}{2011}]{cox2011}
{Cox} P.,  et~al., 2011, \mn@doi [\apj] {10.1088/0004-637X/740/2/63}, \href
  {https://ui.adsabs.harvard.edu/abs/2011ApJ...740...63C} {740, 63}

\bibitem[\protect\citeauthoryear{{Danielson} et~al.,}{{Danielson}
  et~al.}{2017}]{Danielson2017}
{Danielson} A.~L.~R.,  et~al., 2017, \mn@doi [\apj] {10.3847/1538-4357/aa6caf},
  \href {https://ui.adsabs.harvard.edu/abs/2017ApJ...840...78D} {840, 78}

\bibitem[\protect\citeauthoryear{{Dav{\'e}}, {Finlator}  \&
  {Oppenheimer}}{{Dav{\'e}} et~al.}{2011}]{Dave2011}
{Dav{\'e}} R.,  {Finlator} K.,   {Oppenheimer} B.~D.,  2011, \mn@doi [\mnras]
  {10.1111/j.1365-2966.2011.19132.x}, \href
  {https://ui.adsabs.harvard.edu/abs/2011MNRAS.416.1354D} {416, 1354}

\bibitem[\protect\citeauthoryear{{De Vis} et~al.,}{{De Vis}
  et~al.}{2017}]{DeVis2017b}
{De Vis} P.,  et~al., 2017, \mn@doi [\mnras] {10.1093/mnras/stx981}, \href
  {https://ui.adsabs.harvard.edu/abs/2017MNRAS.471.1743D} {471, 1743}

\bibitem[\protect\citeauthoryear{{De Vis} et~al.,}{{De Vis}
  et~al.}{2019}]{devis2019}
{De Vis} P.,  et~al., 2019, \mn@doi [\aap] {10.1051/0004-6361/201834444}, \href
  {https://ui.adsabs.harvard.edu/abs/2019A&A...623A...5D} {623, A5}

\bibitem[\protect\citeauthoryear{{Di Teodoro} \& {Fraternali}}{{Di Teodoro} \&
  {Fraternali}}{2015}]{teodoro2015}
{Di Teodoro} E.~M.,  {Fraternali} F.,  2015, \mn@doi [\mnras]
  {10.1093/mnras/stv1213}, \href
  {https://ui.adsabs.harvard.edu/abs/2015MNRAS.451.3021D} {451, 3021}

\bibitem[\protect\citeauthoryear{{Downes} \& {Solomon}}{{Downes} \&
  {Solomon}}{1998}]{downes1998}
{Downes} D.,  {Solomon} P.~M.,  1998, \mn@doi [\apj] {10.1086/306339}, \href
  {https://ui.adsabs.harvard.edu/abs/1998ApJ...507..615D} {507, 615}

\bibitem[\protect\citeauthoryear{{Dunne}, {Eales}  \& {Edmunds}}{{Dunne}
  et~al.}{2003}]{Dunne2003}
{Dunne} L.,  {Eales} S.~A.,   {Edmunds} M.~G.,  2003, \mn@doi [\mnras]
  {10.1046/j.1365-8711.2003.06440.x}, \href
  {https://ui.adsabs.harvard.edu/abs/2003MNRAS.341..589D} {341, 589}

\bibitem[\protect\citeauthoryear{{Dunne} et~al.,}{{Dunne}
  et~al.}{2011}]{dunne2011}
{Dunne} L.,  et~al., 2011, \mn@doi [\mnras] {10.1111/j.1365-2966.2011.19363.x},
  \href {https://ui.adsabs.harvard.edu/abs/2011MNRAS.417.1510D} {417, 1510}

\bibitem[\protect\citeauthoryear{{Dunne}, {Maddox}, {Vlahakis}  \&
  {Gomez}}{{Dunne} et~al.}{2021}]{dunne2021a}
{Dunne} L.,  {Maddox} S.~J.,  {Vlahakis} C.,   {Gomez} H.~L.,  2021, \mn@doi
  [\mnras] {10.1093/mnras/staa3526}, \href
  {https://ui.adsabs.harvard.edu/abs/2021MNRAS.501.2573D} {501, 2573}

\bibitem[\protect\citeauthoryear{{Dye} et~al.,}{{Dye} et~al.}{2015}]{dye2015}
{Dye} S.,  et~al., 2015, \mn@doi [\mnras] {10.1093/mnras/stv1442}, \href
  {https://ui.adsabs.harvard.edu/abs/2015MNRAS.452.2258D} {452, 2258}

\bibitem[\protect\citeauthoryear{{Dye} et~al.,}{{Dye} et~al.}{2018}]{dye2018}
{Dye} S.,  et~al., 2018, \mn@doi [\mnras] {10.1093/mnras/sty513}, \href
  {https://ui.adsabs.harvard.edu/abs/2018MNRAS.476.4383D} {476, 4383}

\bibitem[\protect\citeauthoryear{{Eales} \& {Edmunds}}{{Eales} \&
  {Edmunds}}{1996}]{eales1996}
{Eales} S.~A.,  {Edmunds} M.~G.,  1996, \mn@doi [\mnras]
  {10.1093/mnras/280.4.1167}, \href
  {https://ui.adsabs.harvard.edu/abs/1996MNRAS.280.1167E} {280, 1167}

\bibitem[\protect\citeauthoryear{{Eales} et~al.,}{{Eales}
  et~al.}{2010}]{eales2010}
{Eales} S.,  et~al., 2010, \mn@doi [\pasp] {10.1086/653086}, \href
  {https://ui.adsabs.harvard.edu/abs/2010PASP..122..499E} {122, 499}

\bibitem[\protect\citeauthoryear{{Eales} et~al.,}{{Eales}
  et~al.}{2012}]{eales2012}
{Eales} S.,  et~al., 2012, \mn@doi [\apj] {10.1088/0004-637X/761/2/168}, \href
  {https://ui.adsabs.harvard.edu/abs/2012ApJ...761..168E} {761, 168}

\bibitem[\protect\citeauthoryear{{Eales} et~al.,}{{Eales}
  et~al.}{2018}]{eales2018}
{Eales} S.,  et~al., 2018, \mn@doi [\mnras] {10.1093/mnras/stx2548}, \href
  {https://ui.adsabs.harvard.edu/abs/2018MNRAS.473.3507E} {473, 3507}

\bibitem[\protect\citeauthoryear{{Engel} et~al.,}{{Engel}
  et~al.}{2010}]{engel2010}
{Engel} H.,  et~al., 2010, \mn@doi [\apj] {10.1088/0004-637X/724/1/233}, \href
  {https://ui.adsabs.harvard.edu/abs/2010ApJ...724..233E} {724, 233}

\bibitem[\protect\citeauthoryear{{Enia} et~al.,}{{Enia}
  et~al.}{2018}]{enia2018}
{Enia} A.,  et~al., 2018, \mn@doi [\mnras] {10.1093/mnras/sty021}, \href
  {https://ui.adsabs.harvard.edu/abs/2018MNRAS.475.3467E} {475, 3467}

\bibitem[\protect\citeauthoryear{{Fakhouri}, {Ma}  \&
  {Boylan-Kolchin}}{{Fakhouri} et~al.}{2010}]{Fakhouri2010}
{Fakhouri} O.,  {Ma} C.-P.,   {Boylan-Kolchin} M.,  2010, \mn@doi [\mnras]
  {10.1111/j.1365-2966.2010.16859.x}, \href
  {https://ui.adsabs.harvard.edu/abs/2010MNRAS.406.2267F} {406, 2267}

\bibitem[\protect\citeauthoryear{{Fan}, {Knudsen}, {Han}  \& {Tan}}{{Fan}
  et~al.}{2019}]{fan19}
{Fan} L.,  {Knudsen} K.~K.,  {Han} Y.,   {Tan} Q.-h.,  2019, \mn@doi [\apj]
  {10.3847/1538-4357/ab5059}, \href
  {https://ui.adsabs.harvard.edu/abs/2019ApJ...887...74F} {887, 74}

\bibitem[\protect\citeauthoryear{{Forbes}, {Krumholz}, {Burkert}  \&
  {Dekel}}{{Forbes} et~al.}{2014}]{Forbes2014a}
{Forbes} J.~C.,  {Krumholz} M.~R.,  {Burkert} A.,   {Dekel} A.,  2014, \mn@doi
  [\mnras] {10.1093/mnras/stt2294}, \href
  {https://ui.adsabs.harvard.edu/abs/2014MNRAS.438.1552F} {438, 1552}

\bibitem[\protect\citeauthoryear{{Forrest} et~al.,}{{Forrest}
  et~al.}{2020a}]{forrest2020b}
{Forrest} B.,  et~al., 2020a, \mn@doi [\apjl] {10.3847/2041-8213/ab5b9f}, \href
  {https://ui.adsabs.harvard.edu/abs/2020ApJ...890L...1F} {890, L1}

\bibitem[\protect\citeauthoryear{{Forrest} et~al.,}{{Forrest}
  et~al.}{2020b}]{forrest2020a}
{Forrest} B.,  et~al., 2020b, \mn@doi [\apj] {10.3847/1538-4357/abb819}, \href
  {https://ui.adsabs.harvard.edu/abs/2020ApJ...903...47F} {903, 47}

\bibitem[\protect\citeauthoryear{{Fudamoto} et~al.,}{{Fudamoto}
  et~al.}{2017}]{fudamoto2017}
{Fudamoto} Y.,  et~al., 2017, \mn@doi [\mnras] {10.1093/mnras/stx1956}, \href
  {https://ui.adsabs.harvard.edu/abs/2017MNRAS.472.2028F} {472, 2028}

\bibitem[\protect\citeauthoryear{{Gallazzi}, {Charlot}, {Brinchmann}, {White}
  \& {Tremonti}}{{Gallazzi} et~al.}{2005}]{gallazzi2005}
{Gallazzi} A.,  {Charlot} S.,  {Brinchmann} J.,  {White} S. D.~M.,   {Tremonti}
  C.~A.,  2005, \mn@doi [\mnras] {10.1111/j.1365-2966.2005.09321.x}, \href
  {https://ui.adsabs.harvard.edu/abs/2005MNRAS.362...41G} {362, 41}

\bibitem[\protect\citeauthoryear{{Geach} \& {Papadopoulos}}{{Geach} \&
  {Papadopoulos}}{2012}]{geach2012}
{Geach} J.~E.,  {Papadopoulos} P.~P.,  2012, \mn@doi [\apj]
  {10.1088/0004-637X/757/2/156}, \href
  {https://ui.adsabs.harvard.edu/abs/2012ApJ...757..156G} {757, 156}

\bibitem[\protect\citeauthoryear{{Genzel}, {Baker}, {Tacconi}, {Lutz}, {Cox},
  {Guilloteau}  \& {Omont}}{{Genzel} et~al.}{2003}]{genzel2003}
{Genzel} R.,  {Baker} A.~J.,  {Tacconi} L.~J.,  {Lutz} D.,  {Cox} P.,
  {Guilloteau} S.,   {Omont} A.,  2003, \mn@doi [\apj] {10.1086/345718}, \href
  {https://ui.adsabs.harvard.edu/abs/2003ApJ...584..633G} {584, 633}

\bibitem[\protect\citeauthoryear{{Genzel} et~al.,}{{Genzel}
  et~al.}{2011}]{genzel2011}
{Genzel} R.,  et~al., 2011, \mn@doi [\apj] {10.1088/0004-637X/733/2/101}, \href
  {https://ui.adsabs.harvard.edu/abs/2011ApJ...733..101G} {733, 101}

\bibitem[\protect\citeauthoryear{{Genzel} et~al.,}{{Genzel}
  et~al.}{2017}]{genzel2017}
{Genzel} R.,  et~al., 2017, \mn@doi [\nat] {10.1038/nature21685}, \href
  {https://ui.adsabs.harvard.edu/abs/2017Natur.543..397G} {543, 397}

\bibitem[\protect\citeauthoryear{{Girelli}, {Bolzonella}  \&
  {Cimatti}}{{Girelli} et~al.}{2019}]{girelli2019}
{Girelli} G.,  {Bolzonella} M.,   {Cimatti} A.,  2019, \mn@doi [\aap]
  {10.1051/0004-6361/201834547}, \href
  {https://ui.adsabs.harvard.edu/abs/2019A&A...632A..80G} {632, A80}

\bibitem[\protect\citeauthoryear{{Glazebrook} et~al.,}{{Glazebrook}
  et~al.}{2017}]{glazebrook2017}
{Glazebrook} K.,  et~al., 2017, \mn@doi [\nat] {10.1038/nature21680}, \href
  {https://ui.adsabs.harvard.edu/abs/2017Natur.544...71G} {544, 71}

\bibitem[\protect\citeauthoryear{{Hodge} \& {da Cunha}}{{Hodge} \& {da
  Cunha}}{2020}]{hodge2020}
{Hodge} J.~A.,  {da Cunha} E.,  2020, \mn@doi [Royal Society Open Science]
  {10.1098/rsos.200556}, \href
  {https://ui.adsabs.harvard.edu/abs/2020RSOS....700556H} {7, 200556}

\bibitem[\protect\citeauthoryear{{Hodge}, {Carilli}, {Walter}, {de Blok},
  {Riechers}, {Daddi}  \& {Lentati}}{{Hodge} et~al.}{2012}]{hodge2012}
{Hodge} J.~A.,  {Carilli} C.~L.,  {Walter} F.,  {de Blok} W.~J.~G.,  {Riechers}
  D.,  {Daddi} E.,   {Lentati} L.,  2012, \mn@doi [\apj]
  {10.1088/0004-637X/760/1/11}, \href
  {https://ui.adsabs.harvard.edu/abs/2012ApJ...760...11H} {760, 11}

\bibitem[\protect\citeauthoryear{{Hopwood} et~al.,}{{Hopwood}
  et~al.}{2011}]{hopwood2011}
{Hopwood} R.,  et~al., 2011, \mn@doi [\apjl] {10.1088/2041-8205/728/1/L4},
  \href {https://ui.adsabs.harvard.edu/abs/2011ApJ...728L...4H} {728, L4}

\bibitem[\protect\citeauthoryear{{Huynh} et~al.,}{{Huynh}
  et~al.}{2017}]{huynh2017}
{Huynh} M.~T.,  et~al., 2017, \mn@doi [\mnras] {10.1093/mnras/stx156}, \href
  {https://ui.adsabs.harvard.edu/abs/2017MNRAS.467.1222H} {467, 1222}

\bibitem[\protect\citeauthoryear{{Inoue}}{{Inoue}}{2003}]{Inoue2003}
{Inoue} A.~K.,  2003, \mn@doi [\pasj] {10.1093/pasj/55.5.901}, \href
  {https://ui.adsabs.harvard.edu/abs/2003PASJ...55..901I} {55, 901}

\bibitem[\protect\citeauthoryear{{Ivison}, {Smail}, {Papadopoulos}, {Wold},
  {Richard}, {Swinbank}, {Kneib}  \& {Owen}}{{Ivison}
  et~al.}{2010}]{ivison2010}
{Ivison} R.~J.,  {Smail} I.,  {Papadopoulos} P.~P.,  {Wold} I.,  {Richard} J.,
  {Swinbank} A.~M.,  {Kneib} J.~P.,   {Owen} F.~N.,  2010, \mn@doi [\mnras]
  {10.1111/j.1365-2966.2010.16322.x}, \href
  {https://ui.adsabs.harvard.edu/abs/2010MNRAS.404..198I} {404, 198}

\bibitem[\protect\citeauthoryear{{Ivison}, {Papadopoulos}, {Smail}, {Greve},
  {Thomson}, {Xilouris}  \& {Chapman}}{{Ivison} et~al.}{2011}]{ivison2011}
{Ivison} R.~J.,  {Papadopoulos} P.~P.,  {Smail} I.,  {Greve} T.~R.,  {Thomson}
  A.~P.,  {Xilouris} E.~M.,   {Chapman} S.~C.,  2011, \mn@doi [\mnras]
  {10.1111/j.1365-2966.2010.18028.x}, \href
  {https://ui.adsabs.harvard.edu/abs/2011MNRAS.412.1913I} {412, 1913}

\bibitem[\protect\citeauthoryear{{Ivison}, {Richard}, {Biggs}, {Zwaan},
  {Falgarone}, {Arumugam}, {van der Werf}  \& {Rujopakarn}}{{Ivison}
  et~al.}{2020}]{ivison2020}
{Ivison} R.~J.,  {Richard} J.,  {Biggs} A.~D.,  {Zwaan} M.~A.,  {Falgarone} E.,
   {Arumugam} V.,  {van der Werf} P.~P.,   {Rujopakarn} W.,  2020, \mn@doi
  [\mnras] {10.1093/mnrasl/slaa046}, \href
  {https://ui.adsabs.harvard.edu/abs/2020MNRAS.495L...1I} {495, L1}

\bibitem[\protect\citeauthoryear{{James}, {Dunne}, {Eales}  \&
  {Edmunds}}{{James} et~al.}{2002}]{james2002}
{James} A.,  {Dunne} L.,  {Eales} S.,   {Edmunds} M.~G.,  2002, \mn@doi
  [\mnras] {10.1046/j.1365-8711.2002.05660.x}, \href
  {https://ui.adsabs.harvard.edu/abs/2002MNRAS.335..753J} {335, 753}

\bibitem[\protect\citeauthoryear{{Jarugula} et~al.,}{{Jarugula}
  et~al.}{2019}]{jarugula2019}
{Jarugula} S.,  et~al., 2019, \mn@doi [\apj] {10.3847/1538-4357/ab290d}, \href
  {https://ui.adsabs.harvard.edu/abs/2019ApJ...880...92J} {880, 92}

\bibitem[\protect\citeauthoryear{{Kaasinen} et~al.,}{{Kaasinen}
  et~al.}{2019}]{kaasinen2019}
{Kaasinen} M.,  et~al., 2019, \mn@doi [\apj] {10.3847/1538-4357/ab253b}, \href
  {https://ui.adsabs.harvard.edu/abs/2019ApJ...880...15K} {880, 15}

\bibitem[\protect\citeauthoryear{{Kassiola} \& {Kovner}}{{Kassiola} \&
  {Kovner}}{1993}]{kassiola1993}
{Kassiola} A.,  {Kovner} I.,  1993, \mn@doi [\apj] {10.1086/173325}, \href
  {https://ui.adsabs.harvard.edu/abs/1993ApJ...417..450K} {417, 450}

\bibitem[\protect\citeauthoryear{{Kennicutt}}{{Kennicutt}}{1998}]{Kennicutt1998}
{Kennicutt} Robert~C. J.,  1998, \mn@doi [\apj] {10.1086/305588}, \href
  {https://ui.adsabs.harvard.edu/abs/1998ApJ...498..541K} {498, 541}

\bibitem[\protect\citeauthoryear{{Kennicutt} \& {Evans}}{{Kennicutt} \&
  {Evans}}{2012}]{kennicutt2012}
{Kennicutt} R.~C.,  {Evans} N.~J.,  2012, \mn@doi [\araa]
  {10.1146/annurev-astro-081811-125610}, \href
  {https://ui.adsabs.harvard.edu/abs/2012ARA&A..50..531K} {50, 531}

\bibitem[\protect\citeauthoryear{{Kroupa} \& {Weidner}}{{Kroupa} \&
  {Weidner}}{2003}]{kroupa2003}
{Kroupa} P.,  {Weidner} C.,  2003, \mn@doi [\apj] {10.1086/379105}, \href
  {https://ui.adsabs.harvard.edu/abs/2003ApJ...598.1076K} {598, 1076}

\bibitem[\protect\citeauthoryear{{Lang}, {Hogg}  \& {Schlegel}}{{Lang}
  et~al.}{2016}]{lang2016}
{Lang} D.,  {Hogg} D.~W.,   {Schlegel} D.~J.,  2016, \mn@doi [\aj]
  {10.3847/0004-6256/151/2/36}, \href
  {https://ui.adsabs.harvard.edu/abs/2016AJ....151...36L} {151, 36}

\bibitem[\protect\citeauthoryear{{Liu} et~al.,}{{Liu} et~al.}{2017}]{liu2017}
{Liu} L.,  et~al., 2017, \mn@doi [\apj] {10.3847/1538-4357/aa81b4}, \href
  {https://ui.adsabs.harvard.edu/abs/2017ApJ...846....5L} {846, 5}

\bibitem[\protect\citeauthoryear{{Madau} \& {Dickinson}}{{Madau} \&
  {Dickinson}}{2014}]{madau2014}
{Madau} P.,  {Dickinson} M.,  2014, \mn@doi [\araa]
  {10.1146/annurev-astro-081811-125615}, \href
  {https://ui.adsabs.harvard.edu/abs/2014ARA&A..52..415M} {52, 415}

\bibitem[\protect\citeauthoryear{{Maiolino} \& {Mannucci}}{{Maiolino} \&
  {Mannucci}}{2019}]{maiolino2019}
{Maiolino} R.,  {Mannucci} F.,  2019, \mn@doi [\aapr]
  {10.1007/s00159-018-0112-2}, \href
  {https://ui.adsabs.harvard.edu/abs/2019A&ARv..27....3M} {27, 3}

\bibitem[\protect\citeauthoryear{{Mannucci}, {Cresci}, {Maiolino}, {Marconi}
  \& {Gnerucci}}{{Mannucci} et~al.}{2010}]{mannucci2010}
{Mannucci} F.,  {Cresci} G.,  {Maiolino} R.,  {Marconi} A.,   {Gnerucci} A.,
  2010, \mn@doi [\mnras] {10.1111/j.1365-2966.2010.17291.x}, \href
  {https://ui.adsabs.harvard.edu/abs/2010MNRAS.408.2115M} {408, 2115}

\bibitem[\protect\citeauthoryear{{Mattsson} \& {Andersen}}{{Mattsson} \&
  {Andersen}}{2012}]{Mattsson2012}
{Mattsson} L.,  {Andersen} A.~C.,  2012, \mn@doi [\mnras]
  {10.1111/j.1365-2966.2012.20574.x}, \href
  {http://adsabs.harvard.edu/abs/2012MNRAS.423...38M} {423, 38}

\bibitem[\protect\citeauthoryear{{McBride}, {Fakhouri}  \& {Ma}}{{McBride}
  et~al.}{2009}]{McBride2009}
{McBride} J.,  {Fakhouri} O.,   {Ma} C.-P.,  2009, \mn@doi [\mnras]
  {10.1111/j.1365-2966.2009.15329.x}, \href
  {https://ui.adsabs.harvard.edu/abs/2009MNRAS.398.1858M} {398, 1858}

\bibitem[\protect\citeauthoryear{{McMullin}, {Waters}, {Schiebel}, {Young}  \&
  {Golap}}{{McMullin} et~al.}{2007}]{mcmullin2007}
{McMullin} J.~P.,  {Waters} B.,  {Schiebel} D.,  {Young} W.,   {Golap} K.,
  2007, in {Shaw} R.~A.,  {Hill} F.,   {Bell} D.~J.,  eds,  Astronomical
  Society of the Pacific Conference Series Vol. 376, Astronomical Data Analysis
  Software and Systems XVI. p.~127

\bibitem[\protect\citeauthoryear{{Merlin} et~al.,}{{Merlin}
  et~al.}{2019}]{merlin2019}
{Merlin} E.,  et~al., 2019, \mn@doi [\mnras] {10.1093/mnras/stz2615}, \href
  {https://ui.adsabs.harvard.edu/abs/2019MNRAS.490.3309M} {490, 3309}

\bibitem[\protect\citeauthoryear{{Millard} et~al.,}{{Millard}
  et~al.}{2020}]{millard2020}
{Millard} J.~S.,  et~al., 2020, \mn@doi [\mnras] {10.1093/mnras/staa609}, \href
  {https://ui.adsabs.harvard.edu/abs/2020MNRAS.494..293M} {494, 293}

\bibitem[\protect\citeauthoryear{{Mizukoshi} et~al.,}{{Mizukoshi}
  et~al.}{2021}]{mizukoshi2021}
{Mizukoshi} S.,  et~al., 2021, arXiv e-prints, \href
  {https://ui.adsabs.harvard.edu/abs/2021arXiv210508894M} {p. arXiv:2105.08894}

\bibitem[\protect\citeauthoryear{{Moss}}{{Moss}}{2020}]{moss2020}
{Moss} A.,  2020, \mn@doi [\mnras] {10.1093/mnras/staa1469}, \href
  {https://ui.adsabs.harvard.edu/abs/2020MNRAS.496..328M} {496, 328}

\bibitem[\protect\citeauthoryear{{Narayanan} et~al.,}{{Narayanan}
  et~al.}{2015}]{Narayanan2015}
{Narayanan} D.,  et~al., 2015, \mn@doi [\nat] {10.1038/nature15383}, \href
  {https://ui.adsabs.harvard.edu/abs/2015Natur.525..496N} {525, 496}

\bibitem[\protect\citeauthoryear{{Neeleman}, {Prochaska}, {Kanekar}  \&
  {Rafelski}}{{Neeleman} et~al.}{2020}]{neeleman2020}
{Neeleman} M.,  {Prochaska} J.~X.,  {Kanekar} N.,   {Rafelski} M.,  2020,
  \mn@doi [\nat] {10.1038/s41586-020-2276-y}, \href
  {https://ui.adsabs.harvard.edu/abs/2020Natur.581..269N} {581, 269}

\bibitem[\protect\citeauthoryear{{Negrello} et~al.,}{{Negrello}
  et~al.}{2017}]{negrello2017}
{Negrello} M.,  et~al., 2017, \mn@doi [\mnras] {10.1093/mnras/stw2911}, \href
  {https://ui.adsabs.harvard.edu/abs/2017MNRAS.465.3558N} {465, 3558}

\bibitem[\protect\citeauthoryear{{Neistein} \& {Dekel}}{{Neistein} \&
  {Dekel}}{2008}]{Neistein2008}
{Neistein} E.,  {Dekel} A.,  2008, \mn@doi [\mnras]
  {10.1111/j.1365-2966.2007.12570.x}, \href
  {https://ui.adsabs.harvard.edu/abs/2008MNRAS.383..615N} {383, 615}

\bibitem[\protect\citeauthoryear{{Nightingale}, {Dye}  \&
  {Massey}}{{Nightingale} et~al.}{2018}]{nightingale2018}
{Nightingale} J.~W.,  {Dye} S.,   {Massey} R.~J.,  2018, \mn@doi [\mnras]
  {10.1093/mnras/sty1264}, \href
  {https://ui.adsabs.harvard.edu/abs/2018MNRAS.478.4738N} {478, 4738}

\bibitem[\protect\citeauthoryear{{Omont} et~al.,}{{Omont}
  et~al.}{2013}]{omont2013}
{Omont} A.,  et~al., 2013, \mn@doi [\aap] {10.1051/0004-6361/201220811}, \href
  {https://ui.adsabs.harvard.edu/abs/2013A&A...551A.115O} {551, A115}

\bibitem[\protect\citeauthoryear{{Papadopoulos} \& {Greve}}{{Papadopoulos} \&
  {Greve}}{2004}]{papadop2004b}
{Papadopoulos} P.~P.,  {Greve} T.~R.,  2004, \mn@doi [\apjl] {10.1086/426059},
  \href {https://ui.adsabs.harvard.edu/abs/2004ApJ...615L..29P} {615, L29}

\bibitem[\protect\citeauthoryear{{Papadopoulos}, {Thi}  \&
  {Viti}}{{Papadopoulos} et~al.}{2004}]{papadop2004}
{Papadopoulos} P.~P.,  {Thi} W.~F.,   {Viti} S.,  2004, \mn@doi [\mnras]
  {10.1111/j.1365-2966.2004.07762.x}, \href
  {https://ui.adsabs.harvard.edu/abs/2004MNRAS.351..147P} {351, 147}

\bibitem[\protect\citeauthoryear{{Papadopoulos}, {van der Werf}, {Xilouris},
  {Isaak}, {Gao}  \& {M{\"u}hle}}{{Papadopoulos} et~al.}{2012a}]{papadop2012}
{Papadopoulos} P.~P.,  {van der Werf} P.~P.,  {Xilouris} E.~M.,  {Isaak} K.~G.,
   {Gao} Y.,   {M{\"u}hle} S.,  2012a, \mn@doi [\mnras]
  {10.1111/j.1365-2966.2012.21001.x}, \href
  {https://ui.adsabs.harvard.edu/abs/2012MNRAS.426.2601P} {426, 2601}

\bibitem[\protect\citeauthoryear{{Papadopoulos}, {van der Werf}, {Xilouris},
  {Isaak}, {Gao}  \& {M{\"u}hle}}{{Papadopoulos} et~al.}{2012b}]{papadop2012b}
{Papadopoulos} P.~P.,  {van der Werf} P.~P.,  {Xilouris} E.~M.,  {Isaak} K.~G.,
   {Gao} Y.,   {M{\"u}hle} S.,  2012b, \mn@doi [\mnras]
  {10.1111/j.1365-2966.2012.21001.x}, \href
  {https://ui.adsabs.harvard.edu/abs/2012MNRAS.426.2601P} {426, 2601}

\bibitem[\protect\citeauthoryear{{Papadopoulos}, {van der Werf}, {Xilouris},
  {Isaak}  \& {Gao}}{{Papadopoulos} et~al.}{2012c}]{papadop2012a}
{Papadopoulos} P.~P.,  {van der Werf} P.,  {Xilouris} E.,  {Isaak} K.~G.,
  {Gao} Y.,  2012c, \mn@doi [\apj] {10.1088/0004-637X/751/1/10}, \href
  {https://ui.adsabs.harvard.edu/abs/2012ApJ...751...10P} {751, 10}

\bibitem[\protect\citeauthoryear{{Peng}, {Ho}, {Impey}  \& {Rix}}{{Peng}
  et~al.}{2002}]{peng2002}
{Peng} C.~Y.,  {Ho} L.~C.,  {Impey} C.~D.,   {Rix} H.-W.,  2002, \mn@doi [\aj]
  {10.1086/340952}, \href
  {https://ui.adsabs.harvard.edu/abs/2002AJ....124..266P} {124, 266}

\bibitem[\protect\citeauthoryear{{Peng}, {Maiolino}  \& {Cochrane}}{{Peng}
  et~al.}{2015}]{peng2015}
{Peng} Y.,  {Maiolino} R.,   {Cochrane} R.,  2015, \mn@doi [\nat]
  {10.1038/nature14439}, \href
  {https://ui.adsabs.harvard.edu/abs/2015Natur.521..192P} {521, 192}

\bibitem[\protect\citeauthoryear{{P{\'e}roux} \& {Howk}}{{P{\'e}roux} \&
  {Howk}}{2020}]{peroux2020}
{P{\'e}roux} C.,  {Howk} J.~C.,  2020, \mn@doi [\araa]
  {10.1146/annurev-astro-021820-120014}, \href
  {https://ui.adsabs.harvard.edu/abs/2020ARA&A..58..363P} {58, 363}

\bibitem[\protect\citeauthoryear{{Pineda}, {Langer}, {Velusamy}  \&
  {Goldsmith}}{{Pineda} et~al.}{2013}]{pineda2013}
{Pineda} J.~L.,  {Langer} W.~D.,  {Velusamy} T.,   {Goldsmith} P.~F.,  2013,
  \mn@doi [\aap] {10.1051/0004-6361/201321188}, \href
  {https://ui.adsabs.harvard.edu/abs/2013A&A...554A.103P} {554, A103}

\bibitem[\protect\citeauthoryear{{Planck Collaboration} et~al.,}{{Planck
  Collaboration} et~al.}{2011}]{planck2011}
{Planck Collaboration} et~al., 2011, \mn@doi [\aap]
  {10.1051/0004-6361/201116479}, \href
  {https://ui.adsabs.harvard.edu/abs/2011A&A...536A..19P} {536, A19}

\bibitem[\protect\citeauthoryear{{Planck Collaboration} et~al.,}{{Planck
  Collaboration} et~al.}{2016}]{planck2016}
{Planck Collaboration} et~al., 2016, \mn@doi [\aap]
  {10.1051/0004-6361/201525830}, \href
  {https://ui.adsabs.harvard.edu/abs/2016A&A...594A..13P} {594, A13}

\bibitem[\protect\citeauthoryear{{R{\'e}my-Ruyer} et~al.,}{{R{\'e}my-Ruyer}
  et~al.}{2014}]{remyruyer2014}
{R{\'e}my-Ruyer} A.,  et~al., 2014, \mn@doi [\aap]
  {10.1051/0004-6361/201322803}, \href
  {https://ui.adsabs.harvard.edu/abs/2014A&A...563A..31R} {563, A31}

\bibitem[\protect\citeauthoryear{{Reuter} et~al.,}{{Reuter}
  et~al.}{2020}]{reuter2020}
{Reuter} C.,  et~al., 2020, \mn@doi [\apj] {10.3847/1538-4357/abb599}, \href
  {https://ui.adsabs.harvard.edu/abs/2020ApJ...902...78R} {902, 78}

\bibitem[\protect\citeauthoryear{{Riechers} et~al.,}{{Riechers}
  et~al.}{2013}]{Riechers2013}
{Riechers} D.~A.,  et~al., 2013, \mn@doi [\nat] {10.1038/nature12050}, \href
  {https://ui.adsabs.harvard.edu/abs/2013Natur.496..329R} {496, 329}

\bibitem[\protect\citeauthoryear{{Romano}, {Matteucci}, {Zhang}, {Papadopoulos}
   \& {Ivison}}{{Romano} et~al.}{2017}]{Romano2017}
{Romano} D.,  {Matteucci} F.,  {Zhang} Z.~Y.,  {Papadopoulos} P.~P.,   {Ivison}
  R.~J.,  2017, \mn@doi [\mnras] {10.1093/mnras/stx1197}, \href
  {https://ui.adsabs.harvard.edu/abs/2017MNRAS.470..401R} {470, 401}

\bibitem[\protect\citeauthoryear{{Rowan-Robinson}}{{Rowan-Robinson}}{2000}]{rowanrobinson2000}
{Rowan-Robinson} M.,  2000, \mn@doi [\mnras]
  {10.1046/j.1365-8711.2000.03588.x}, \href
  {https://ui.adsabs.harvard.edu/abs/2000MNRAS.316..885R} {316, 885}

\bibitem[\protect\citeauthoryear{{Rowlands}, {Gomez}, {Dunne},
  {Arag{\'o}n-Salamanca}, {Dye}, {Maddox}, {da Cunha}  \& {van der
  Werf}}{{Rowlands} et~al.}{2014}]{rowlands2014}
{Rowlands} K.,  {Gomez} H.~L.,  {Dunne} L.,  {Arag{\'o}n-Salamanca} A.,  {Dye}
  S.,  {Maddox} S.,  {da Cunha} E.,   {van der Werf} P.,  2014, \mn@doi
  [\mnras] {10.1093/mnras/stu605}, \href
  {https://ui.adsabs.harvard.edu/abs/2014MNRAS.441.1040R} {441, 1040}

\bibitem[\protect\citeauthoryear{{Rybak}, {McKean}, {Vegetti}, {Andreani}  \&
  {White}}{{Rybak} et~al.}{2015a}]{rybak2015a}
{Rybak} M.,  {McKean} J.~P.,  {Vegetti} S.,  {Andreani} P.,   {White} S.~D.~M.,
   2015a, \mn@doi [\mnras] {10.1093/mnrasl/slv058}, \href
  {https://ui.adsabs.harvard.edu/abs/2015MNRAS.451L..40R} {451, L40}

\bibitem[\protect\citeauthoryear{{Rybak}, {Vegetti}, {McKean}, {Andreani}  \&
  {White}}{{Rybak} et~al.}{2015b}]{rybak2015b}
{Rybak} M.,  {Vegetti} S.,  {McKean} J.~P.,  {Andreani} P.,   {White} S.~D.~M.,
   2015b, \mn@doi [\mnras] {10.1093/mnrasl/slv092}, \href
  {https://ui.adsabs.harvard.edu/abs/2015MNRAS.453L..26R} {453, L26}

\bibitem[\protect\citeauthoryear{{Salpeter}}{{Salpeter}}{1955}]{Salpeter1955}
{Salpeter} E.~E.,  1955, \mn@doi [\apj] {10.1086/145971}, \href
  {https://ui.adsabs.harvard.edu/abs/1955ApJ...121..161S} {121, 161}

\bibitem[\protect\citeauthoryear{{Santini} et~al.,}{{Santini}
  et~al.}{2014}]{santini2014}
{Santini} P.,  et~al., 2014, \mn@doi [\aap] {10.1051/0004-6361/201322835},
  \href {https://ui.adsabs.harvard.edu/abs/2014A&A...562A..30S} {562, A30}

\bibitem[\protect\citeauthoryear{{Sault}, {Teuben}  \& {Wright}}{{Sault}
  et~al.}{2011}]{sault2011}
{Sault} R.~J.,  {Teuben} P.,   {Wright} M. C.~H.,  2011, {MIRIAD: Multi-channel
  Image Reconstruction, Image Analysis, and Display} (\mn@eprint {ascl}
  {1106.007})

\bibitem[\protect\citeauthoryear{{Schneider} \& {Sluse}}{{Schneider} \&
  {Sluse}}{2013}]{Schneider2013}
{Schneider} P.,  {Sluse} D.,  2013, \mn@doi [\aap]
  {10.1051/0004-6361/201321882}, \href
  {https://ui.adsabs.harvard.edu/abs/2013A&A...559A..37S} {559, A37}

\bibitem[\protect\citeauthoryear{{Scoville} et~al.,}{{Scoville}
  et~al.}{2014}]{scoville2014}
{Scoville} N.,  et~al., 2014, \mn@doi [\apj] {10.1088/0004-637X/783/2/84},
  \href {https://ui.adsabs.harvard.edu/abs/2014ApJ...783...84S} {783, 84}

\bibitem[\protect\citeauthoryear{{Scoville} et~al.,}{{Scoville}
  et~al.}{2016}]{scoville2016}
{Scoville} N.,  et~al., 2016, \mn@doi [\apj] {10.3847/0004-637X/820/2/83},
  \href {https://ui.adsabs.harvard.edu/abs/2016ApJ...820...83S} {820, 83}

\bibitem[\protect\citeauthoryear{{Scoville} et~al.,}{{Scoville}
  et~al.}{2017}]{scoville2017}
{Scoville} N.,  et~al., 2017, \mn@doi [\apj] {10.3847/1538-4357/aa61a0}, \href
  {https://ui.adsabs.harvard.edu/abs/2017ApJ...837..150S} {837, 150}

\bibitem[\protect\citeauthoryear{{Serjeant}}{{Serjeant}}{2012}]{serjeant2012}
{Serjeant} S.,  2012, \mn@doi [\mnras] {10.1111/j.1365-2966.2012.20761.x},
  \href {https://ui.adsabs.harvard.edu/abs/2012MNRAS.424.2429S} {424, 2429}

\bibitem[\protect\citeauthoryear{{Smith} et~al.,}{{Smith}
  et~al.}{2013}]{smith2013}
{Smith} D.~J.~B.,  et~al., 2013, \mn@doi [\mnras] {10.1093/mnras/stt1737},
  \href {https://ui.adsabs.harvard.edu/abs/2013MNRAS.436.2435S} {436, 2435}

\bibitem[\protect\citeauthoryear{{Solomon} \& {Vanden Bout}}{{Solomon} \&
  {Vanden Bout}}{2005}]{solomon2005}
{Solomon} P.~M.,  {Vanden Bout} P.~A.,  2005, \mn@doi [\araa]
  {10.1146/annurev.astro.43.051804.102221}, \href
  {https://ui.adsabs.harvard.edu/abs/2005ARA&A..43..677S} {43, 677}

\bibitem[\protect\citeauthoryear{{Spilker} et~al.,}{{Spilker}
  et~al.}{2014}]{spilker2014}
{Spilker} J.~S.,  et~al., 2014, \mn@doi [\apj] {10.1088/0004-637X/785/2/149},
  \href {https://ui.adsabs.harvard.edu/abs/2014ApJ...785..149S} {785, 149}

\bibitem[\protect\citeauthoryear{{Stewart}, {Blyth}  \& {de Blok}}{{Stewart}
  et~al.}{2014}]{stewart2014}
{Stewart} I.~M.,  {Blyth} S.~L.,   {de Blok} W.~J.~G.,  2014, \mn@doi [\aap]
  {10.1051/0004-6361/201423602}, \href
  {https://ui.adsabs.harvard.edu/abs/2014A&A...567A..61S} {567, A61}

\bibitem[\protect\citeauthoryear{{Stutzki} et~al.,}{{Stutzki}
  et~al.}{1997}]{stutzki1997}
{Stutzki} J.,  et~al., 1997, \mn@doi [\apjl] {10.1086/310514}, \href
  {https://ui.adsabs.harvard.edu/abs/1997ApJ...477L..33S} {477, L33}

\bibitem[\protect\citeauthoryear{{Swinbank} et~al.,}{{Swinbank}
  et~al.}{2010}]{Swinbank2010}
{Swinbank} A.~M.,  et~al., 2010, \mn@doi [\nat] {10.1038/nature08880}, \href
  {https://ui.adsabs.harvard.edu/abs/2010Natur.464..733S} {464, 733}

\bibitem[\protect\citeauthoryear{{Swinbank} et~al.,}{{Swinbank}
  et~al.}{2015}]{swinbank2015}
{Swinbank} A.~M.,  et~al., 2015, \mn@doi [\apjl] {10.1088/2041-8205/806/1/L17},
  \href {https://ui.adsabs.harvard.edu/abs/2015ApJ...806L..17S} {806, L17}

\bibitem[\protect\citeauthoryear{{Tacconi} et~al.,}{{Tacconi}
  et~al.}{2008}]{Tacconi2008}
{Tacconi} L.~J.,  et~al., 2008, \mn@doi [\apj] {10.1086/587168}, \href
  {https://ui.adsabs.harvard.edu/abs/2008ApJ...680..246T} {680, 246}

\bibitem[\protect\citeauthoryear{{Tacconi} et~al.,}{{Tacconi}
  et~al.}{2018}]{tacconi2018}
{Tacconi} L.~J.,  et~al., 2018, \mn@doi [\apj] {10.3847/1538-4357/aaa4b4},
  \href {https://ui.adsabs.harvard.edu/abs/2018ApJ...853..179T} {853, 179}

\bibitem[\protect\citeauthoryear{{Talia} et~al.,}{{Talia}
  et~al.}{2018}]{tali18}
{Talia} M.,  et~al., 2018, \mn@doi [\mnras] {10.1093/mnras/sty481}, \href
  {https://ui.adsabs.harvard.edu/abs/2018MNRAS.476.3956T} {476, 3956}

\bibitem[\protect\citeauthoryear{{Toft} et~al.,}{{Toft}
  et~al.}{2014}]{Toft2014}
{Toft} S.,  et~al., 2014, \mn@doi [\apj] {10.1088/0004-637X/782/2/68}, \href
  {https://ui.adsabs.harvard.edu/abs/2014ApJ...782...68T} {782, 68}

\bibitem[\protect\citeauthoryear{{Tomassetti}, {Porciani}, {Romano-Diaz},
  {Ludlow}  \& {Papadopoulos}}{{Tomassetti} et~al.}{2014}]{tomassetti2014}
{Tomassetti} M.,  {Porciani} C.,  {Romano-Diaz} E.,  {Ludlow} A.~D.,
  {Papadopoulos} P.~P.,  2014, \mn@doi [\mnras] {10.1093/mnrasl/slu137}, \href
  {https://ui.adsabs.harvard.edu/abs/2014MNRAS.445L.124T} {445, L124}

\bibitem[\protect\citeauthoryear{{Toomre}}{{Toomre}}{1964}]{toomre1964}
{Toomre} A.,  1964, \mn@doi [\apj] {10.1086/147861}, \href
  {https://ui.adsabs.harvard.edu/abs/1964ApJ...139.1217T} {139, 1217}

\bibitem[\protect\citeauthoryear{{Valentino} et~al.,}{{Valentino}
  et~al.}{2020}]{valentino2020}
{Valentino} F.,  et~al., 2020, \mn@doi [\apj] {10.3847/1538-4357/ab64dc}, \href
  {https://ui.adsabs.harvard.edu/abs/2020ApJ...889...93V} {889, 93}

\bibitem[\protect\citeauthoryear{{Vieira} et~al.,}{{Vieira}
  et~al.}{2013}]{vieira2013}
{Vieira} J.~D.,  et~al., 2013, \mn@doi [\nat] {10.1038/nature12001}, \href
  {https://ui.adsabs.harvard.edu/abs/2013Natur.495..344V} {495, 344}

\bibitem[\protect\citeauthoryear{{Walter}, {Riechers}, {Cox}, {Neri},
  {Carilli}, {Bertoldi}, {Weiss}  \& {Maiolino}}{{Walter}
  et~al.}{2009}]{Walter2009}
{Walter} F.,  {Riechers} D.,  {Cox} P.,  {Neri} R.,  {Carilli} C.,  {Bertoldi}
  F.,  {Weiss} A.,   {Maiolino} R.,  2009, \mn@doi [\nat]
  {10.1038/nature07681}, \href
  {https://ui.adsabs.harvard.edu/abs/2009Natur.457..699W} {457, 699}

\bibitem[\protect\citeauthoryear{{Warren} \& {Dye}}{{Warren} \&
  {Dye}}{2003}]{warren2003}
{Warren} S.~J.,  {Dye} S.,  2003, \mn@doi [\apj] {10.1086/375132}, \href
  {https://ui.adsabs.harvard.edu/abs/2003ApJ...590..673W} {590, 673}

\bibitem[\protect\citeauthoryear{{Wechsler} \& {Tinker}}{{Wechsler} \&
  {Tinker}}{2018}]{Wechsler2018}
{Wechsler} R.~H.,  {Tinker} J.~L.,  2018, \mn@doi [\araa]
  {10.1146/annurev-astro-081817-051756}, \href
  {https://ui.adsabs.harvard.edu/abs/2018ARA&A..56..435W} {56, 435}

\bibitem[\protect\citeauthoryear{{Whiting}}{{Whiting}}{2012}]{whit12}
{Whiting} M.~T.,  2012, \mn@doi [\mnras] {10.1111/j.1365-2966.2012.20548.x},
  \href {https://ui.adsabs.harvard.edu/abs/2012MNRAS.421.3242W} {421, 3242}

\bibitem[\protect\citeauthoryear{{Yang}, {Gao}, {Omont}, {Liu}, {Isaak},
  {Downes}, {van der Werf}  \& {Lu}}{{Yang} et~al.}{2013}]{yang2013}
{Yang} C.,  {Gao} Y.,  {Omont} A.,  {Liu} D.,  {Isaak} K.~G.,  {Downes} D.,
  {van der Werf} P.~P.,   {Lu} N.,  2013, \mn@doi [\apjl]
  {10.1088/2041-8205/771/2/L24}, \href
  {https://ui.adsabs.harvard.edu/abs/2013ApJ...771L..24Y} {771, L24}

\bibitem[\protect\citeauthoryear{{Yang} et~al.,}{{Yang}
  et~al.}{2016}]{yang2016}
{Yang} C.,  et~al., 2016, \mn@doi [\aap] {10.1051/0004-6361/201628160}, \href
  {https://ui.adsabs.harvard.edu/abs/2016A&A...595A..80Y} {595, A80}

\bibitem[\protect\citeauthoryear{{Yang} et~al.,}{{Yang}
  et~al.}{2017}]{yang2017}
{Yang} C.,  et~al., 2017, \mn@doi [\aap] {10.1051/0004-6361/201731391}, \href
  {https://ui.adsabs.harvard.edu/abs/2017A&A...608A.144Y} {608, A144}

\bibitem[\protect\citeauthoryear{{Yoshida}, {Stoehr}, {Springel}  \&
  {White}}{{Yoshida} et~al.}{2002}]{Yoshida2002}
{Yoshida} N.,  {Stoehr} F.,  {Springel} V.,   {White} S. D.~M.,  2002, \mn@doi
  [\mnras] {10.1046/j.1365-8711.2002.05661.x}, \href
  {https://ui.adsabs.harvard.edu/abs/2002MNRAS.335..762Y} {335, 762}

\bibitem[\protect\citeauthoryear{{Zanella} et~al.,}{{Zanella}
  et~al.}{2018}]{zanella2018}
{Zanella} A.,  et~al., 2018, \mn@doi [\mnras] {10.1093/mnras/sty2394}, \href
  {https://ui.adsabs.harvard.edu/abs/2018MNRAS.481.1976Z} {481, 1976}

\bibitem[\protect\citeauthoryear{{Zhang}, {Papadopoulos}, {Ivison}, {Galametz},
  {Smith}  \& {Xilouris}}{{Zhang} et~al.}{2016}]{Zhang2016}
{Zhang} Z.-Y.,  {Papadopoulos} P.~P.,  {Ivison} R.~J.,  {Galametz} M.,  {Smith}
  M.~W.~L.,   {Xilouris} E.~M.,  2016, \mn@doi [Royal Society Open Science]
  {10.1098/rsos.160025}, \href
  {https://ui.adsabs.harvard.edu/abs/2016RSOS....360025Z} {3, 160025}

\bibitem[\protect\citeauthoryear{{Zhang}, {Romano}, {Ivison}, {Papadopoulos}
  \& {Matteucci}}{{Zhang} et~al.}{2018}]{Zhang2018}
{Zhang} Z.-Y.,  {Romano} D.,  {Ivison} R.~J.,  {Papadopoulos} P.~P.,
  {Matteucci} F.,  2018, \mn@doi [\nat] {10.1038/s41586-018-0196-x}, \href
  {https://ui.adsabs.harvard.edu/abs/2018Natur.558..260Z} {558, 260}

\bibitem[\protect\citeauthoryear{{Zhukovska}}{{Zhukovska}}{2014}]{Zhukovska2014}
{Zhukovska} S.,  2014, \mn@doi [\aap] {10.1051/0004-6361/201322989}, \href
  {http://adsabs.harvard.edu/abs/2014A%26A...562A..76Z} {562, A76}

\bibitem[\protect\citeauthoryear{{da Cunha}, {Charlot}  \& {Elbaz}}{{da Cunha}
  et~al.}{2008}]{dacunha2008}
{da Cunha} E.,  {Charlot} S.,   {Elbaz} D.,  2008, \mn@doi [\mnras]
  {10.1111/j.1365-2966.2008.13535.x}, \href
  {https://ui.adsabs.harvard.edu/abs/2008MNRAS.388.1595D} {388, 1595}

\bibitem[\protect\citeauthoryear{{da Cunha} et~al.,}{{da Cunha}
  et~al.}{2013}]{daCunha2013}
{da Cunha} E.,  et~al., 2013, \mn@doi [\apj] {10.1088/0004-637X/766/1/13},
  \href {https://ui.adsabs.harvard.edu/abs/2013ApJ...766...13D} {766, 13}

\bibitem[\protect\citeauthoryear{{van der Werf} et~al.,}{{van der Werf}
  et~al.}{2011}]{vanderwerf2011}
{van der Werf} P.~P.,  et~al., 2011, \mn@doi [\apjl]
  {10.1088/2041-8205/741/2/L38}, \href
  {https://ui.adsabs.harvard.edu/abs/2011ApJ...741L..38V} {741, L38}

\makeatother
\end{thebibliography}



\appendix

\section{Stellar mass estimation}
\label{sec_app1}

Our estimate of the stellar mass of ID141 follows the procedure used by \cite{hopwood2011} who used {\tt MAGPHYS} to fit an SED to observed optical, mid-IR and far-IR photometry. In this paper, we have used version 2 of {\tt MAGPHYS} \citep[see][]{battisti2020}.

\begin{figure*}
	\includegraphics[width=14cm]{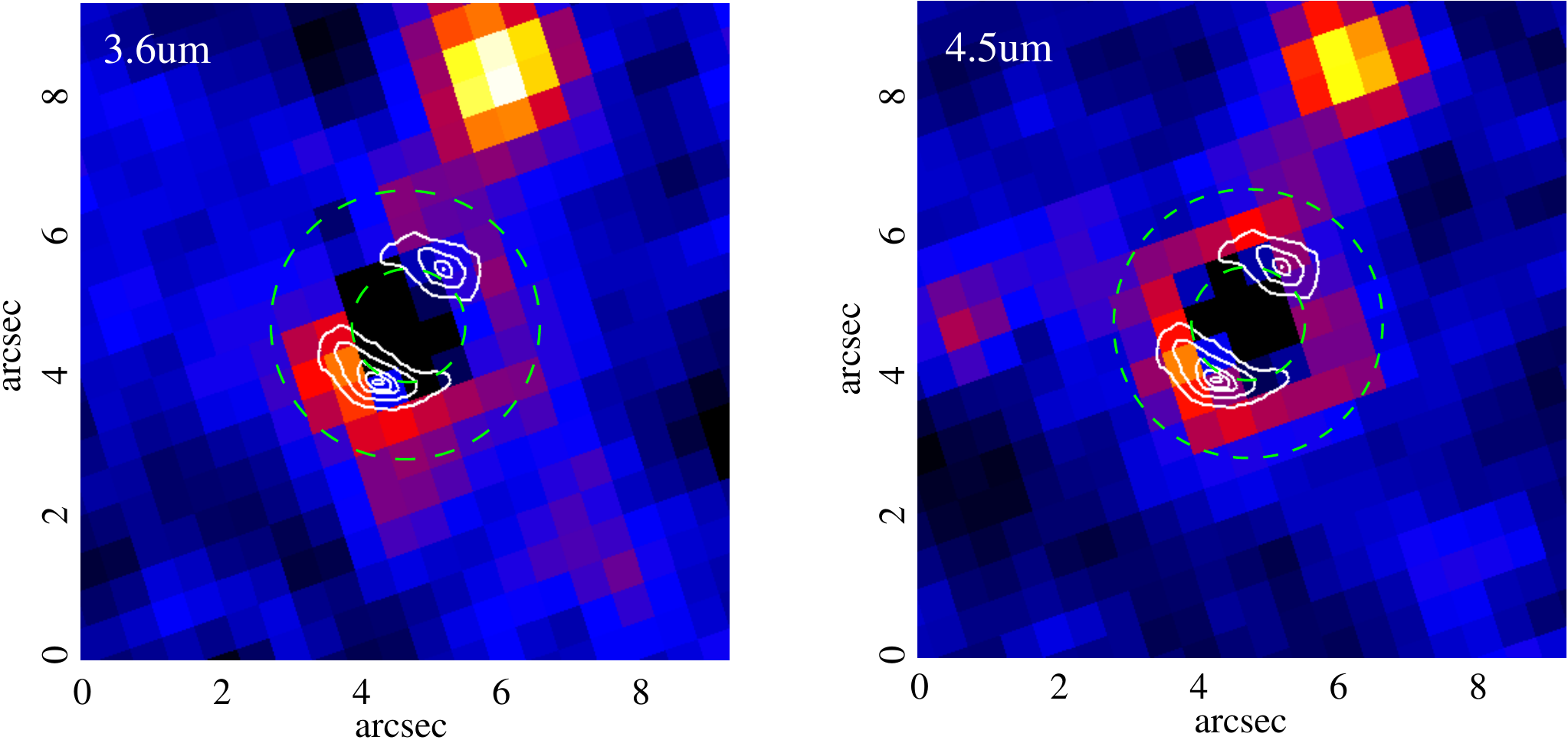}
    \caption{Residual flux remaining at 3.6\,$\mu$m and 4.5\,$\mu$m after subtraction of the model lens light profiles generated using {\tt GALFIT} (see main text). The white contours show the ALMA band 4 continuum flux and the green dashed lines denote the annulus within which the mid-IR flux was extracted.}
    \label{irac_resids}
\end{figure*}

The dominant constraints on the SED in the mid-IR region are provided by Spitzer Space Telescope data which were obtained under proposal ID 80156 using the Infra Red Array Camera (IRAC) operating at 3.6\,$\mu$m and 4.5\,$\mu$m. To obtain a measure of the source emission, it was first necessary to remove the lens emission since the two are blended due to the relatively large IRAC PSF. This was done by first determining the surface brightness profiles of the two foreground galaxies using the $K_{\rm s}$-band data presented in \cite{bussmann2012} using {\tt GALFIT} \citep{peng2002}. Subtracting the model profiles of these two well-resolved foreground galaxies which lie well within the lensed source light observed by ALMA left only negligible residuals and no evidence of any source emission. We then used these profiles to construct model surface brightness profiles of the lens galaxies as they would be observed by IRAC at 3.6\,$\mu$m and 4.5\,$\mu$m. To achieve this, we extracted the PSF at both wavelengths using five stars in the field of view of the IRAC observations and convolved these with the $K_{\rm s}$-band profiles. Subtracting these model light profiles from the observed IRAC profiles left residuals which resemble the lensed source image as seen by ALMA (see Fig.~\ref{irac_resids}). 

To test whether these residuals were the result of a poorly-estimated PSF or because the mid-IR does not closely follow the near-IR light, we repeated the procedure for three galaxies of similar flux and morphology selected at random within the field of view common to both the $K_{\rm s}$-band and IRAC observations. Removing the model IRAC profiles left no significant residuals in any of the three galaxies we tested. We therefore assumed that the residuals left from removing the lens profiles were indeed the source image although we can not rule out the possibility that the light profile of the lens in the near-IR is different to that in the mid-IR in the ID141 system. To obtain a final measure of the lensed source IRAC flux, we summed the residuals within an annulus surrounding the observed residual flux as shown in Fig.~\ref{irac_resids}. This resulted in fluxes of $8.7\pm2.2\,\mu$Jy and $8.1\pm1.8\,\mu$Jy at 3.6\,$\mu$m and 4.5\,$\mu$m respectively. The errors quoted here are estimated from the image noise and do not include any uncertainty arising from lens light removal.

\begin{figure*}
	\includegraphics[width=17.8cm]{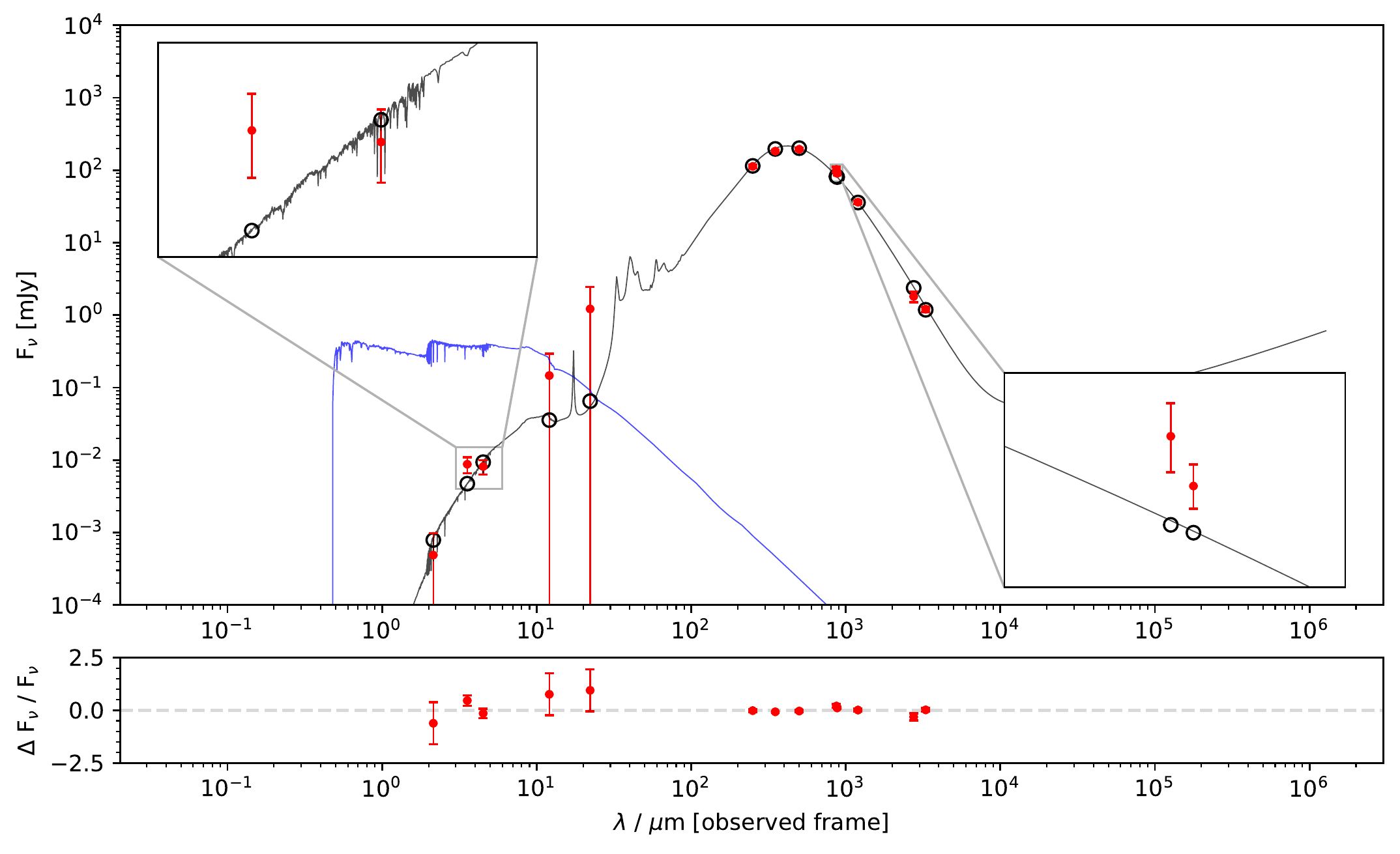}
    \caption{{\tt MAGPHYS} SED (black line) fitted to Keck $K_{\rm s}$ band, Spitzer-IRAC (3.6\,$\mu$m and 4.5\,$\mu$m), WISE and far-IR photometry as described in main text. The unattenuated spectrum is shown as the blue line. Red points and black circles show observed and predicted fluxes respectively. The bottom plot shows the residuals calculated as observed minus predicted flux. Note that all photometry in this plot is as observed, i.e., not lens de-magnified.}
    \label{magphys_sed}
\end{figure*}

To constrain the far-IR part of the SED fitted by {\tt MAGPHYS}, we used the photometry as discussed in Section \ref{sec_sed}. The optical/near-IR part of the SED was constrained using the Keck $K_{\rm s}$-band data. Since the lensed source was not detected in the lens-subtracted $K_{\rm s}$-band image, we set the flux in this band to be three times the standard deviation of the background flux and assumed a generous $1\sigma$ error equal to this value. In this way, the $K_{\rm s}$ flux acts like an upper limit in the SED fitting.  Finally, we also included Wide-field Infrared Survey Explorer fluxes at 12\,$\mu$m and 22\,$\mu$m \citep[WISE;][]{lang2016}. Since the WISE fluxes are an unknown combination of the lens and source light, we attributed 50 per cent of the flux in each passband to the source and took a $1\sigma$ error equal to this value so that again, the WISE data act like upper limits. To convert the quantities derived by {\tt MAGPHYS} from observed (i.e., lensed) to intrinsic quantities of the unlensed source, we assumed a magnification of 5.8, equal to the mean of the continuum magnifications given in Table~\ref{tab_magns}. The resulting stellar mass we obtained is $3.4^{+1.8}_{-1.7} \times 10^{11}$\,M$_\odot$ where the quoted $1\sigma$ errors are taken from the distribution of masses output by {\tt MAGPHYS}. Fig.~\ref{magphys_sed} shows the corresponding {\tt MAGPHYS} SED.

\section{Chemical Evolution Modelling}
\label{app_chem_evol}

In this Section, we describe a one-zone chemical evolution model that we developed to compare the evolution of the interstellar dust, gas and stellar mass of a ID141-like system. The model is based on \citet{DeVis2017b}: the galaxy starts as a cloud of gas, which is converted into stars via the initial mass function and ongoing star formation. It includes a realistic description of the amount of gas flowing in (accretion from cosmic web) and out (supernovae and AGN) from the galaxy.

The dust prescription accounts for dust formation by stars (both low-intermediate mass stars in their asymptotic giant branch phase and massive-star supernovae). It also includes dust grain growth in the diffuse and dense environments of the galaxy, thought to be a dominant production route of dust in galaxies above a metallicity threshold \citep{Asano2013,dunne2011,Zhukovska2014,DeVis2017b}. Dust is removed from the ISM via astration (locked up in stellar remnants), dust destruction by SN shocks and outflows.  

As detailed in the main text, we used two star formation histories, one with a fixed continuous star formation rate and one with a starburst. For the latter, we assumed that we are currently observing a starburst at 40\,per\,cent of the way through a burst lasting $200\rm \,Myr$. This is motivated by the fact that the duration of high-$z$ starbursts are thought to range from 50-250\,Myr \citep[e.g.][]{Danielson2017}; typical duty cycles of high-redshift SMGs range from 40-100\,Myr \citep{Coppin2008,Toft2014,Narayanan2015}.  

The three chosen IMFs adhere to the form $d N/d \log m \propto m^{-\alpha}$ where $m$ is stellar mass. The Salpeter IMF is described by $\alpha=1.7$. Both the Chabrier and top-heavy IMF of \citet{Zhang2018} are broken power laws with exponents $\alpha_{\rm Chab}=[0.4, 1.35]$ and $\alpha_{\rm TH}=[0.3,1.1]$ (see Table~\ref{tab:imfs}).  Both IMFs are normalised to unity between stellar masses of 0.1\,M$_\odot$ and 100\,M$_\odot$.

\begin{table}
    \centering
    \begin{tabular}{l|c|c|c|c|c|c|} \\ \hline
        IMF & $\alpha_0$ & $\alpha_1$ & $m_0$ & $m_1$ & $m_2$ \\ \hline
         Salpeter & 1.7 & 1.7 & 0.1 & 0.5 & 100 \\
         Chabrier & 0.4 & 1.35 & 0.4& 1.0 & 100 \\
         Top heavy (this work) & 0.3 & 1.1 & 0.1 & 0.5 & 100 \\ \hline
    \end{tabular}
    \caption{The IMFs used in the model to show extremes of the gas and dust evolution. The IMF slopes $\alpha$ are in terms of mass, and $m_0$ and $m_2$ are the upper and lower limits of stellar masses for which the IMF is normalised to unity.  $m_1$ indicates the mass at which the slope may change. The top-heavy IMF used here is from \citet{Zhang2018}, see also \citet{Romano2017,Cai2020}. This is a less extreme form than the \citet{baugh2005} top-heavy IMF, where $\alpha = 0$ across the mass range.}
    \label{tab:imfs}
\end{table}

Following \citet{Zhukovska2014}, we set the gas inflow rate to $I(t) \propto e^{-t/\tau}$ where $\tau$ is the infall timescale.  We normalise this such that the galaxy accretes a total mass of gas $M_{\rm acc}$ over its lifetime (see below).  To check that this is a realistic level of inflow, we compared with the halo baryonic mass accretion rate expected for a massive star-forming galaxy such as ID141 in the following way. Extreme star-forming galaxies with SFRs $>1000$\,M$_{\odot}\,\rm yr^{-1}$ at $z \sim 4$ exist in dark matter halo masses of a few $\times 10^{12}$\,M$_{\odot}$ \citep{Behroozi2013}. The mass of the halo with redshift $M_h(z)$ can be estimated using the prescriptions in \citet{McBride2009} and \citet{Fakhouri2010} where $M_h(z) = M_h(z=0) \times (1+z)^{\beta}e^{-\gamma z}$. $M_h(z=0)$ is the halo mass the galaxy resides in at $z=0$, $\beta$ and $\gamma$ are set to 0.1 and 0.69 following \citet{McBride2009}. The accretion rate for a given halo mass and redshift can be estimated using the relationships from \citet{Forbes2014a} (their Eqns. 15 and 22) and \citet{Neistein2008} where $\dot{M_{\rm ext}} ({\rm M}_{\odot}\,yr^{-1}) = \epsilon_{\rm in} f_b \dot{M_h(z)}$. $f_b$ is the baryonic fraction (0.17) and $\epsilon_{\rm in}$ denotes an efficiency. As an example, a galaxy with $M_h(z=0) = 1\times 10^{14}$\,M$_{\odot}$ \citep{Narayanan2015} would have a mass accretion rate of $\dot{M_{\rm ext}} \sim 700$\,M$_{\odot}\,yr^{-1}$ at $z=4.2$, and $\sim 400$\,M$_{\odot}\,yr^{-1}$ at $z=7$. A current day halo mass of $10^{13}$\,M$_{\odot}$ would have a mass accretion rate of roughly 8 times lower at these redshifts\footnote{Similar results are obtained using the halo mass accretion prescriptions in \citet{Behroozi2013,Wechsler2018,Belfiore2019}.}. For the top heavy IMF with conSF SFR prescription, accretion of a total mass of $5\times 10^{11}$\,M$_{\odot}$ of gas over the lifetime of the galaxy in our model, produces very similar inflow rates to those predicted from the halo mass growth described above. For the Salpeter IMF, the mass accretion rates need to be approximately twice as high to sustain the higher SFRs.

For the star formation histories, we also account for the different redshifts at which different mass halos undergo their peak SFE. Generally, this is at higher redshifts for higher mass halos. We scale the star formation efficiency by redshift such that the peak efficiency occurs at $z=3-5$ but decreases at higher and lower redshifts.  For the halo described above, \citet{Behroozi2013} (their Fig.~8) shows that this varies from $(0.4 - 1)\times \epsilon(z=4)$.  We fit a polynomial to their function.  

We set the dimensionless parameter controlling the rate of dust grain growth in the dense interstellar medium to 7000. Recent work by \citet{DeVis2017b} and De Vis et al. {\it in review} suggests an appropriate range of values for the rate of grain growth is 4000-7000. As discussed in the main text, this was allowed to vary from 700-15,000 in our analysis of ID141. In addition, we allow 60\,per\,cent of the metals in the cloud to be `available' for dust formation in order to produce a higher dust mass (lower gas to dust ratio). This factor is predicted to be $\sim$0.5 in nearby spirals (De Vis et al. {\it in press)}. The grain growth parameter depends on the galaxy metal mass fraction, the current dust-to-metal ratio and the SFE, as well as the fraction of the ISM in dense clouds (we set this to 0.6, \citealt{Inoue2003,Yoshida2002}). Finally, we let the galaxy evolve until (i) the stellar mass reaches $5\times 10^{11}$\,M$_{\odot}$ where we assume an outflow clears the galaxy of the ISM \citep{Romano2017}, (ii) in the case of the starburst model, we stop the model when the burst has ended or (iii) when the gas fraction $f_g$ falls to less than 3\,per\,cent.


\bsp	
\label{lastpage}
\end{document}